\documentclass[twocolumn,pra,aps,showpacs,longbibliography]{revtex4-1}
\usepackage{bm}
\usepackage{graphicx}
\usepackage{amsbsy}
\usepackage{color}
\usepackage[utf8x]{inputenc}
\usepackage{epstopdf}
\usepackage{amsmath,amssymb}

\graphicspath{{pix/}}
\usepackage[caption=false]{subfig}
\def \LNI{{I^{(1)}_{\rm ncl}}}
\def \LNII{{I^{(2)}_{\rm ncl}}}
\def \EI{{I_{\rm ent}}}
\def \GNI{{I_{\rm ncl}}}
\def \LNj{{I^{(j)}_{\rm ncl}}}

\def\<{\langle}
\def\>{\rangle}

\begin{document}

\title{Interplay of nonclassicality and entanglement of two-mode Gaussian
fields generated in optical parametric processes}
\author{Ievgen I. Arkhipov}
\email{ievgen.arkhipov01@upol.cz}
\address{RCPTM, Joint Laboratory of Optics of Palack\'y University and
Institute of Physics of CAS, Palack\'y University, 17. listopadu
12, 771 46 Olomouc, Czech Republic}

\author{Jan Pe\v{r}ina Jr.}
\address{Joint Laboratory of Optics of Palack\'y University and
Institute of Physics of CAS, Institute of Physics of the Czech
Academy of Science, 17. listopadu 50a, 771 46 Olomouc, Czech
Republic}

\author{Jan Pe\v{r}ina}
\address{Joint Laboratory of Optics of Palack\'y University and
Institute of Physics of CAS, Palack\'y University, 17. listopadu
12, 771 46 Olomouc, Czech Republic}

\author{Adam Miranowicz}
\address{Faculty of Physics, Adam Mickiewicz University, PL-61-614 Poznan, Poland}

\begin{abstract}
The behavior of general nonclassical two-mode Gaussian states at a
beam splitter is investigated. Single-mode nonclassicality as well
as two-mode entanglement of both  input and output states are
analyzed suggesting their suitable quantifiers. These quantifiers
are derived from local and global invariants of linear unitary
two-mode transformations such that the sum of input (or output)
local nonclassicality measures and entanglement measure gives a
global invariant. This invariant quantifies the global
nonclassicality resource. Mutual transformations of local
nonclassicalities and entanglement induced by the beam splitter
are analyzed considering incident noisy twin beams, single-mode
noisy squeezed vacuum states, and states encompassing both squeezed states and
twin beams. A rich tapestry of interesting nonclassical output
states is predicted.
\end{abstract}

\maketitle

\section{Introduction}

The nonclassical properties of light have been for a long time the
main topic of interest in quantum optics. The question whether a
given quantum state is nonclassical (i.e., cannot be treated by
the classical statistical theory) has been considered as one of
the most important problems since the early days of quantum
physics~\cite{Einstein05,Einstein35,Schrodinger35} (for a review
see, e.g., Refs.~\cite{Perina1994Book,MandelBook,GlauberBook}). For
optical fields, a commonly accepted criterion for distinguishing
nonclassical states from the classical ones is expressed as
follows~\cite{DodonovBook,VogelBook,MandelBook,Perina1991Book}: a
quantum state is nonclassical if its Glauber-Sudarshan
$P$~function fails to have all the properties of a probability
density. We recall that the Glauber-Sudarshan $P$~function for an
$M$-mode bosonic state $\hat\rho$ can be defined
as~\cite{Glauber63,Sudarshan63}:
\begin{eqnarray}
  \hat\rho &=& \int P(\bm{\alpha,\alpha}^*)|\bm{\alpha}\>\< \bm{\alpha}|{\rm d}^2\bm{\alpha},
  \label{Pfunction}
\end{eqnarray}
where $|\bm{\alpha}\>= \prod_{m=1}^M|\alpha_m\>$ is given in terms
of the $m$th-mode coherent state $|\alpha_m\>$, which is the
eigenstate of the $m$th-mode annihilation operator $\hat a_m$,
$\bm{\alpha}$ denotes complex multivariable
$(\alpha_1,\alpha_2,...,\alpha_M)$, and ${\rm d}^2
\bm{\alpha}=\prod_{m}{\rm d}^2\alpha_m$. It is worth noting that
the negativity of the $P$ function is necessary and sufficient
for nonclassicality, while the singularity or irregularity of the
$P$ function is only a sufficient condition (i.e., it is a
nonclassical witness). Thus, if the $P$ function is more singular
or more irregular than Dirac's $\delta$-function for a given
state, then it is also nonpositive (semidefinite) in the formalism
of generalized functions. A standard example of such irregular
functions is the $P$ function for an $n$-photon Fock state (with
$n=1,2,...$), which is given by the $n$th derivative of
$\delta(\alpha)$.

Based on this definition of nonclassicality, various operational
criteria (also called witnesses) have been described for testing
the nonclassicality of
single-mode~\cite{DodonovBook,VogelBook,Richter02} and
multi-mode~\cite{Miranowicz10,Bartkowiak11,Allevi2013}] fields.
Their derivations are based either on the fields
moments~\cite{Vogel08,Miranowicz10,Allevi2013} or exploit the
Bochner theorem written for the characteristic function of the
Glauber-Sudarshan $P$~function~\cite{Ryl2015}. A direct
reconstruction of the quasi-distributions of integrated
intensities is a sufficient but not necessary condition of the
nonclassicality of the detected fields by the
definition~\cite{Haderka2005a,PerinaJr2012,PerinaJr2013a}. We note
that nonclassicality criteria derived from the majorization theory
have also been found useful~\cite{Lee1990a,Verma10}.

Entanglement between two optical fields is one of the most
frequently studied forms of nonclassical light. Such light emerges
in various two-mode  or multimode nonlinear optical processes, e.g., in
spontaneous parametric down-conversion. In this process, pairs of
photons composed of the signal and  idler modes are created at the
expense of the annihilated pump photons. This pairwise character
of emitted light lies in the heart of entanglement here. The
process of spontaneous parametric down-conversion has its
degenerate variant called second-subharmonic generation, where
 both photons in a pair are
emitted into the same optical mode. This gives raise to phase
squeezing of the second-subharmonic field composed of, in general,
many photon pairs. The squeezed light is also considered
nonclassical as it has its phase fluctuations suppressed
below the classical limit. The nonclassicality in both cases has the
same origin which is pairing of photons. On the other side, the
emitted photon pairs can be manipulated by linear optics. In
detail, two photons from one pair present in the same mode of a
squeezed state of light can be split (on a beam splitter) and
contribute to the entanglement of the output fields. Also, two photons
from a pair incident on different input ports of a beam splitter
can ``stick together'' (bunch) and leave the beam splitter in the same
output port (as testified in the Hong-Ou-Mandel
interferometer~\cite{Hong1987}). The interconnection of these two
types of fields by the means of linear optics has already been
shown by Braunstein~\cite{Braunstein05} { and later elaborated
by Adesso \cite{Adesso06} } for arbitrarily strong Gaussian
states. This behavior poses a natural question whether it is
possible to introduce a physical quantity that quantifies ``a
nonclassicality resource'' present during the creation of both
types of fields { and later conserved during linear-optical
transformations.}

The answer to this question is intimately related to the
quantifiers of {\em entanglement} and {\em local nonclassicality}.
Several measures were proposed to quantify the entanglement in
both  discrete and continuous
domains~\cite{Vidal02,Plenio05,Horodecki09review,Peres96,Horodecki97,Marian08}.
The negativity (or its logarithmic variant) is considered,
probably, as the most useful at present. On the other hand, the
Lee nonclassicality depth~\cite{Lee1990a} is conventionally used
to quantify the nonclassicality of optical fields. Alternatively,
the nonclassicality of an optical field can be transcribed to
entanglement using a beam splitter and quantified via an
entanglement measure~\cite{Asboth05,Brunelli15}. For a comparative study of
these two nonclassicality measures see, e.g., recent
Refs.~\cite{Miran15a,Miran15b}.

We note that, apart from the local nonclassicalities of two parts
of a bipartite state, also  {\em global nonclassicality} can
naturally be defined. All these three quantities have been
analyzed in Ref.~\cite{arkhipov15} for intense multi-mode twin
beams with the following result: whenever a twin beam is entangled
it is globally nonclassical. On the other hand, its signal and
idler constituents are multi-thermal and so locally classical. A
general approach for describing the relation between the
entanglement and global nonclassicality of two-mode states has
been proposed in Ref.~\cite{Vogel14}.

Returning back to our question, we look for an invariant with
respect to linear-unitary transformations (conserving the overall
number of photons) that comprises both the entanglement and local
nonclassicalities. This question has recently been addressed in
Ref.~\cite{Zubairy15} considering beam-splitter transformations
and a quantity composed of the logarithmic negativity and the
logarithm of nonclassicality depth. However, the introduced
quantity has been found useful only under very specific
conditions~\cite{arkhipov_letter16}.

\begin{figure}              
 \includegraphics[width=0.48\textwidth]{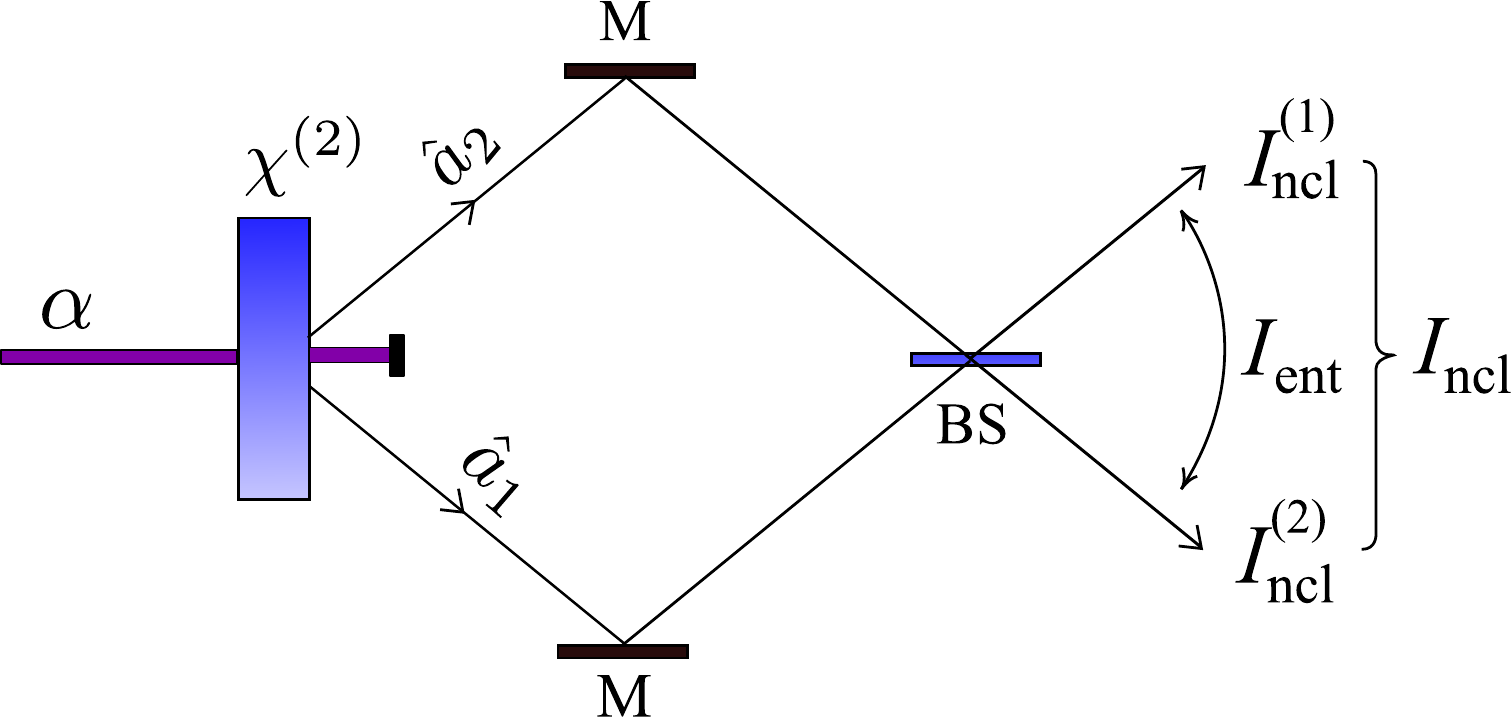}
  \caption{(Color online) Diagram showing the main goal of this
  paper: The local ($\LNI$ and $\LNII$) and global ($\GNI$)
  nonclassicality invariants are analyzed in relation with the
  entanglement, described by the invariant $\EI$, for the light
  generated by the optical parametric process (described by the
second-order susceptibility $\chi^{(2)}$) and then combined at a
  beam splitter BS with varying transmissivity $T$. Here, $\alpha$
is the amplitude of a classical pump field, $\hat a_1$ and $\hat a_2$ are the
annihilation operators of the generated light, and $M$ denotes a mirror.}
\label{fig1}
\end{figure}

In this paper we construct such an invariant for general two-mode
Gaussian states arising in nonlinear processes described by the
second order susceptibility $ \chi^{(2)}$. The processes of
spontaneous parametric down-conversion and second-subharmonic
generation represent their most important examples. As
schematically shown in Fig.~\ref{fig1}, the found invariant is
decomposable into three parts characterizing in turn entanglement
and two local nonclassicalities. The entanglement indicator is
shown to be a monotone of the logarithmic negativity similarly to
the newly defined nonclassicality measure that is a monotone of
the Lee nonclassicality depth under any linear unitary
transformation.

The obtained results are potentially interesting for manipulations
with nonclassicality in quantum engineering that have become substantial ingredients of a
growing number of applications of quantum
technologies~\cite{NielsenBook,Horodecki09review,Polzik92,Ralph98,Wolfgramm10}.

The paper is organized as follows. In Sec.~II, a model comprising
parametric down-conversion and second-subharmonic generation is
developed. A suitable nonclassicality invariant is
suggested using local and global invariants of two-mode Gaussian
fields. Its decomposition into an entanglement quantifier and
local nonclassicality quantifiers is also discussed. Twin beams as
they behave on a beam splitter are discussed in Sec.~III. In
Sec.~IV, a single-mode squeezed state on a beam splitter is
analyzed. Section~V is devoted to the behavior of two single-mode
squeezed states interfering on a beam splitter. States having both
`twin-beam' and squeezed components are investigated in Sec.~VI.
Conclusions are drawn in Sec.~VII. Quasidistributions related to
the normal and symmetric ordering of operators are discussed in the
Appendix.

\section{Gaussian states generated in \protect{$ \chi^{(2)} $} interactions and their invariants}

\begin{figure}              
\includegraphics[width=0.28\textwidth]{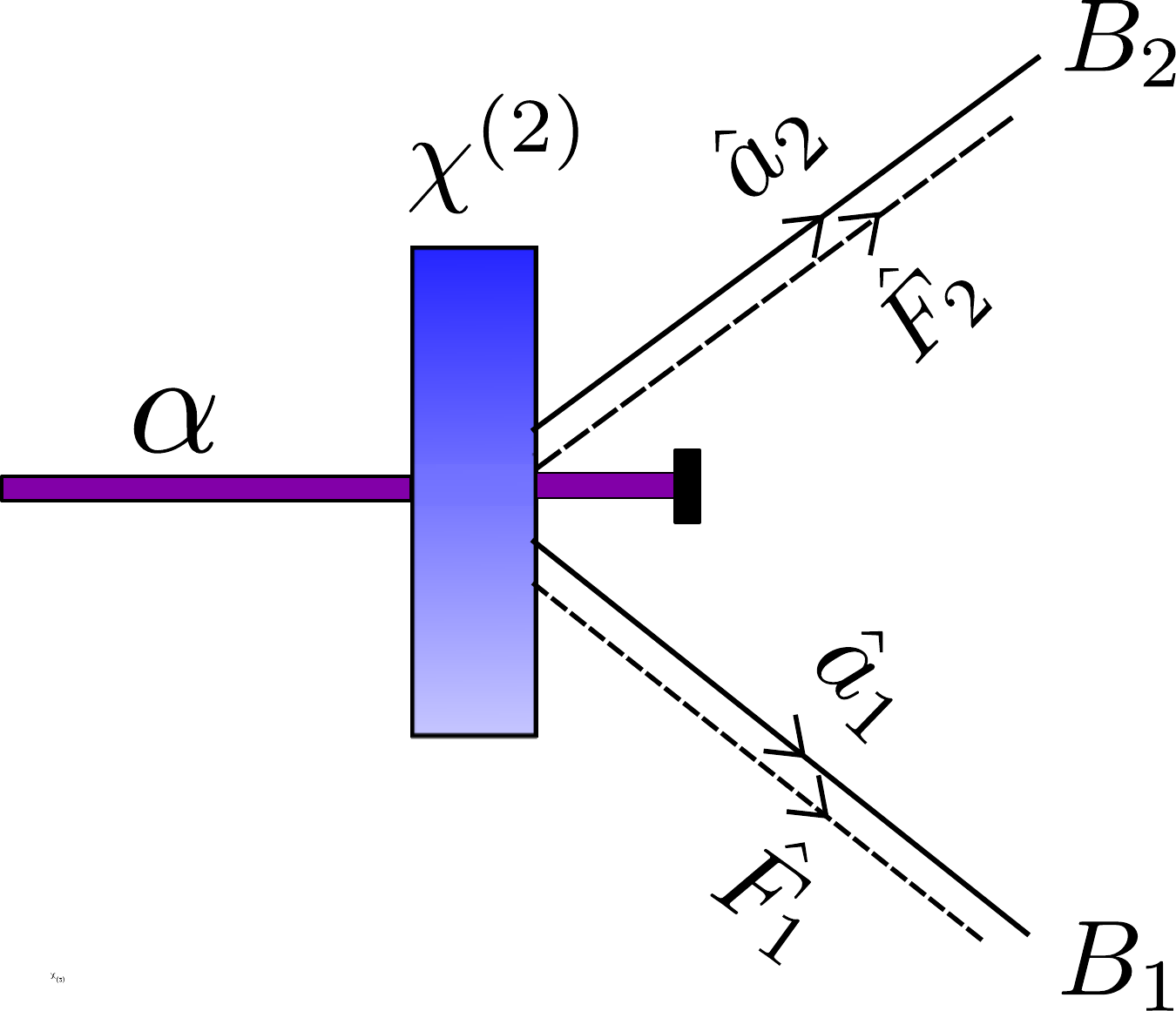}
 \caption{(Color online) Diagram of the optical parametric process
  described by Eq.~(\ref{hamil}): the classical pump field, with
  complex amplitude $\alpha$, generates the signal and idler modes
  described by the annihilation operators $\hat a_j$ and affected by
  the noise stochastic operators $\hat F_j$, $ j=1,2 $. For simplicity,
  the pump-field amplitude $\alpha$ is incorporated into the
  coupling constants $g_{ij}$. The mean photon number in the signal
  (idler) mode influenced by the noise is denoted by $B_1$ ($B_2$).
  In Sec.~III, $B_1=B_{\rm p}+B_{\rm s}$ and $B_2=B_{\rm p}+B_{\rm
  i}$, where $B_{\rm p} = \sinh^2(g_{12}t)$ is the mean number of
  generated photon pairs and $B_{\rm s} = \langle\hat
  F^\dagger_{1}\hat F_{1}\rangle$ ($ B_{\rm i} =\langle\hat
  F^\dagger_{2}\hat F_{2}\rangle $) is the mean number of signal
  (idler) noise photons. In Secs.~IV and V, $B_1=\tilde B^{\rm
  s}_{\rm p}+B_{\rm s}$ and $B_2=\tilde B_{\rm p}^{\rm i}+B_{\rm
  i}$, where $\tilde B_{\rm p}^{\rm s}$ $ (\tilde B_{\rm p}^{\rm
  i})$ is the mean number of squeezed photons in the signal (idler)
  mode.}
\label{fig2}
\end{figure}

We consider a nonlinear interaction Hamiltonian $ \hat H_{\rm int}
$ describing both parametric down-conversion and
second-subharmonic generation that provide photon
pairs~\cite{Perina1991Book} (for the scheme, see Fig.~\ref{fig2}):
\begin{equation}\label{hamil}           
 \hat H_{\mathrm{int}} = -\hbar g_{12}^{\ast}\hat a_{1}\hat a_{2}-\hbar g_{11}^{\ast}\hat a_1^2-
 \hbar g_{22}^{\ast}\hat a_2^2+ {\rm h.c.}
\end{equation}
In Eq.~(\ref{hamil}), the symbols $\hat a_{1}$ $(\hat
a^{\dagger}_{1})$ and $\hat a_{2}$ $(\hat a^{\dagger}_{2})$
represent the annihilation (creation) operators of the fields 1
and 2, $g_{12}$ is a nonlinear coupling constant characterizing
parametric down-conversion and $ g_{ii} $ stands for a nonlinear
coupling constant of the second-subharmonic generation in the
$i$th mode described by the second-order susceptibility $\chi^{(2)}$ of a medium.
 Symbol h.c. represents the Hermitian conjugated
terms. Due to the presence of noise in real nonlinear processes we
also consider the Langevin forces $\hat L_j $ arising in the
interaction with the reservoir chaotic oscillators characterized by
means of noise photon numbers $\langle n_{d}\rangle$. This leads to
damping processes described by the damping constants $\gamma_j$.

The Heisenberg-Langevin operator equations corresponding to the
Hamiltonian $ \hat H_{\mathrm{int}} $ are derived in the following
matrix form:
\begin{eqnarray} \label{hle}    
 \frac{d\hat{\bf a}}{dt} &=& {\bf M} \hat{\bf a} + \hat{\bf
 L}
\end{eqnarray}
using the vectors $\hat{\bf a} = \left(\hat a_1,\hat
a_1^{\dagger},\hat a_2,\hat a_2^{\dagger}\right)^T$ and $\hat{\bf
L} = \left(\hat L_1,\hat L_1^{\dagger},\hat L_2,\hat
L_2^{\dagger}\right)^T$, and the matrix
\begin{equation}    
 {\bf M}= \left[\begin{array}{cccc}
- \gamma_{1}/2 &2ig_{11} & 0& ig_{12} \\
-2ig_{11} & -\gamma_1/2 & -ig_{12} &0 \\
0 &ig_{12} &-\gamma_2/2& 2ig_{22} \\
 -ig_{12} &  0 & -2ig_{22} & -\gamma_2/2
\end{array}\right].
\end{equation}
The Langevin operators $\hat L_1 $ and $\hat L_2 $ introduced in
Eq.~(\ref{hle}) obey the following relations:
\begin{eqnarray}            
\langle\hat L_{i}(t)\rangle &=& \langle\hat
L^{\dagger}_{i}(t)\rangle= 0, \nonumber \\
 \langle\hat L^{\dagger}_{i}(t)\hat L_{j}(t')\rangle &=&
\delta_{ij}\langle n_{d}\rangle\delta(t-t'), \nonumber \\
\langle\hat L_{i}(t)\hat L^{\dagger}_{j}(t')\rangle
&=&\delta_{ij}\big( \langle n_{d}\rangle +1\big)\delta(t-t'),
\end{eqnarray}
where $\delta_{ij}$ stands for the Kronecker symbol and $\delta$ denotes
the Dirac delta function.

The solution of Eq.~(\ref{hle}) for the operators $ \hat a_1 $ and
$ \hat a_2 $ is conveniently written in the following matrix form
using suitable evolution matrices $ {\bf U} $ and $ {\bf V} $ and a
stochastic operator vector $ \hat{\bf F} $ (for details, see,
e.g.~\cite{PerinaJr2000}):
\begin{eqnarray} \label{heissol}            
 \left[ \begin{array}{c} \hat a_{1}(t) \cr \hat a_{2}(t) \end{array} \right] &=&
 {\bf U}(t) \left[ \begin{array}{c} \hat a_{1}(0) \cr \hat a_{2}(0) \end{array} \right]
 +  {\bf V}(t) \left[ \begin{array}{c} \hat a_{1}^\dagger(0) \cr \hat a_{2}^\dagger(0) \end{array} \right]
 + \hat{\bf F}(t) .
\end{eqnarray}
Specific forms of the general evolution matrices $ {\bf U} $ and $
{\bf V} $ are discussed in the sections below. The elements of the
stochastic operator vector $ \hat{\bf F} \equiv
(\hat{F}_1,\hat{F}_2) $ are derived as linear combinations of the
Langevin forces $ \hat{L}_j $ and $ \hat{L}_j^\dagger $ that
reflect the `deterministic' solution described by the matrices $
{\bf U} $ and $ {\bf V} $ \cite{PerinaJr2000}.

Statistical properties of the emitted fields, in a given state
$\hat\rho$, are described by the Glauber-Sudarshan $P$~function,
given by Eq.~(\ref{Pfunction}), or, equivalently, by the normal
quantum characteristic function $ C_{\mathcal N} $ defined as
\begin{equation}    
C_{\mathcal N}(\beta_{1},\beta_{2}) = \left\langle\exp
(\beta_{1}\hat a^{\dagger}_{1}+\beta_{2}\hat
a^{\dagger}_{2})\exp(-\beta^{\ast}_{1}\hat a_{1}
-\beta^{\ast}_{2}\hat a_{2})\right\rangle, \label{CF_normal}
\end{equation}
where symbol $ \langle\dots\rangle $ denotes quantum averaging
including both  system and reservoir. Using the solution given in
Eq.~(\ref{heissol}) and the initial vacuum states in both fields,
the normal characteristic function $ C_{\mathcal N} $ attains the
following form:
\begin{eqnarray}\label{NCF}       
 &C_{\mathcal N}(\beta_{1},\beta_{2})&=\exp\left[-
  B_{1} \vert\beta_{1}\vert^{2}
  -B_{2}\vert\beta_{2}\vert^{2}+\left(\frac{C_1}{2}\beta^{\ast2}_{1} \right. \nonumber \right. \nonumber \\
 &  &\left. \left. +\frac{C_2}{2}\beta^{\ast2}_{2}+
 D_{12}\beta_{1}^*\beta_{2}^* + \bar{D}_{12}\beta_{1}\beta_{2}^*
 + {\rm c.c.} \right)\right],\quad
\end{eqnarray}
where the auxiliary functions are defined as follows:
\begin{eqnarray} \label{qnc}   
 B_j&=&\langle\Delta\hat a^{\dagger}_{j}\Delta\hat a_{j}\rangle= \sum_{k=1,2}|V_{jk}|^2 + \langle\hat F^{\dagger}_{j}\hat F_{j}\rangle, \nonumber \\
 C_j&=&\langle(\Delta\hat a_{j})^2\rangle= \sum_{k=1,2} U_{jk} V_{jk} +\langle\hat F^{2}_{j}\rangle, \nonumber \\
 D_{12} &=&\langle\Delta\hat a_{1}\Delta\hat a_{2}\rangle= \sum_{k=1,2} U_{1k}V_{2k} +\langle\hat F_{1}\hat F_{2}\rangle, \nonumber \\
  \bar{D}_{12} &=& -\langle\Delta\hat a_{1}^{\dagger}\Delta\hat
 a_{2}\rangle=-\sum_{k=1,2} V_{1k}^*V_{2k} -\langle\hat
  F_{1}^{\dagger}\hat F_{2}\rangle.
\end{eqnarray}
\begin{widetext}
\begin{center}
\begin{table}              
 \begin{tabular}{ c | c c c c c}
  Ordering & Quasidistribution &  & Characteristic & & Covariance matrix\\
           &                   &  & function       & &  of a Gaussian state\\
  \hline
  Normal   & $ P(\alpha_1,\alpha_2) \equiv W^{(s=1)}(\alpha_1,\alpha_2) $
    & $ \Longleftrightarrow $ & $ C_{\cal N}(\beta_1,\beta_2) $ &
    $ \longleftrightarrow $ & $ A_{\cal N} $ \\
   & $ \Downarrow \uparrow $ & & $ \Downarrow \Uparrow $ & & $ \Downarrow \Uparrow $\\
  Symmetric  & $ W(\alpha_1,\alpha_2) \equiv W^{(s=0)}(\alpha_1,\alpha_2) $
    & $ \Longleftrightarrow $ & $ C_{\cal S}(\beta_1,\beta_2) $ &
    $ \longleftrightarrow $ & $ A_{\cal S} $
 \end{tabular}
 \caption{Schematic diagram for the
 relations between (a) two quasiprobability distributions
 (quasidistributions), i.e., the Glauber-Sudarshan $P$ and Wigner
 $W$ functions for a given two-mode state $\hat\rho$, (b)
 characteristic functions $C_{\cal N}$ and $C_{\cal S}$, and (c)
 covariance matrices $A_{\cal N}$ and $A_{\cal S}$ assuming here
 that $\hat\rho$ is a Gaussian state for normal and symmetric
 orderings, respectively. Their interrelations (as marked by left-right
 arrows) are given in Appendix~A. The single arrow indicates that
 the calculation of the $P$ function from the Wigner function is
 more complicated (it can be done via the relation between $C_{\cal
 S}$ and $C_{\cal N}$) than the trivial calculation of the Wigner
 function from the $P$ function (as marked by double arrow).}
\label{table1}
\end{table}
\end{center}
\end{widetext}

The normal characteristic function given in Eq.~(\ref{NCF}) can
conveniently be rewritten into its matrix form $ C_{\cal
N}(\mbox{\boldmath$ \beta $})= \exp (\mbox{\boldmath$ \beta
$}^{\dagger}{\bf A}_{\cal N} \mbox{\boldmath$ \beta $}/2)$ using the
covariance matrix $ {\bf A}_{\cal N} $ related to the normal
ordering~\cite{PerinaKrepelka11} (for different possibilities in
describing the generated fields, see Table~\ref{table1}):
\begin{eqnarray}\label{CM}   
 {\bf A}_{\cal N} = \left[\begin{array}{cccc}
  - B_{1} & C_1 & {\bar{D}}_{12}^{\ast}& D_{12} \\
  C_{1}^{\ast} & -B_1 & D_{12}^{\ast} & {\bar{D}}_{12} \\
  {\bar{D}}_{12} &D_{12} &-B_2& C_2 \\
   D_{12}^{\ast} &  {\bar{D}}_{12}^{\ast} & C_{2}^{\ast} & -B_2
  \end{array}\right] , \\ \nonumber
\end{eqnarray}
and the column vector $ \mbox{\boldmath$ \beta $} =
(\beta_1,\beta_1^*,\beta_2,\beta_2^*)^T$.

The covariance matrix $ {\bf A}_{\cal N} $ related to the normal
ordering determines the {\em global nonclassicality} of a
two-mode Gaussian state via the Lee nonclassicality depth $\tau$. The
nonclassicality depth $ \tau $ is defined with the help of the
maximal positive eigenvalue $\lambda_+({\bf A}_{\cal N}) $ of the
matrix $ {\bf A}_{\cal N} $ as follows:
\begin{equation}\label{ndw}     
\tau = {\rm {max}}[0,\lambda_+({\bf A}_{\cal N})].
\end{equation}
We note that the nonclassicality depth $ \tau $, according to its
definition~\cite{Lee1990a}, gives the amount of noise photons
present equally in both modes and needed to conceal the
nonclassical character of the state.

The covariance matrix $ {\bf A}_{\cal N} $ of the two-mode field
can be written in the following block form:
\begin{eqnarray}   
 {\bf A}_{\cal N} &=& \left[ \begin{array}{cc} {\bf B}_1 & {\bf
  D}_{12}  \cr {\bf D}_{21} & {\bf B}_2 \end{array} \right] ,
\label{11} \\
 {\bf B}_j &=& \left[ \begin{array}{cc} -B_j & C_j
  \cr C_j^* & -B_j \end{array} \right] , \hspace{3mm} j=1,2,
  \nonumber \\
 {\bf D}_{12} &=& \left[ \begin{array}{cc} \bar{D}_{12}^* & {D}_{12}
  \cr {D}_{12}^* & \bar{D}_{12} \end{array} \right],
   \nonumber \\
 {\bf D}_{21} &=& \left[ \begin{array}{cc} \bar{D}_{12} & {D}_{12}
  \cr {D}_{12}^* & \bar{D}_{12}^*  \end{array} \right].
  \nonumber
\end{eqnarray}
This form points out at the existence of three local invariants $
I_j $, $ j=1,2,3 $, that do not change under any local linear
unitary transformation applied in mode 1 or 2. The local
invariants $ I_j $ are expressed as:
\begin{equation}  
 I_1 = {\rm det}({\bf B}_1),  \hspace{3mm} I_2 = {\rm det}({\bf B}_2),  \hspace{3mm}
 I_3 = {\rm det}({\bf D_{\textnormal {12}}}).
\label{12}
\end{equation}
Moreover, there exist two global invariants $ I $ and $ \Delta $
preserved under arbitrary linear unitary transformations and
applied to both modes:
\begin{equation}  
 I = {\rm det}({\bf A}_{\cal N}), \hspace{3mm} \Delta =
 I_1+I_2+2I_3.
\label{13}
\end{equation}
Whereas the global invariant $ I $ encompasses the whole complex
structure of the matrix $ {\bf A}_{\cal N} $ and, as such, is not
useful in our considerations, the global invariant $ \Delta $
reflects the block structure of the matrix $ {\bf A}_{\cal N} $ and
lies in the center of our attention.

Moreover, the global invariant $ \Delta $ includes the additive
local invariants $ I_1 $ and $ I_2 $ that indicate the
nonclassical behavior of the reduced states of modes 1 and 2,
respectively. Indeed, the determinants defining these invariants
occur in the Fourier transform of the normal characteristic
functions of the reduced states directly related to their local
Glauber-Sudarshan $P$ functions. If a determinant fails to be
positive then the corresponding Glauber-Sudarshan $P$ function
does not exist as a nonnegative function. Thus, the value of
determinant $ I_j $ can be used to quantify the {\em local
nonclassicality} of the reduced state in mode $ j $ as it is a
monotone of the local Lee nonclassicality depth $ \tau_j $. The
local Lee nonclassicality depth $ \tau_j $ is defined along the
formula (\ref{ndw}) that provides the relation:
\begin{equation}\label{taul}        
 \tau_j= {\rm max}(0,|C_j|-B_{j}), \hspace{3mm} j=1,2.
\end{equation}
Using Eq.~(\ref{taul}) we arrive at the monotonic relation between
the local nonclassicality depth $ \tau_j $ and local
nonclassicality invariant (NI) $ I_j $ if we assume $ \tau_j $ to
be continuous:
\begin{equation}\label{15}
 I_j = -\tau_j\left(\tau_j+2B_j\right).     
\end{equation}
We can redefine the local symplectic invariant in Eq.~(\ref{15})
as $I_{\rm ncl}^{(j)}=-I_j$ in order to deal with positive values
when quantifying the local nonclassicality. We note that not only
the positive values of this local NI $ \LNI $ are useful for
quantifying the local nonclassicality, also the negative values of
this invariant are important as they quantify the ``robustness''
of the classicality of a local state.

Returning back to the last term $ I_3 $ in the global invariant $
\Delta $, this term describes solely the mutual quantum
correlations between the fields 1 and 2. As such, it has to play a
crucial role in the description of the entanglement between two
fields. To reveal and quantify this entanglement and the role of
local invariant $ I_3 $ here, we apply for a while the phase space
$ (x,p) $ approach for describing the fields in the symmetric
ordering of field operators corresponding to the Wigner formalism
(see Table~1 and then the Appendix). The reason is technical and is
given by the fact that we know how to derive the covariance matrix
of a Gaussian state obtained by the partial transposition of the
original state. According to Simon~\cite{Simon00}, the partial
transposition means to replace $ p $ by $ -p $. The covariance
matrix of the partially transposed state then provides us the
logarithmic negativity $ E_N $ that is a commonly used measure for
the entanglement. Moreover, it provides as an entanglement measure
useful in our considerations.

In detail, the covariance matrix $ {\bf A}_{\cal S} $ expressed in
the symmetric ordering is obtained in its block structure as
follows:
\begin{eqnarray}\label{CMps}   
 {\bf A}_{\cal S} &=& \left[ \begin{array}{cc} {\bf B}_{{\cal S}1} &
  {\bf D}_{\cal S} \cr {\bf D}^T_{\cal S} & {\bf B}_{{\cal S}2} \end{array} \right] ,
\label{16} \\
 {\bf B}_{{\cal S}j} &=& \left[ \begin{array}{cc} B_{j}+{\rm Re}(C_j)+ 1/2  &
  {\rm Im}(C_j) \cr
  {\rm Im}(C_j) & B_{j}-{\rm Re}(C_j)+ 1/2 \end{array} \right] , \nonumber \\
 & & \hspace{20mm}  \hspace{3mm} j=1,2,
  \nonumber \\
 {\bf D}_{\cal S} &=& \left[ \begin{array}{cc} {\rm Re}(D_{12}-\bar
  D_{12}) &  {\rm Im}(D_{12}+\bar D_{12}) \cr
  {\rm Im}(D_{12}-\bar D_{12}) & - {\rm Re}(D_{12}+\bar D_{12}) \end{array} \right].
  \nonumber
\end{eqnarray}
The covariance matrix $ {\bf A}_{\cal S} $, similarly as its
normally-ordered counterpart, has three local invariants $
I_{{\cal S}j} $, $ j=1,2,3 $, and two global ones denoted as $
I_{\cal S} $ and $ \Delta_{\cal S} $:
\begin{eqnarray}\label{inv}     
 & I_{{\cal S}1} ={\rm det}({\bf B}_{{\cal S}1}),\quad I_{{\cal S}2} ={\rm det}({\bf B}_{{\cal S}2}), \hspace{3mm} I_{{\cal S}3}={\rm det}({\bf D}_{\cal S}),&  \nonumber \\
 & I_{\cal S}={\rm det}({\bf A}_{\cal S}), \hspace{3mm} \Delta_{\cal S} = I_{{\cal S}1}+I_{{\cal S}2}+2I_{{\cal S}3}.&
\end{eqnarray}
Moreover, the comparison of the formulas for the invariants $ I_3 $
and $ I_{{\cal S}3} $ shows that $ I_3 = I_{{\cal S}3} $.

Following Refs.~\cite{Marian01,Simon00,Marian08}, the entanglement criterion can be
expressed through the positivity of the entanglement indicator
(EI) $ \EI $ defined in terms of the invariants in Eq.~(\ref{inv})
as follows:
\begin{equation}\label{S}       
\EI =\frac{1}{4} (I_{{\cal S}1}+I_{{\cal S}2}-2I_{{\cal S}3}) -
I_{\cal S}- \frac{1}{16}.
\end{equation}
As we show below the EI $ \EI $ is a monotonous function of
logarithmic negativity $ E_N $, which can be derived from the
symplectic eigenvalue $\tilde d_-$ of the partially transposed
(PT) matrix $ {\bf A}_{\cal S}^{\rm PT} $ along the formula~\cite{Olivares12} (see Fig.~3)
\begin{equation}\label{logneg}  
 E_N={\rm {max}}[0,- \ln(2\tilde d_-) ].
\end{equation}
According to Eq.~(\ref{logneg}), a state is entangled if $\tilde
d_- < 1/2$. In turn, the symplectic eigenvalue $\tilde d_-$ is
found as:
\begin{equation}\label{symd}    
 \tilde d_- = \frac{1}{\sqrt{2}}\sqrt{\tilde\Delta_{\cal S}-\sqrt{\tilde\Delta_{\cal S}^2-4I_{\cal S}}}\, ,
\end{equation}
where $ \tilde\Delta_{\cal S} = I_{{\cal S}1} +  I_{{\cal S}2} - 2
I_{{\cal S}3} $. Combining Eqs.~(\ref{S}) and (\ref{symd}) we
arrive at the relation between the symplectic eigenvalue $\tilde
d_-$ and entanglement indicator $ \EI $:
\begin{equation}\label{symdf}   
 \tilde d_- = \frac{1}{\sqrt{2}}\sqrt{I'-\sqrt{I'^2-4I_{\cal S}}}\,,
\end{equation}
where $I' = 4I_{\cal S}+4\EI +\frac{1}{4}$.

Assuming the global invariant $ I_{\cal S} $ is fixed, the
relation (\ref{symdf}) shows that the larger is the entanglement
indicator $\EI $, the smaller is the symplectic eigenvalue $
\tilde d_- $ and, according to formula (\ref{logneg}), also the
larger is the logarithmic negativity $ E_N $. As a consequence,
the entanglement indicator $ \EI $ represents an alternative to
the logarithmic negativity $ E_N $ in quantifying entanglement. We
illustrate the monotonous dependence of the logarithmic negativity $
E_N $ on the entanglement indicator $\EI $ in Fig.~\ref{fig2}. We note
that a simple analytical formula between the logarithmic
negativity $ E_N $ and entanglement indicator $ \EI $ is derived
for pure states ($I_{\cal S}=1/16$) assuming $ \EI>0 $:
\begin{equation}\label{NvsS}  
E_N= \ln(2\sqrt{\EI}+\sqrt{1+4\EI}).
\end{equation}
\begin{figure}              
\includegraphics[width=0.45\textwidth]{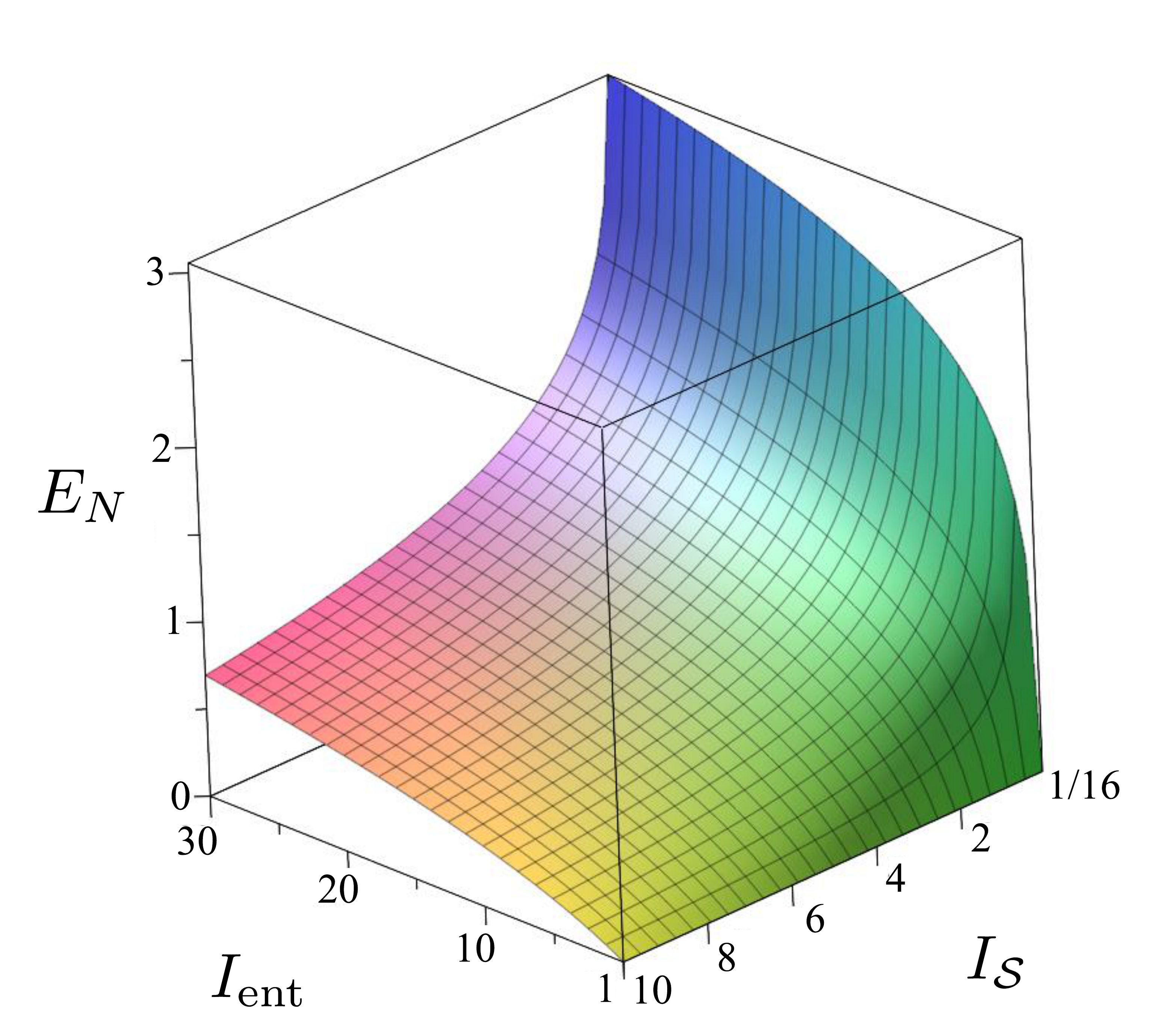}
 \caption{(Color online) Logarithmic negativity $E_N$ as a function
  of entanglement indicator $\EI$, given by Eq.~(\ref{S}), and
  global nonclassicality invariant $I_{\cal S}$, given by
  Eq.~(\ref{inv}).}
\label{fig3}
\end{figure}

As we look for a relation among the local invariants $\LNI $ and $
\LNII $ and the entanglement indicator $ \EI $ (see
Fig.~\ref{fig1}), we eliminate the invariants $ I_3 = I_{{\cal
S}3} $ from Eqs.~({13}) and (\ref{S}) by their comparing. This
leaves us with the relation:
\begin{equation}\label{law}     
\LNI +\LNII + 2\EI = \frac{1}{2}\Delta_{\cal S} -\Delta-2I_{\cal S}- \frac{1}{8}.
\end{equation}
As only the global invariants occur at the r.h.s. of
Eq.~(\ref{law}), the relation $ \LNI +\LNII + 2\EI $ is invariant
under any global linear unitary transformation.

Equation~(\ref{law}) can be transformed into the central result of
our paper, if we define a new quantity $\GNI$, which is a global
nonclassicality invariant:
\begin{equation}
  \GNI = \LNI +\LNII + 2\EI,
\label{25}
\end{equation}
In the derivation of this equation, it is useful to recall the
property that the local determinants for the normally-ordered CF,
$I_3$, and the symmetrically-ordered CF, $I_{{\cal S} 3}$, are
equal $I_3=I_{{\cal S} 3}$, and given by Eqs.~(\ref{13}) and
(\ref{inv}). Thus, we have
\begin{eqnarray}
\GNI&=&\LNI +\LNII + 2\EI \nonumber \\
&=&-I_1-I_2+\frac{1}{2} (I_{{\cal S}1}+I_{{\cal S}2}-2I_{{\cal
S}3}) -
2I_{\cal S}- \frac{1}{8} \nonumber \\
&=&-I_1-I_2-2I_{{\cal S}3}+\frac{1}{2} (I_{{\cal S}1}+I_{{\cal
S}2}+2I_{{\cal S}3}) -
2I_{\cal S}- \frac{1}{8} \nonumber \\
&=&-\Delta+\frac{1}{2}\Delta_{\cal S}-2I_{\cal S}- \frac{1}{8}.
\end{eqnarray}

Equation~(\ref{25}) means that the local nonclassicality
invariants $ \LNI $ and $ \LNII $ together with the entanglement
indicator $ \EI $ form the global NI $ \GNI $. Any linear unitary
transformation in general modifies both the local NIs $ \LNI $ and
$ \LNII $ and the entanglement invariant $\EI$ only in such a way
that it preserves the value of the global NI $ \GNI $. Whenever $
\GNI $ is positive, the analyzed state is nonclassical due to the
local nonclassicality of the reduced states or its entanglement.
The negative values of the global NI $ \GNI $ do not necessarily
mean that a given state is classical, as we will see below.

In the next sections, we analyze the nonclassicality and entanglement
of several kinds of important quantum states from the point of
view of their transformation by a beam splitter. The division of the
global NI into the EI and the local NIs is in the center of our
attention. In general, six regions differing in the occurrence of
entanglement and local nonclassicalities can be defined (see
Table~\ref{table2}). All these regions are found in the examples
analyzed in the next sections, as indicated in Table~\ref{table2}.
\begin{widetext}
\begin{center}
\begin{table}
\caption{Regions of different entanglement and local
 nonclassicalities observed in the figures of Secs.~III---VI.}
 \begin{tabular}{| c | c | c | c | c |}
  \hline case/region & Entanglement & Nonclassicality of one
  mode & Nonclassicality of another mode & Figures \\ \hline
  I& Yes & Yes & Yes & 6, 10 \\ \hline II & Yes & Yes & No & 6(b) \\
  \hline III &
  Yes & No & No & 6, 10 \\ \hline IV & No & Yes & Yes & 6, 10 \\
  \hline V & No & Yes & No & 6(b)\\ \hline VI & No & No & No & 6, 10 \\
  \hline
 \end{tabular}
\label{table2}
\end{table}
\end{center}
\end{widetext}

We note that an invariant based on the second-order intensity
moments and, as such, describing intensity auto- and
cross-correlations has been suggested in Ref.~\cite{Klyshko1994}
for two-mode fields with specific mode correlations and unitary
transformations. Later, this invariant was experimentally analyzed
in Ref.~\cite{Iskhakov13}. Here, we describe the propagation of
fields through the beam splitter (see Fig.~\ref{fig1}) described
by the real transmissivity $ T $ and the phase $ \phi $ through
the unitary transformation characterized by the matrix $ {\bf U}
$,
\begin{eqnarray}\label{BS}      
 &&{\bf U}=\left(\begin{array}{cccc}
  \sqrt{T} & 0 &-\sqrt{R}e^{i\phi}&0 \\
  0 & \sqrt{T}& 0 & -\sqrt{R}e^{-i\phi}  \\
  \sqrt{R}e^{-i\phi} & 0 &\sqrt{T} & 0\\
  0 & \sqrt{R}e^{i\phi} & 0 & \sqrt{T}\end{array}\right);
\end{eqnarray}
$R=1-T$ is the reflectivity of the beam splitter. The covariance
matrix $ {\bf A}^{\rm out} $ at the output of the beam splitter is
obtained as $ {\bf A}^{\rm out} = {\bf U}^{\dagger}{\bf A}{\bf U}
$.

\section{Twin beam}

These beams are generated by parametric down-conversion from the
vacuum into which photon pairs are ideally emitted. For this
reason, only the terms $B_{1}$, $B_{2}$, and $D_{12}$ in the normal
characteristic function $ C_{\cal N} $ are nonzero. The evolution
matrices $ {\bf U} $ and $ {\bf V} $ in Eq.~(\ref{heissol}) have
the following nonzero elements:
\begin{eqnarray}  
 U_{11}(t)&= &U_{22}(t)=\cosh(gt), \nonumber \\
 V_{12}(t)&=&V_{21}(t)=i\exp(i\theta)\sinh(gt).
\end{eqnarray}
The coefficients $ B_1 $ and $ B_2 $ can be expressed as
$B_1=B_{\rm p}+B_{\rm s}$ and $B_2=B_{\rm p}+B_{\rm i}$, where $
B_{\rm p} = \sinh^2(g_{12}t) $ gives the mean number of generated
photon pairs and $ B_{\rm s} = \langle\hat F^\dagger_{1}\hat
F_{1}\rangle$ ($ B_{\rm i} =\langle\hat F^\dagger_{2}\hat
F_{2}\rangle $) denotes the mean number of signal (idler) noise
photons coming from the reservoir (see Fig.~\ref{fig2}). On the
other hand, the parameter $D_{12}$ characterizing mutual
correlations depends only on the mean number $ B_{\rm p} $ of
photon pairs as $D_{12}=i\sqrt{B_{\rm p}(B_{\rm p}+1)}$
($\theta=0$ is assumed without the loss of generality).

The general formulas  for the local NIs $ I^{(j)}_{\rm ncl}$,
entanglement invariant $\EI $, and the global NI $ \GNI $ attain
the following forms for twin beams:
\begin{eqnarray} \label{gen_twb} 
 \LNI &=& 4TR ( B^2_{\rm p}+B_{\rm p} )  -\big[B_{\rm p}+TB_{\rm s}+RB_{\rm i}\big]^2, \nonumber \\
 \LNII &=& 4TR ( B^2_{\rm p}+B_{\rm p} ) -\big[B_{\rm p}+TB_{\rm i}+RB_{\rm s}\big]^2, \nonumber \\
 \EI &=& -\big[ (B_{\rm s}+B_{\rm i})^2 -(T-R)^2\big] ( B^2_{\rm p}+B_{\rm p} ) \nonumber \\
 & & \mbox{} -2B_{\rm p}B_{\rm s}B_{\rm i}(B_{\rm s}+B_{\rm i})
  - ( B^2_{\rm s}+B_{\rm s} )( B^2_{\rm i}+B_{\rm i} ) \nonumber \\
 & & \mbox{} -TR (B_{\rm s}+B_{\rm i})^2,
\label{28}  \\
 \GNI &=& 2B_{\rm p} - (B_{\rm s}+B_{\rm i})^2 [2( B^2_{\rm p}+B_{\rm p} ) +1] \nonumber \\
 & & \mbox{}- 2B_{\rm p}(1+2B_{\rm s}B_{\rm i})
 (B_{\rm s}+B_{\rm i}) \nonumber \\
 & & \mbox{} -2B_{\rm s}B_{\rm i}(B_{\rm s}+B_{\rm i} +B_{\rm s}B_{\rm i}).
\label{29}
\end{eqnarray}

We first discuss the behavior of noiseless twin beams for which $
B_{\rm s} = B_{\rm i} = 0 $. In this case, the global NI $ \GNI $
equals $ 2B_{\rm p}$ and
\begin{eqnarray}\label{VSpure}  
 I^{(j)}_{\rm ncl} &=& 4TR(B^2_{\rm p}+B_{\rm p})- B^2_{\rm p}, \hspace{3mm} j=1,2, \nonumber \\
 \EI &=& (T-R)^2(B^2_{\rm p}+B_{\rm p}).
\end{eqnarray}
As suggested by the formula in Eq.~(\ref{VSpure}), the local NIs
$I^{(j)}_{\rm ncl}$ can be decomposed into two terms. The negative
term reflects classical thermal statistics of photon pairs in a
twin beam with its photon bunching effect and as such suppresses
the nonclassical behavior of the twin beam. On the other hand, the
positive term refers to squeezing appearing at the
individual output ports of the beam splitter. The squeezing effect
originates in pairing of photons in individual output ports caused
by ``sticking of two photons from a pair together'' (photon bunching) at the beam
splitter~\cite{MandelBook}. Photon pairs with both photons in one
output port contribute to the local nonclassicality of the field in
this port. On the other hand, when two photons from one
photon-pair occur in different output ports, they contribute to
the entanglement. ``A given individual photon pair'' is, thus,
responsible either for the local nonclassicality in one of the
output ports or for their entanglement. Never for both. Propagation
through the beam splitter can, thus, be viewed as the process of
breaking photon pairs (antibunching) arriving at the same input port and gluing (bunching)
of photons from a given pair coming from different input ports.
Whereas the first process disturbs local squeezing and supports
entanglement, the second process strengthens squeezing at the
expense of entanglement. The global NI $ \GNI $ is equal twice the
number $ B_{\rm p} $ of photon pairs and, as such,  indicates an
appropriate choice of this nonclassicality resource quantifier.

\begin{figure}      
\includegraphics[width=0.45\textwidth]{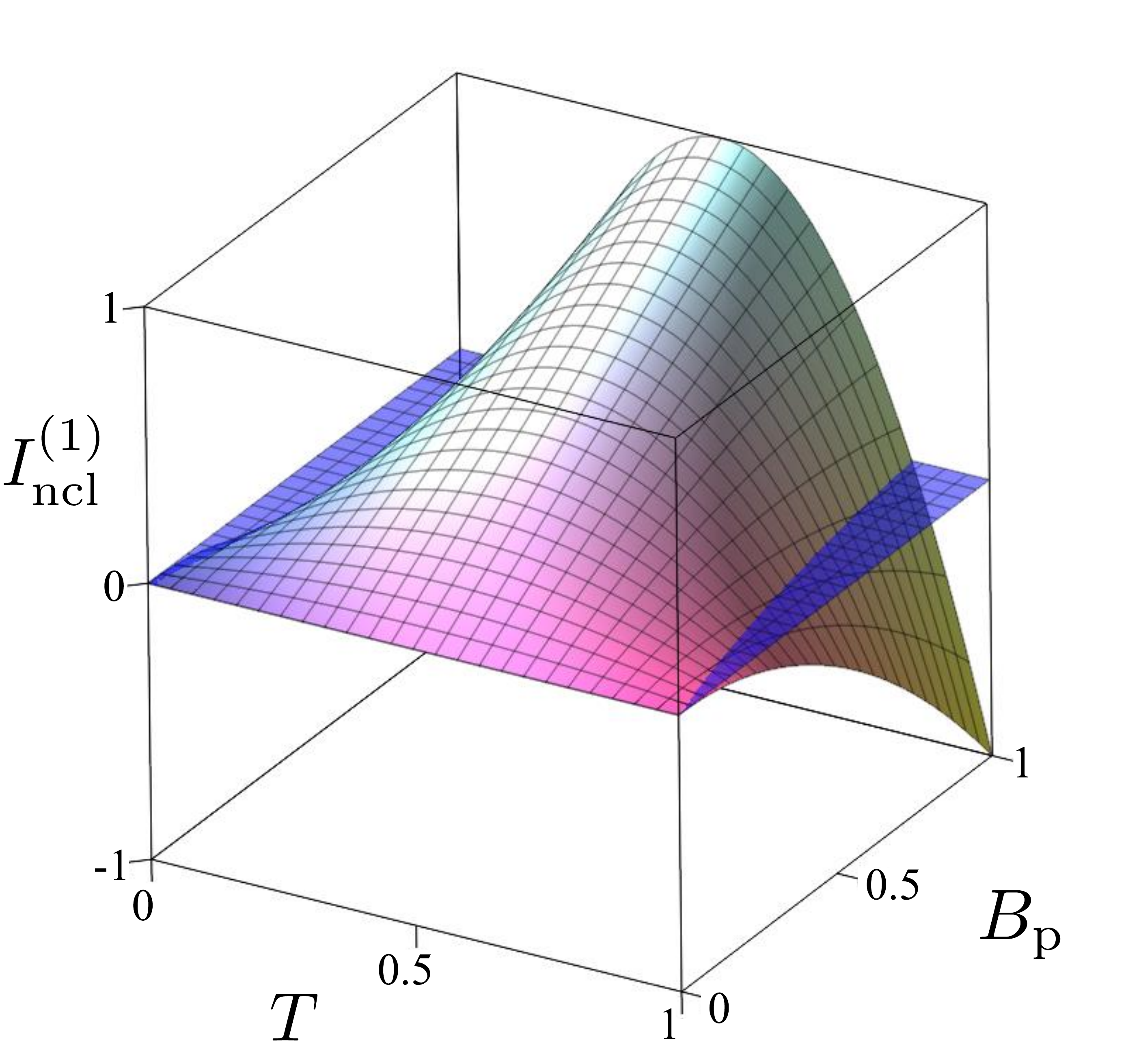}
 \centerline{(a)}
\includegraphics[width=0.45\textwidth]{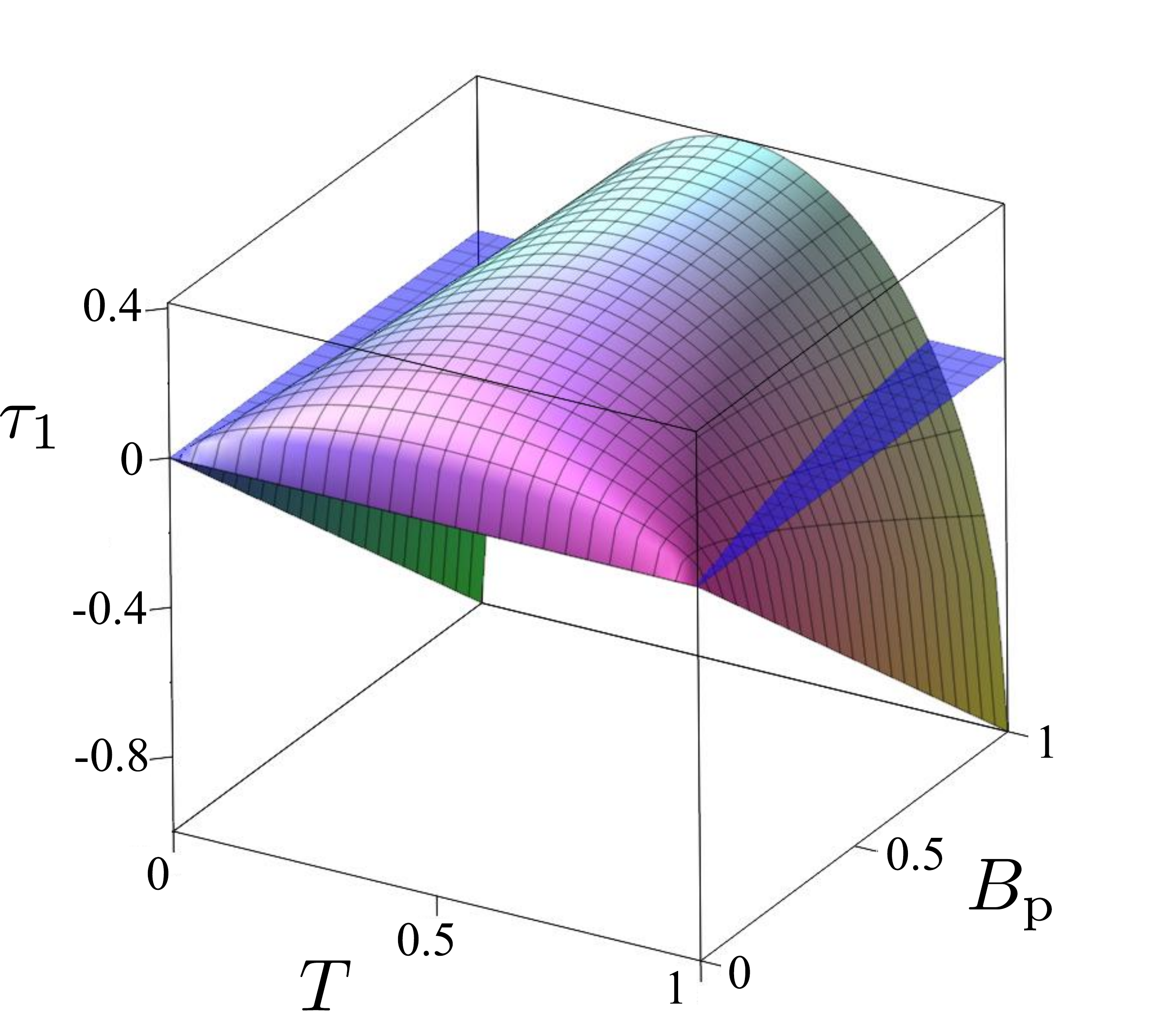}
 \centerline{(b)}
 \caption{(Color online) (a) Local nonclassicality invariant $ \LNI$ and (b) continuous Lee nonclassicality depth
 $\tau_{1}$ (including negative values) at the output port 1 of the beam splitter as a function of the mean
 photon-pair number $B_{\rm p}$ and the beam-splitter transmissivity $T$ for pure twin beam states.
 In panel (a) and (b), the blue dark grey plain surface at $ \LNI=0$ and $\tau_1=0$ shows the boundary between the classical
 and nonclassical domains.}
\label{fig4}
\end{figure}

\begin{figure}          
\includegraphics[width=0.45\textwidth]{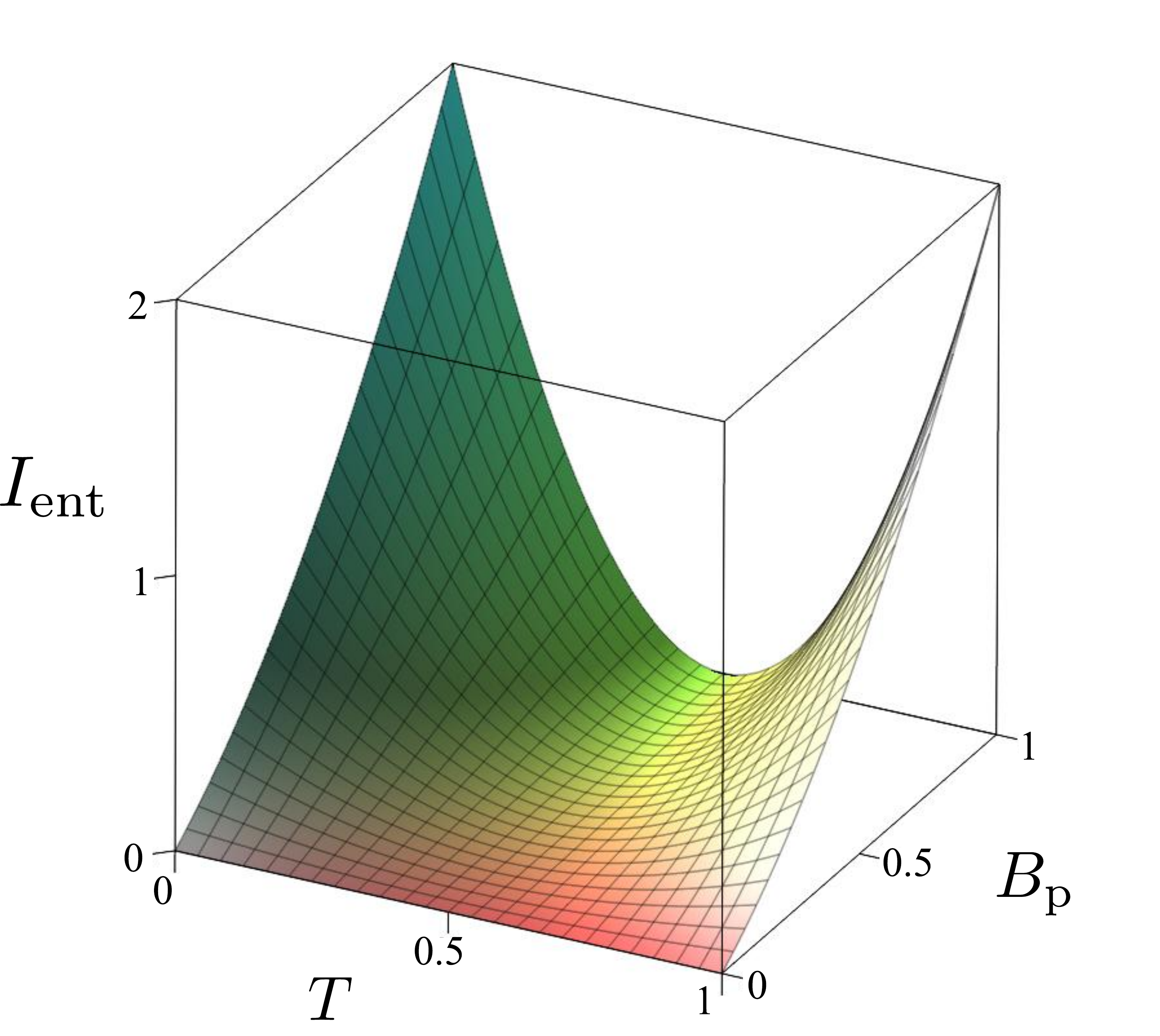}
 \centerline{(a)}
\includegraphics[width=0.45\textwidth]{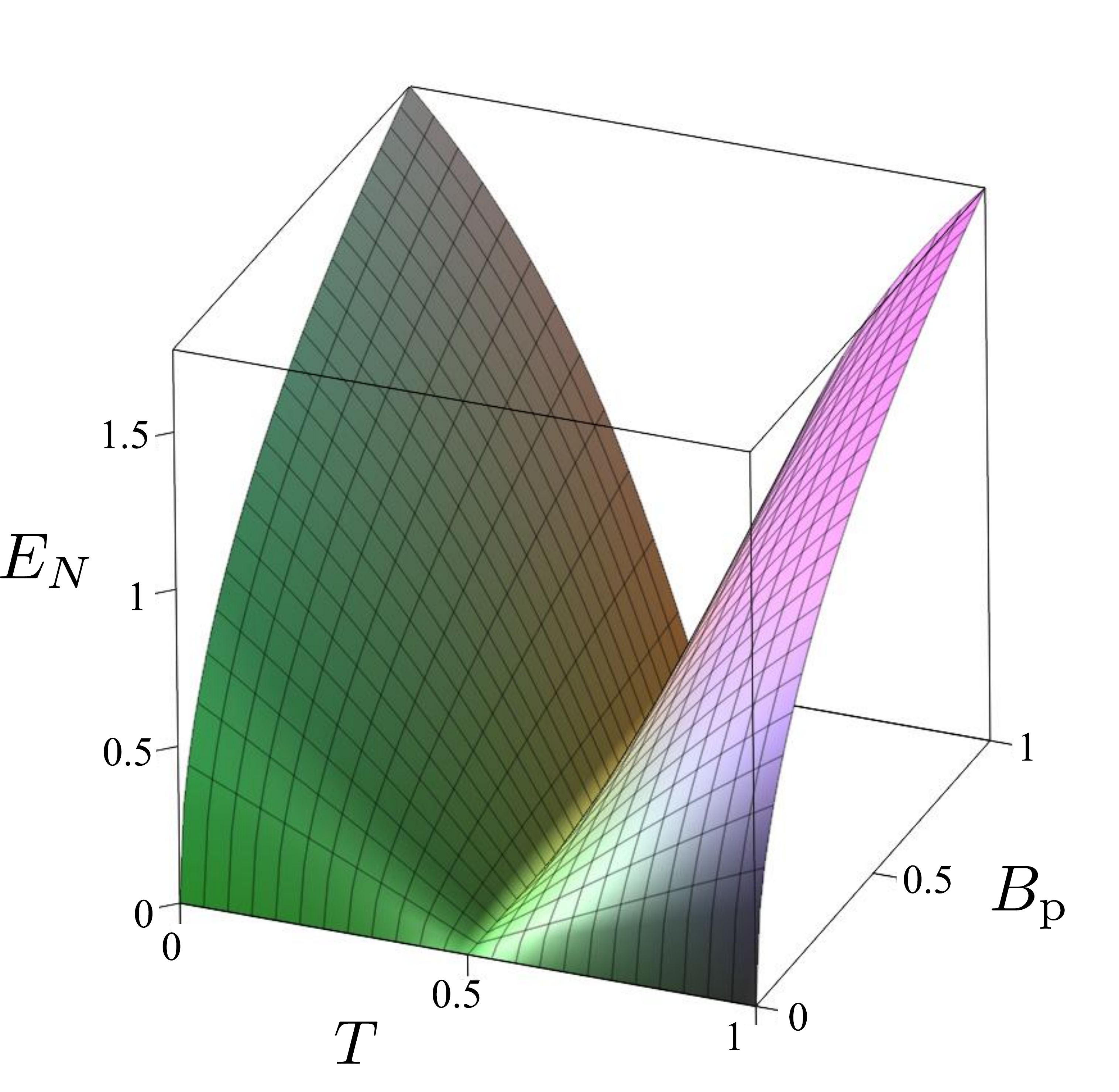}
 \centerline{(b)}
 \caption{(Color online) (a) Entanglement invariant $ E_{I}$  and (b) logarithmic negativity $ E_N $ after the beam splitter
 transformation considered as functions of the mean photon-pair number $B_{\rm p}$ and the beam-splitter transmissivity $T$ for
 pure twin beams states.}
\label{fig5}
\end{figure}

Detailed analysis of the formulas in Eq.~(\ref{VSpure}) shows that
the local marginal states are nonclassical only if the
transmissivity $T$ lies in certain interval around $\frac{1}{2}$:
\begin{equation}                    
T\in\left(\frac{1}{2}-\frac{1}{2\sqrt{B_{\rm
p}+1}},\frac{1}{2}+\frac{1}{2\sqrt{B_{\rm p}+1}}\right).
\end{equation}
It holds that the larger is the mean photon-pair number $ B_{\rm
p} $, the narrower is the interval. The optimal transmissivity $ T
$ maximizing the local NIs $ I^{(j)}_{\rm ncl} $ equals $\frac{1}{2}$. In
this case, the entanglement of the incident twin beam is
completely and equally transferred into the local
nonclassicalities of the two output modes. On the other hand, the
twin beam loses its entanglement only when $ T = \frac{1}{2} $. In this
case, all the incident photon pairs stick together (bunch) at the beam
splitter suppressing completely their entanglement. Hand in hand,
the local NIs $ \LNI = \LNII $ attain their maximal values. This
can be interpreted such that the initial entanglement is
transferred into the squeezing of the marginal output
fields~\cite{Paris97}. These effects are shown in
Figs.~\ref{fig4}(a) and \ref{fig5}(a) for the dependencies of the
local NI $ \LNI $ and EI $ \EI $ on the transmissivity $ T $ and
mean photon-pair number $ B_{\rm p} $. The commonly used the Lee
nonclassicality depth $ \tau_1 $ and the logarithmic negativity $ E_N
$ are shown for comparison in Figs.~\ref{fig4}(b) and
\ref{fig5}(b). We note, that whereas the values of the Lee
nonclassicality depth $ \tau_1 $ cannot exceed $\frac{1}{2}$, the values of
the local NI $ \LNI $ can be arbitrarily large depending on
the intensity of the twin beam.

Now we consider  general noisy twin beams. It has been shown in
Ref.~\cite{arkhipov15} that whenever the overall noise $ B_{\rm s}
+ B_{\rm i} $ exceeds one, the twin beam is unentangled and, thus,
it cannot generate any nonclassical feature. Even if $ B_{\rm s} +
B_{\rm i} < 1 $, the mean photon-pair number $ B_{\rm p} $ has to
be sufficiently large, as given by
\begin{equation}            
 B_{\rm p} > \frac{B_{\rm s}B_{\rm i}}{1-(B_{\rm s}+B_{\rm i})}.
\label{32}
\end{equation}
Then, the incident noisy twin beam is entangled and is capable to
provide its entanglement and local nonclassicality after the beam
splitter. However, the general analysis of Eqs.~(\ref{28}) and
(\ref{29}) leads to the conclusion that the noise only degrades
the non-classical behavior independently whether it is manifested
by local nonclassicality or entanglement. The stronger the noise,
the weaker the non-classical features.

To provide a deeper insight into the role of noise, we analyze two
special cases: in the first one, the noise is equally divided into
both modes of the incident twin beam; while noise occurs only in one
mode of the incident twin beam in the second case.

\begin{figure}         
 \includegraphics[width=0.44\textwidth]{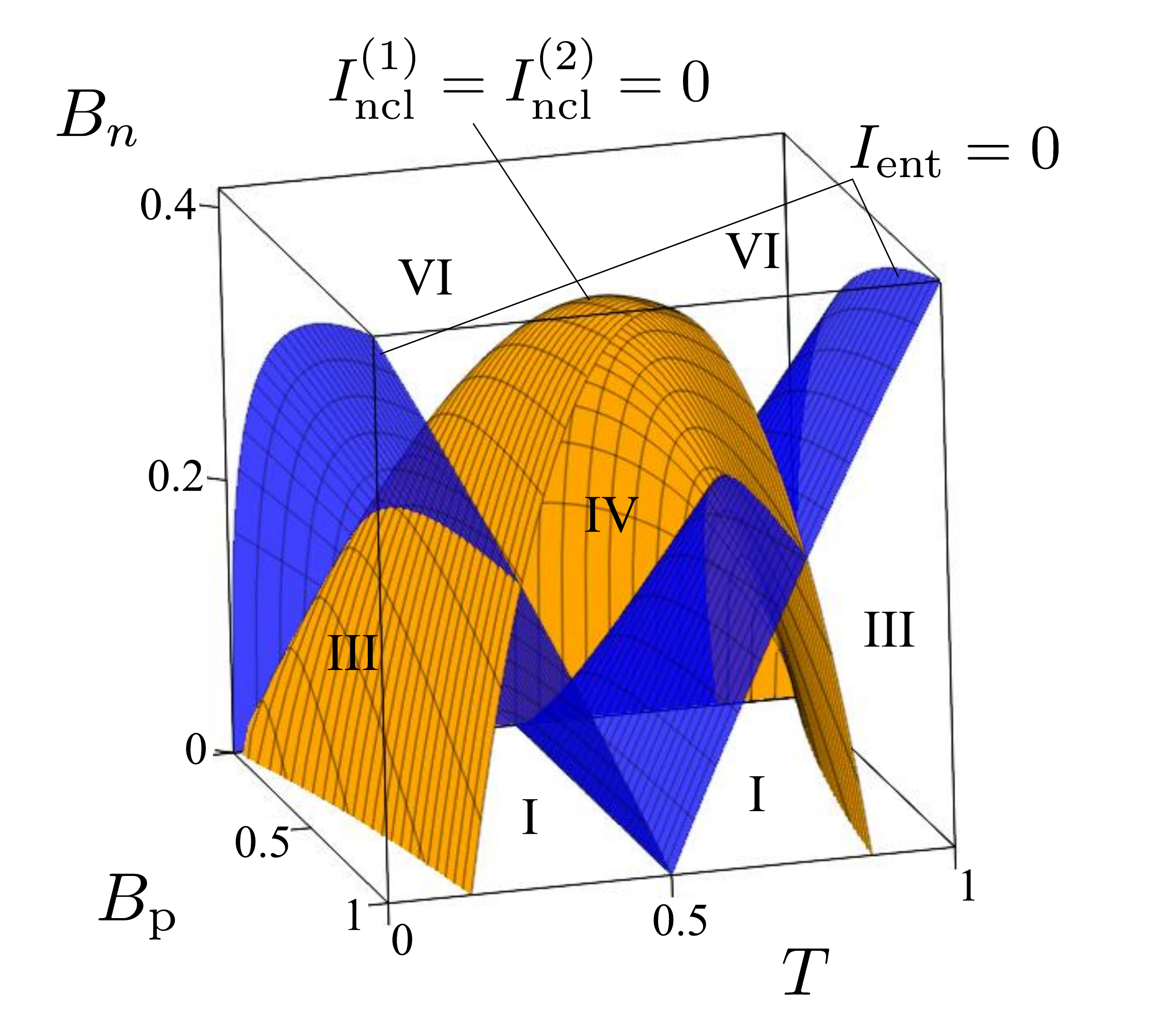}
 \centerline{(a)}
\includegraphics[width=0.48\textwidth]{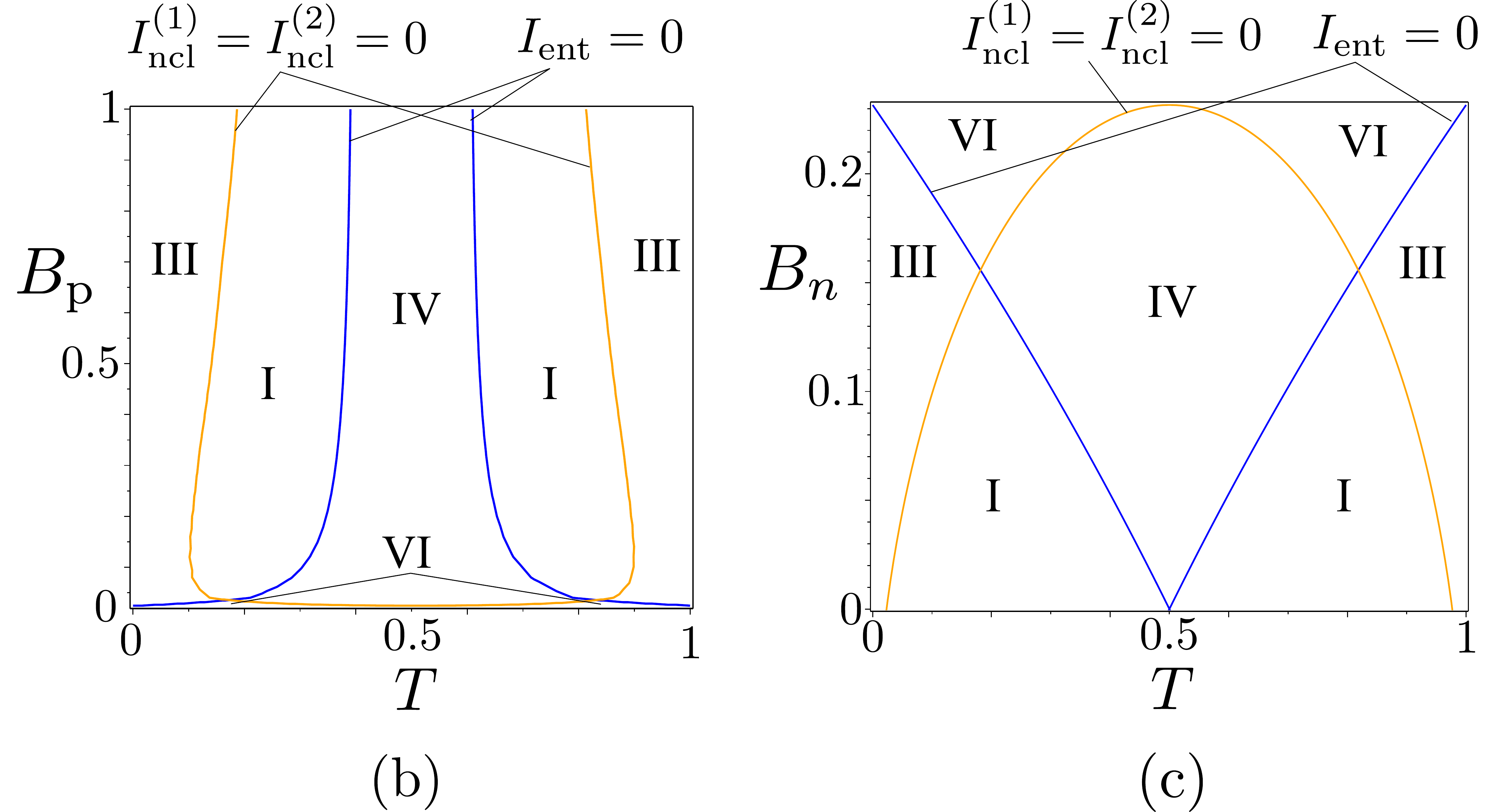}
\caption{(Color online) Diagram (a) shows the nonclassicality and
entanglement invariants for the twin beams states occurring at the
output ports of a beam splitter depending on the mean noise photon
number $ B_{\rm n} $, mean photon-pair number $B_{\rm p}$, and
transmissivity $ T $ according to Eq.~(\ref{gen_twb}) for $ B_{\rm
n}\equiv B_{\rm s}=B_{\rm i}$. The surfaces are plotted at $
\LNI(B_{\rm n},B_{\rm p},T) = 0 $ $\big[$ orange light gray
surface$\big]$, $ \LNII(B_{\rm n},B_{\rm p},T) = 0 $ $\big[$orange
light gray$\big]$ and $E_I(B_{\rm n},B_{\rm p},T) = 0 $ $\big[$
blue dark surface$\big]$ indicating six different regions
specified in the text and Tab.~\ref{table2}. Diagrams~(b) and (c)
show the perpendicular cross-sections of diagram~(a) taken at
chosen values of $B_{\rm n}=0.1$ and $B_{\rm p}=0.1$,
respectively. } \label{fig6}
\end{figure}

When noise occurs in both modes of the incident twin beam ($
B_{\rm n}\equiv B_{\rm s}=B_{\rm i} $), the globally nonclassical
output states can be divided into three groups. They are displayed
in the ``phase diagram'' in Fig.~\ref{fig6}. In this diagram,
the surfaces $ \LNI(B_{\rm n},B_{\rm p},T) = 0 $ and $ \EI(B_{\rm
n},B_{\rm p},T)=0 $ are shown. They identify four different
regions belonging to different groups of states (see
Table~\ref{table2} for details). The states exhibiting both
entanglement and local nonclassicality occur in region I. In
region III, the states are entangled but locally classical. The
locally nonclassical and unentangled states are found in region
IV. In region VI, the unentangled and locally classical states
exist.

\begin{figure} 
 \includegraphics[width=0.44\textwidth]{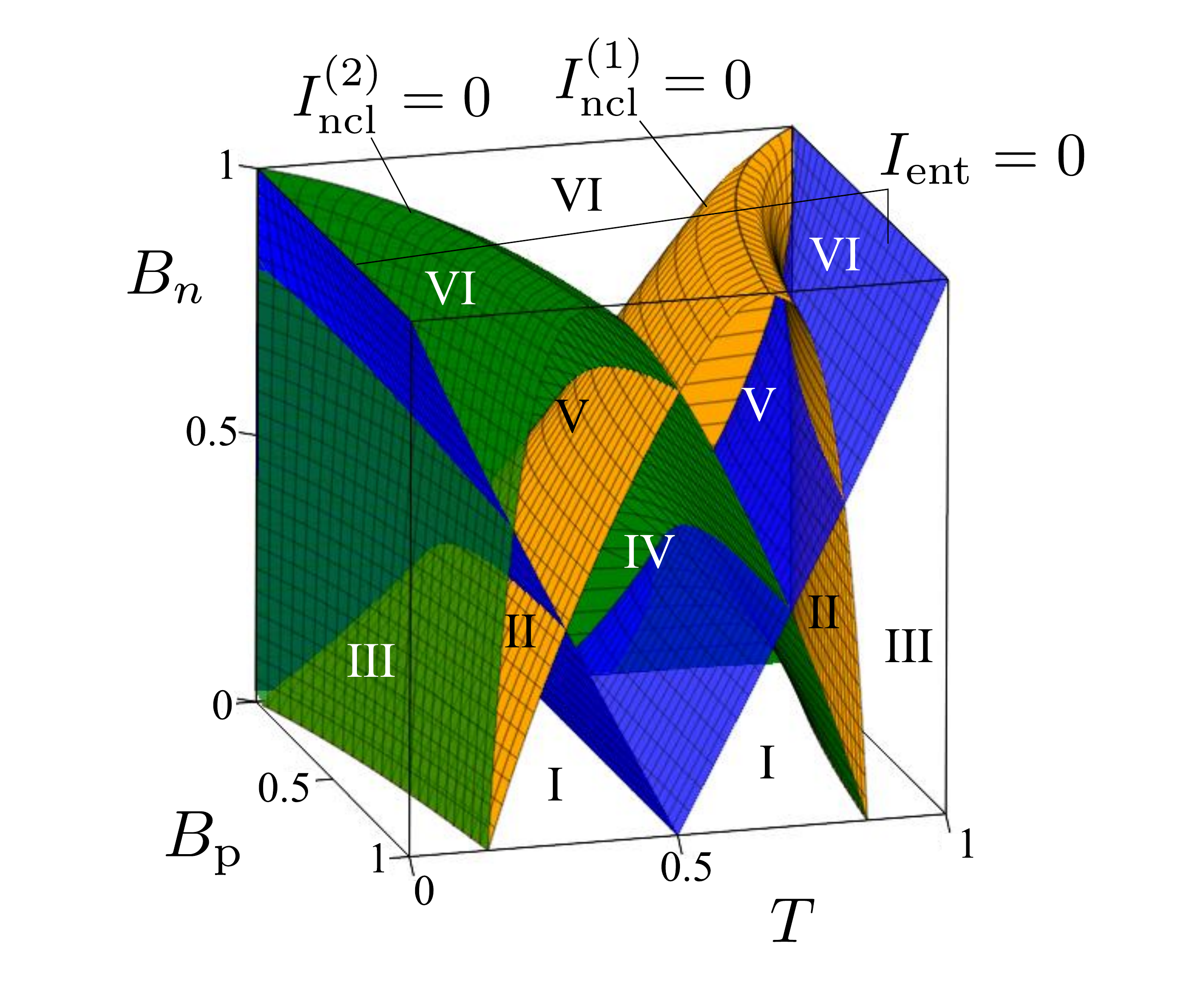}
 \centerline{(a)}
\includegraphics[width=0.48\textwidth]{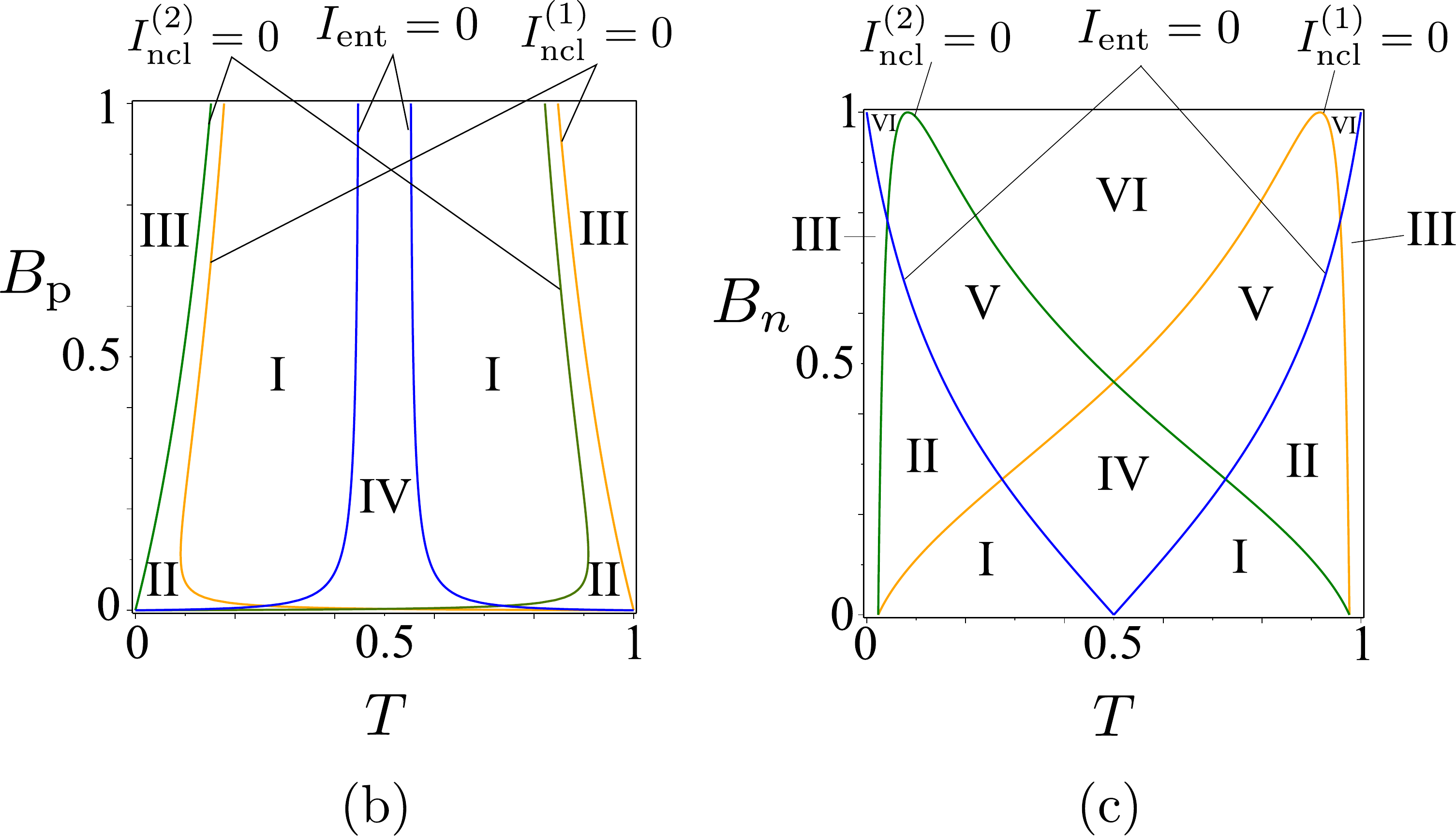}
 \caption{(Color online)
Diagram (a) shows the nonclassicality and entanglement invariants
for the twin beams states occurring at the output ports of a beam
splitter depending on the mean noise photon number $ B_{\rm n} $,
mean photon-pair number $B_{\rm p}$, and transmissivity $ T $
according to Eq.~(\ref{gen_twb}) for $ B_{\rm s}=0 $  and $ B_{\rm
n}=B_{\rm i} $. The surfaces are plotted at $ \LNI(B_{\rm
n},B_{\rm p},T) = 0 $ $\big[$orange light gray surface$\big]$, $
\LNII(B_{\rm n},B_{\rm p},T) = 0 $ $\big[$green dark gray
surfaces$\big]$ and $ E_I(B_{\rm n},B_{\rm p},T) = 0 $ $\big[$blue
dark surface$\big]$ indicating six different regions specified in
the text and Tab.~\ref{table2}. Diagrams~(b) and (c) show the
perpendicular cross-sections of diagram~(a) taken at fixed values
of $B_{\rm n}=0.1$ and $B_{\rm p}=0.1$, respectively. These
cross-sections are analogous to those in Figs.~\ref{fig6}(b)
and~\ref{fig6}(c).} \label{fig7}
\end{figure}

The presence of noise in only one mode of the incident twin beam
($B_{\rm s}=0$, $B_{\rm i}\equiv B_{\rm n} \neq 0 $) leads to
asymmetry between the output modes. This is shown in
Fig.~\ref{fig7}, where the surfaces $ \LNI(B_{\rm n},B_{\rm
p},T) = 0 $ and $ \LNII(B_{\rm n},B_{\rm p},T) = 0 $ behave
differently. The symmetry, with respect to the plane for $ T=\frac{1}{2} $,
which is clearly visible in Fig.~\ref{fig6}, does not exist in
Fig.~\ref{fig7}. As a consequence, two additional groups of
states are found in the diagram. In region V, there are
unentangled states with only one marginal field exhibiting local
nonclassicality. The entangled states with only one locally
nonclassical field are found in region II. In detail, mode 1 (2)
is locally nonclassical for $T<\frac{1}{2} $ ($T>\frac{1}{2}$). We note that the
EI $ \EI $ is not sensitive to the noise asymmetry, as shown by
the surface $ \EI(B_{\rm n},B_{\rm p},T)=0 $ in
Fig.~\ref{fig7}.
It is worth noting that positive values of the GNI $ \GNI $ are
exhibited when either  entanglement or  local
nonclassicality or even both are found. The negative values of the
global NI $ \GNI $ do not necessarily mean classicality. The state
with the negative GNI $ \GNI $ can still be globally nonclassical due
to either  its entanglement or local nonclassicality, but not
both. The diagram in Fig.~\ref{fig6}(a) can serve as an example.
The surface $ \GNI(B_{\rm n},B_{\rm p},T) = 0 $ lies naturally in
between the surfaces $ \LNI(B_{\rm n},B_{\rm p},T) = 0 $, and $
\EI(B_{\rm n},B_{\rm p},T) = 0 $ and its position identifies the
globally nonclassical states with $ \GNI < 0 $.

\section{Squeezed vacuum state with noise}

\begin{figure}  
\includegraphics[width=0.45\textwidth]{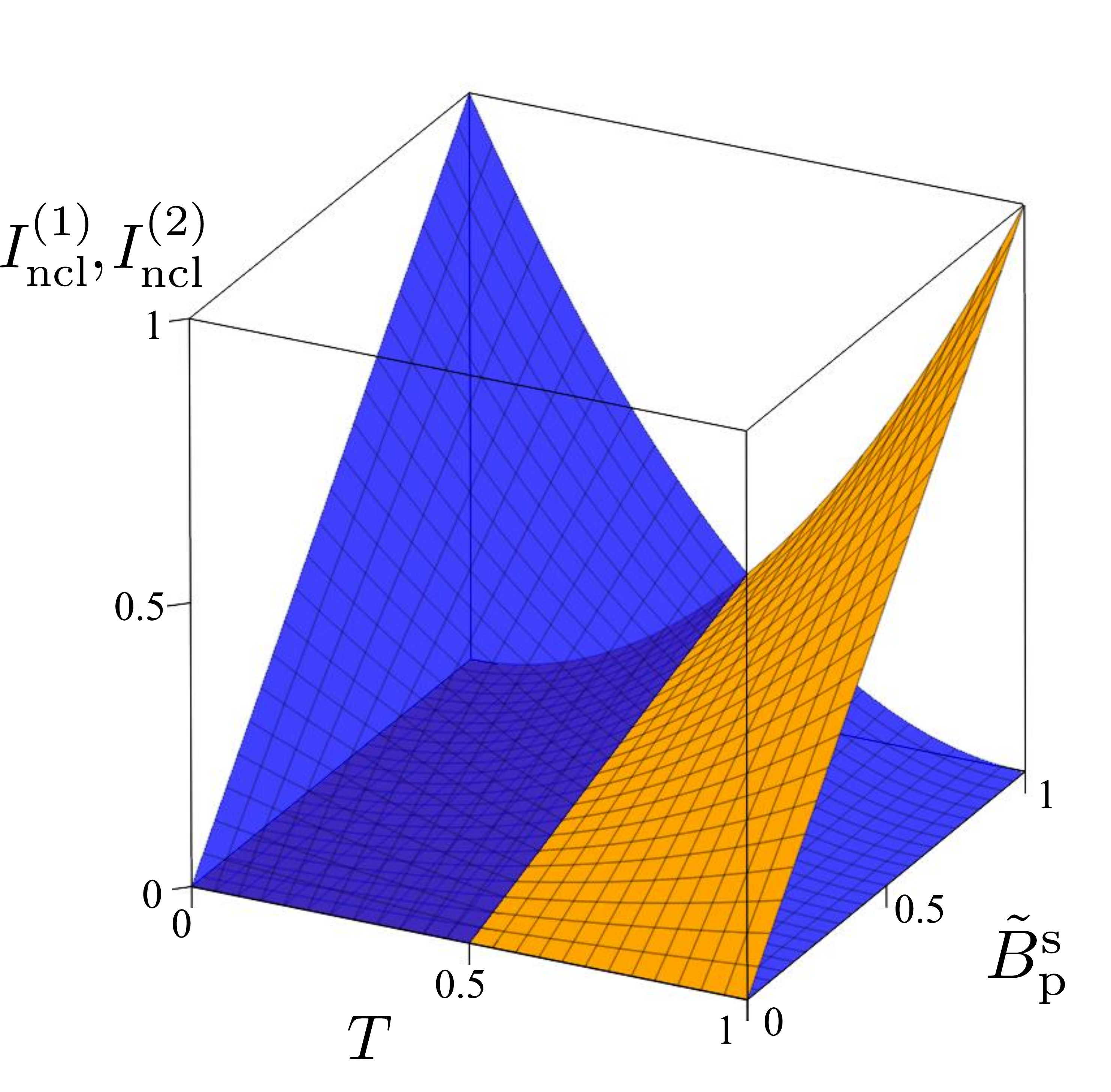}
 \centerline{(a)}
\includegraphics[width=0.45\textwidth]{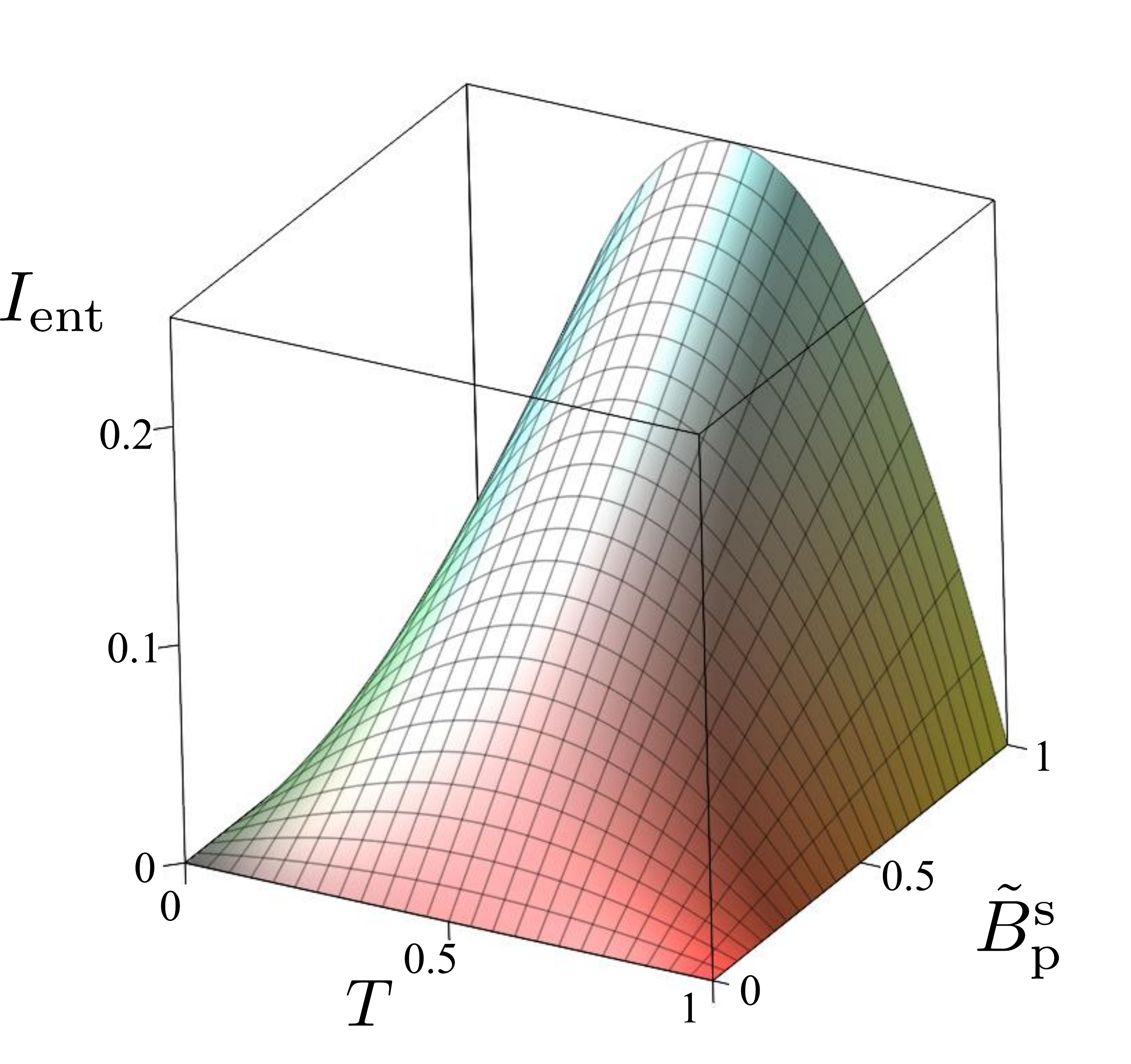}
 \centerline{(b)}
 \caption{(Color online) Invariant nonclassicality parameters: (a)
  the local nonclassicality invariants  $ \LNI $ (orange light gray
  surface) and $ \LNII$ (blue dark gray surface), and (b) the entanglement invariant $
  \EI $ versus the mean number $ \tilde B_{\rm p}^{\rm s}$ of
  squeezed photons and the beam-splitter transmissivity $T$  according to Eq.~(\ref{sq_vac_par}) assuming $
  B_{\rm s} = 0 $.}
\label{fig8}
\end{figure}

Here, we consider a squeezed vacuum state~\cite{MandelBook} mixed
with the noise incident on one input port of the beam splitter,
whereas the second input port is left in the vacuum state. In this
case, the nonzero elements of evolution matrices $ {\bf U} $ and $
{\bf V} $ in Eq.~(\ref{heissol}) are given as ($\theta = 0$ is
assumed):
\begin{eqnarray}  
 U_{11}(t)&=&\cosh(gt), \hspace{3mm} U_{22}(t) = 1, \nonumber \\
  V_{11}(t)&=& i\exp(i\theta)\sinh(gt).
\end{eqnarray}
The non-zero parameters of the normal characteristic function $
C_{\cal N} $ in Eq.~(\ref{CM}) are $ B_1 $ and $ C_1 $ as given
by: $B_1 = \tilde B_{{\rm p}}^{\rm sq} + B_{\rm s}$  and $C_1 =
i\sqrt{ \tilde B_{\rm p}^{\rm sq}(\tilde B_{\rm p}^{\rm sq} +1)}$.
 The symbol  $ \tilde
B_{\rm p}^{\rm sq} $ denotes the mean number of squeezed
photons and the symbol $B_{\rm s} $ stands for the mean number of
the signal noise photons (see also Fig.~\ref{fig2}). The local NIs
$ I^{(j)}_{\rm ncl} $ and EI $ \EI $ are easily expressed in terms
of the global NI $ \GNI $ as follows
\begin{eqnarray}\label{sq_vac_par}        
&  \LNI = T^2 \GNI, \hspace{3mm} \LNII = R^2 \GNI , \hspace{3mm}
 \EI = TR\GNI,&  \nonumber \\
  &\GNI = \tilde B_{\rm p}^{\rm sq}(1-2B_{\rm s}) - B_{\rm s}^2.&
\end{eqnarray}
As the local NIs $\LNI $ and $ \LNII $, as well as the EI $ \EI $
are linearly proportional to the global NI $ \GNI $, the global
nonclassicality of the output states immediately guarantees both
local nonclassicalities and entanglement. This occurs only for the
positive values of the global NI $ \GNI $. According to
Eq.~(\ref{sq_vac_par}), $ \GNI > 0 $ provided that the mean noise
photon number $ B_{\rm s} $ in the signal mode is sufficiently
small:
\begin{equation}                
 B_{\rm s} < \sqrt{\tilde B_{\rm p}^{\rm sq}(\tilde B_{\rm p}^{\rm sq}+1)}-
  \tilde B_{\rm p}^{\rm sq}.
\label{35}
\end{equation}
Following Eq.~(\ref{sq_vac_par}), the mean noise photon number
$B_{\rm s}$ in the signal mode has to be smaller than 1. Also, the
more intense is the squeezed state, the smaller is the number $
B_{\rm s} $ of accepted noise photons. We note that the condition,
given in Eq.~(\ref{sq_vac_par}), can immediately be revealed when
the global Lee nonclassicality depth $ \tau $ is analyzed. As an
illustration, the dependencies of the local NIs $ \LNI $ and $
\LNII $ and the EI $ \EI $ on the beam-splitter transmissivity $ T
$ are plotted in Fig.~\ref{fig8} for the incident noiseless
squeezed states. The greatest values of EI $ \EI $ are reached for
the balanced beam splitter ($ T=\frac{1}{2} $). However, some incident
photon pairs are not broken (i.e., split) by the beam splitter and give raise to
nonzero local nonclassicalities $ \LNI $ and $ \LNII $ even in
this case.

The strength of squeezing in a given mode is commonly characterized by a
principal squeeze variance $ \lambda $~\cite{Luks1988}, which is
here given by
\begin{equation}  
 \lambda_j = 1/2 + B_j - |C_j| .
\label{36}
\end{equation}
When a given output mode $ j=1,2 $ is locally nonclassical, it is
also squeezed, which corresponds to $ \lambda_j < \frac{1}{2} $. According to the
relation between the local NI $ I^{(j)}_{\rm ncl} $ and the
principal squeeze variance $ \lambda_j $ derived by combining
Eqs.~(\ref{15}) and (\ref{36}),
\begin{equation}  
 I^{(j)}_{\rm ncl} = (\frac{1}{2}-\lambda_j)(2B_j+ \frac{1}{2}-\lambda_j),
  \hspace{3mm}
\label{37}
\end{equation}
the smaller is the value of the principal squeeze variance $
\lambda_j $ below $\frac{1}{2}$, the greater is the value of the local NI $
I^{(j)}_{\rm ncl} $.

\section{Two squeezed vacua}

Two independent squeezed states are generated by the Hamiltonian
given in Eq.~(\ref{hamil}) provided that the process of parametric
down-conversion does not occur in the nonlinear medium
($g_{12}=0$). The solution of the evolution governed by the
Hamiltonian (\ref{hamil}) gives us the following nonzero elements
of the evolution matrices $ {\bf U} $ and $ {\bf V} $:
\begin{eqnarray}  
 U_{11} &= \cosh(2g_{11}t), \hspace{3mm} V_{11} &= i\exp(i\kappa_1)\sinh(2g_{11}t), \nonumber \\
 U_{22} &= \cosh(2g_{22}t), \hspace{3mm} V_{22} &=
 i\exp(i\kappa_2)\sinh(2g_{11}t),
\end{eqnarray}
where $ \kappa_1 $ and $ \kappa_2 $ are arbitrary phases. The
nonzero coefficients of the incident covariance matrix $ {\bf
A}_{\cal N} $ are given as $ B_{1,2} = \tilde B_{\rm p}^{\rm s,i}
+ B_{\rm s,i}$ and $ C_{1,2} = \exp(i\theta_{1,2})\sqrt{\tilde
B_{\rm p}^{\rm s,i}(\tilde B_{\rm p}^{\rm s,i}+1)}$, $\theta_j =
\kappa_j+\pi/2$ for $ j=1,2 $, where $\tilde B_{\rm p}^{\rm s}$
$(\tilde B_{\rm p}^{\rm i})$ stands for the mean number of
squeezed photons in the signal (idler) mode, whereas the
corresponding mean signal (idler) noise photon number is denoted
as $ B_{\rm s} $ ($ B_{\rm i} $).

After the beam splitter, the local NIs $ I^{(j)}_{\rm ncl}$, EI $
\EI$ and global NI $ \GNI$ acquire the form:
\begin{widetext}
\begin{eqnarray}\label{eq_two_sq}           
 \LNI &=& T^2\tilde {B_{\rm p}^{\rm s}}(\tilde {B_{\rm p}^{\rm s}}+1) +R^2\tilde {B_{\rm p}^{\rm i}}(\tilde {B_{\rm
  p}^{\rm i}}+1) + TR\bar D_{12}'\cos(\theta_1-\theta_2)  - \left[ T\tilde B_{\rm p}^{\rm s}+R\tilde B_{\rm p}^{\rm i} +TB_{\rm s}+RB_{\rm i}\right]^2,
 \nonumber \\
 \GNI &=& B_1+ B_2- 2B_{\rm s}B_{\rm i}\Big[2B_1(1+\tilde B_{\rm p}^{\rm i})+2\tilde B_{\rm p}^{\rm i}
  (1+B_1)+
  B_{\rm i}(1+2B_1)+B_{\rm s}(1+2B_2)\Big] 
 -2(B_{\rm s}B_1 +B_{\rm i}B_2) -(B_{\rm s} +B_{\rm
  i})^2,
 \nonumber \\
 \EI &=& TR\Big[-\bar D_{12}'\cos(\theta_1-\theta_2)
  + (\tilde {B_{\rm p}^{\rm s}}+\tilde {B_{\rm p}^{\rm i}}+2\tilde {B_{\rm p}^{\rm s}}\tilde {B_{\rm p}^{\rm i}})-(B_{\rm s}+B_{\rm i})^2 - 2(\tilde {B_{\rm p}^{\rm s}}-\tilde {B_{\rm p}^{\rm i}})(B_{\rm s}-B_{\rm i})\Big] \nonumber \\
 & & \mbox{} +B_{\rm s}B_{\rm i}\Big[2\tilde {B_{\rm p}^{\rm s}}(1+B_{\rm i})
  +2\tilde {B_{\rm p}^{2}}(1+B_{\rm s})+4\tilde {B_{\rm p}^{\rm s}}\tilde {B_{\rm p}^{\rm i}}+(1+B_{\rm s})(1+B_{\rm i})\Big], \nonumber \\
\end{eqnarray}
\end{widetext}
where $\bar D_{12}'=2\sqrt{\tilde B_{\rm p}^{\rm s}(\tilde B_{\rm
p}^{\rm s}+1)\tilde B_{\rm p}^{\rm   i}  (\tilde B_{\rm p}^{\rm
i}+1)}$, $B_1= \tilde B_{\rm p}^{\rm s} + B_{\rm s}$, $B_2=B_{\rm
p}^{\rm i} + B_{\rm i}$, and, for simplicity, we assumed $\phi=0$ in
Eq.~(\ref{BS}). The formula for $ \LNII $ is obtained from that for $
\LNI $ in Eq.~(\ref{eq_two_sq}) with the substitution $ {\rm
s} \leftrightarrow {\rm i} $.

\begin{figure}[h!]      
\includegraphics[width=0.45\textwidth]{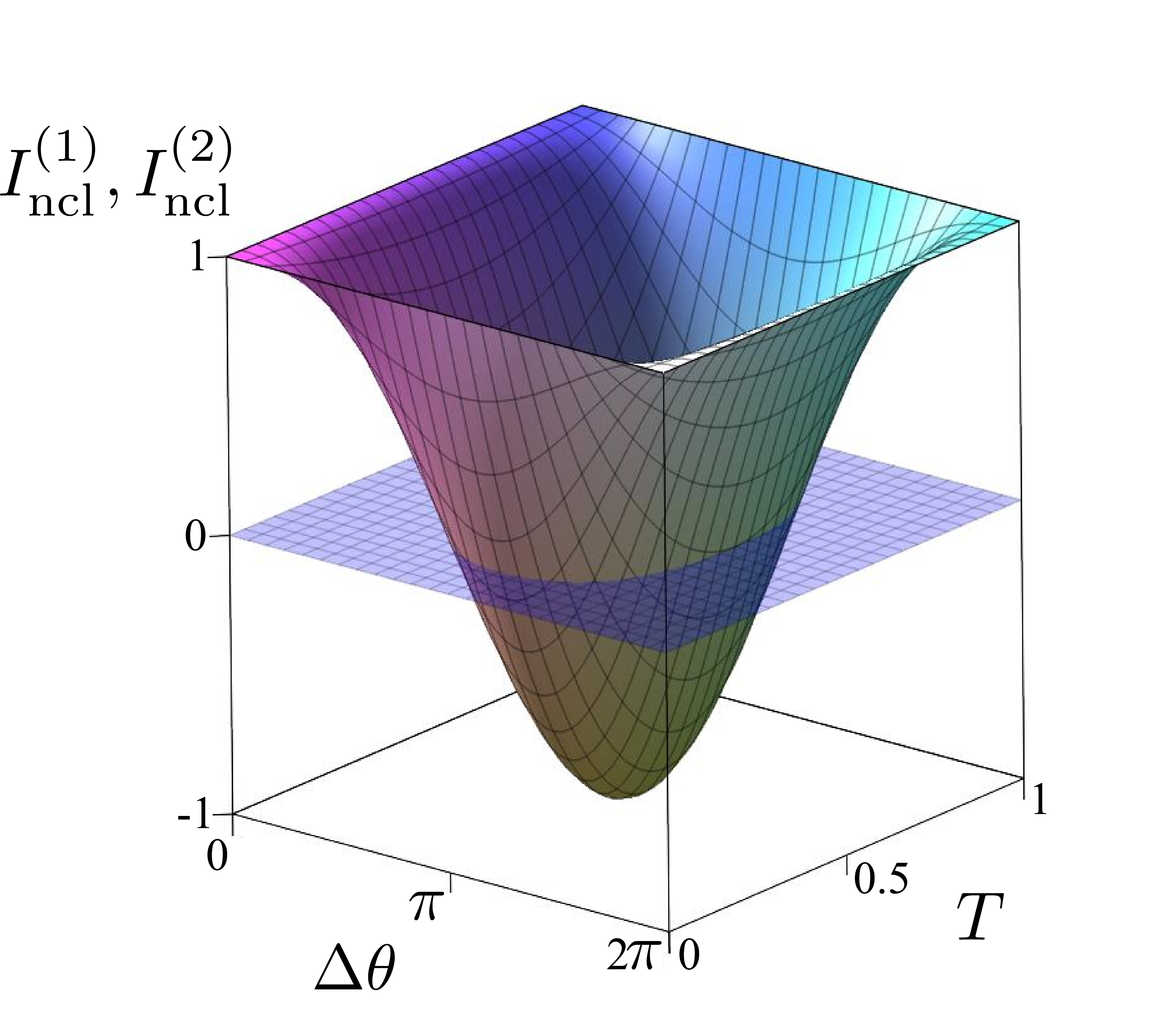}
 \centerline{(a)}
\includegraphics[width=0.45\textwidth]{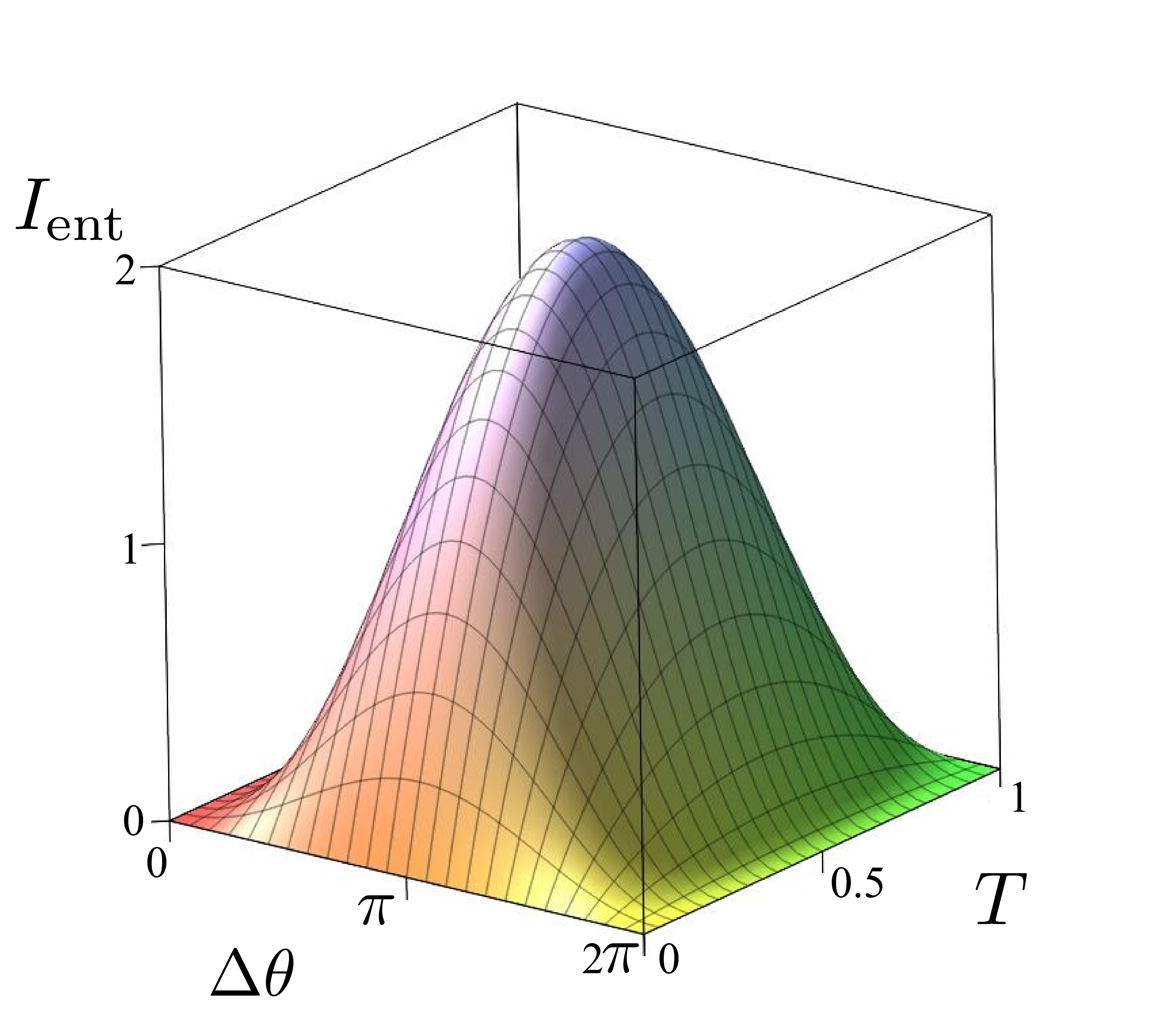}
 \centerline{(b)}
 \caption{(Color online) (a) Local nonclassicality invariants $
  \LNI=\LNII $  and (b) entanglement invariant $ \EI $  versus the
  phase difference $\Delta\theta$ and beam-splitter transmissivity
  $T$ for two noiseless squeezed states  according to Eq.~(\ref{eq_two_sq}); $ \tilde B_{\rm p}^{\rm s}=
  \tilde B_{\rm p}^{\rm i} =1$. In panel (a),  the blue surface at $\LNI=\LNII=0$ shows the boundary
between classical and nonclassical states.}
\label{fig9}
\end{figure}

The global NI $ \GNI $ does not depend on the relative phase
$\Delta\theta = \theta_1-\theta_2$ of two incident squeezed
states, while the local NIs $ I^{(j)}_{\rm ncl} $ and EI $\EI $
change with the relative phase $ \Delta\theta $. The case of two
equally intense incident noiseless squeezed states, as graphically
analyzed in Fig.~\ref{fig9}, shows that the phase difference $
\Delta\theta $ plays a crucial role in distributing the
nonclassicality between the output entanglement and local
nonclassicalities. If the phases $ \theta_1 $ and $ \theta_2 $ are
equal, the incident photon pairs stick (bunch) ideally together due to
the interference at the beam splitter and the incident
locally-nonclassical squeezed states are moved into the output
ports. No photon pair is broken and so no entanglement is
observed. On the other hand, if $ \Delta\theta = \pi $, then some
incident photon pairs are broken and, thus, the output squeezing
(as well as local nonclassicalities) is weaker. The broken photon
pairs give rise to the entanglement. The value of EI $ \EI $ is
maximal for the transmissivity $ T=\frac{1}{2} $. In this case, all the photon
pairs are broken, their signal and idler photons occur in
different output ports and, as a consequence, the ideal conditions
for entanglement generation are met. Hand in hand, the vanishing
local NIs $ I^{(j)}_{\rm ncl} $ are found (see Fig.~\ref{fig9}).

\begin{figure}[h!]  
\includegraphics[width=0.45\textwidth]{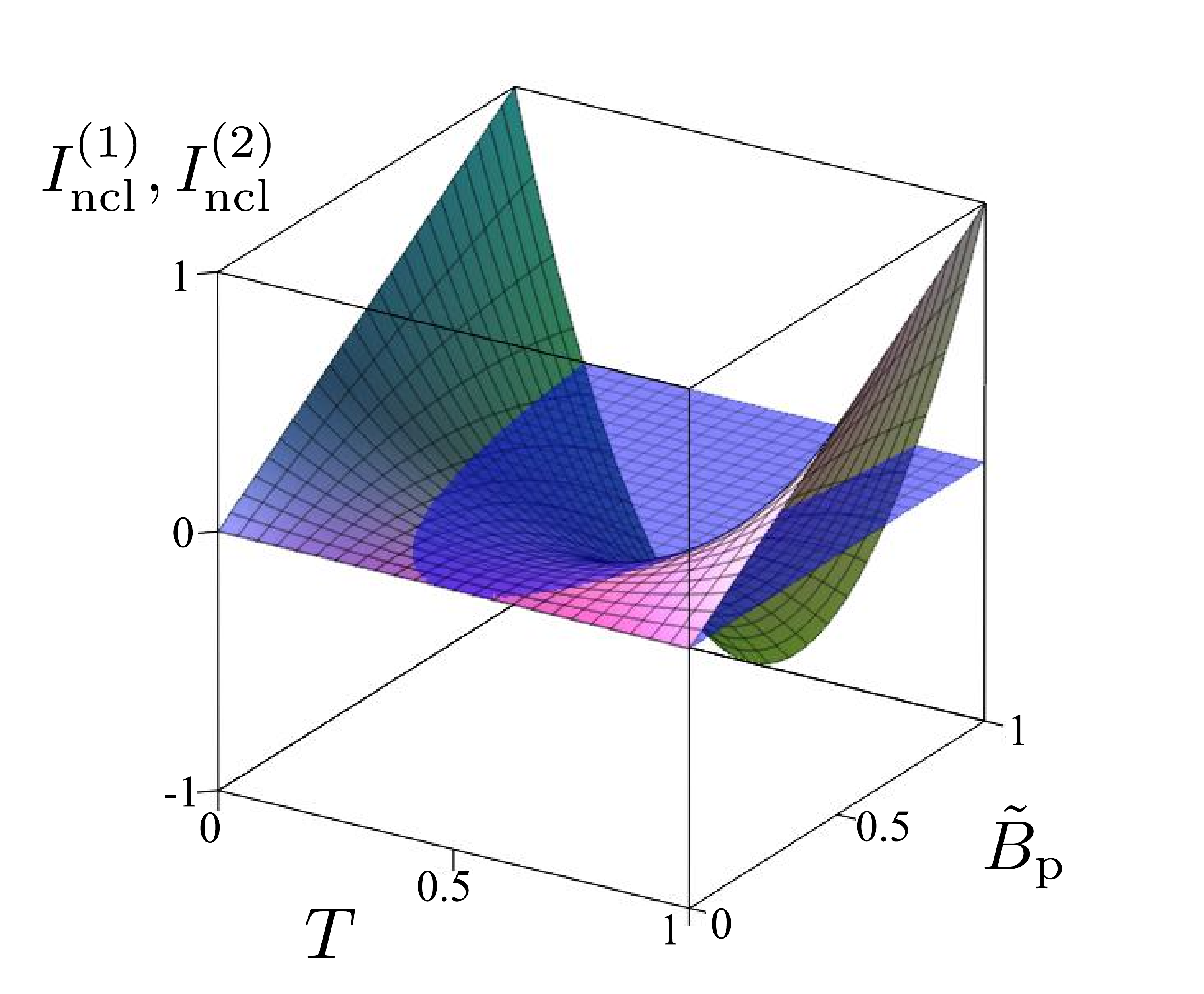}
 \centerline{(a)}
\includegraphics[width=0.45\textwidth]{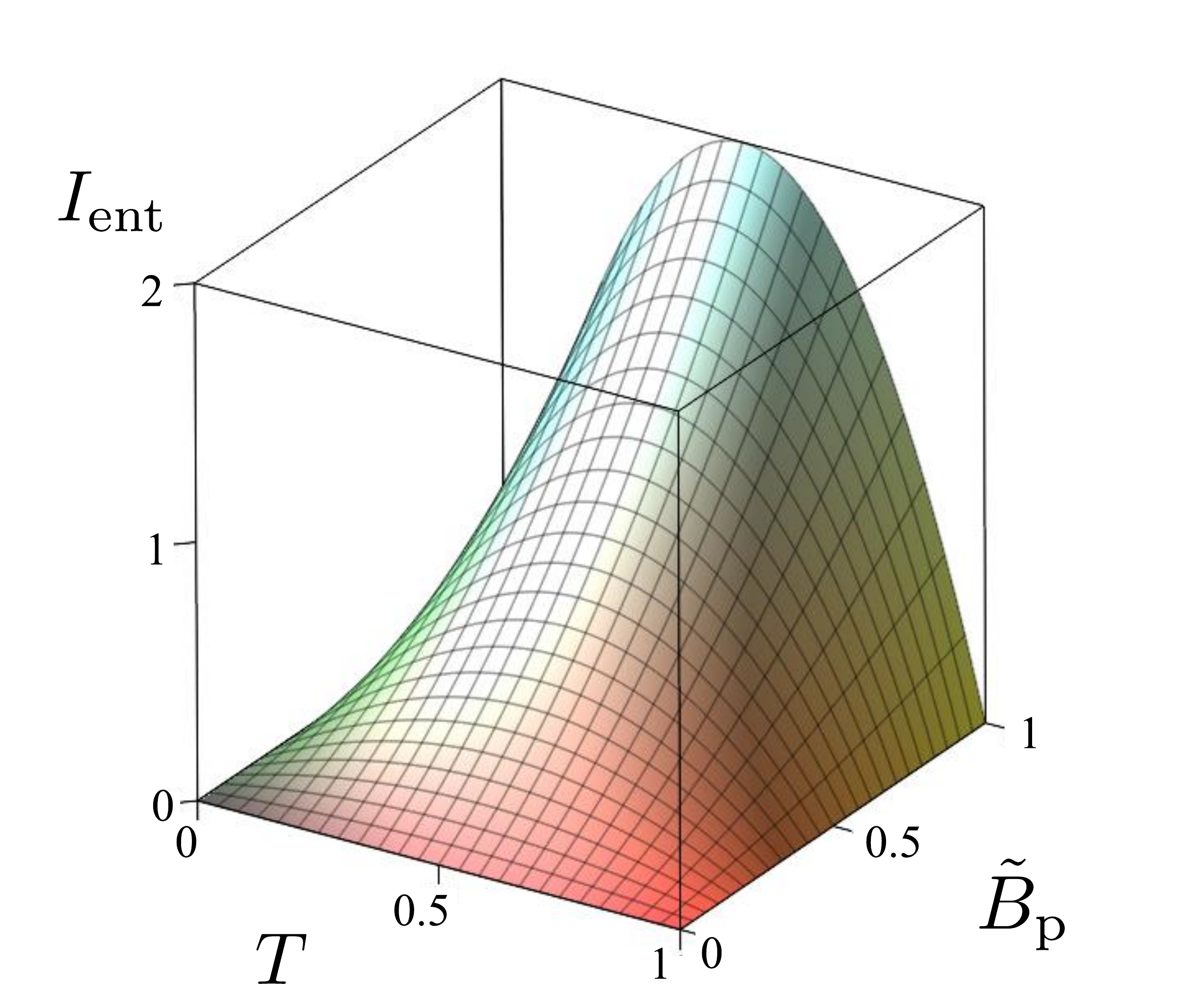}
 \centerline{(b)}
 \caption{(Color online) (a) Local nonclassicality invariant $ \LNI $
  and (b) entanglement invariant $ \EI $ versus the beam-splitter
  transmissivity $T$ and mean number $ \tilde B_{\rm p} $ of
  squeezed photons for two noiseless squeezed states according to Eq.~(\ref{eq_two_sq}); $ \tilde
  B_{\rm p} \equiv
  \tilde B_{\rm p}^{\rm s}= \tilde B_{\rm p}^{\rm i} $;
  $\Delta\theta=\pi$. In panel (a) the blue surface at $\LNI=\LNII=0$ shows the boundary
between classical and nonclassical states.}
\label{fig10}
\end{figure}

It is remarkable that the global NI $ \GNI $ for the equally
intense noiseless squeezed states is given formally by the same
formula as that valid for the noiseless twin beams considering the
mean photon-pair number $ B_{\rm p} $ instead of $\tilde B_{\rm
p}^{\rm s}=\tilde B_{\rm p}^{\rm i}\equiv \tilde B_{\rm p}$.
However, the incident twin beam serves as a source of
locally-nonclassical (squeezed) states, whereas the incident
squeezed states provide entangled states at the output of the beam
splitter. The comparison of graphs in Figs.~\ref{fig4}(a) and
\ref{fig5}(a) with those in Figs.~\ref{fig10}(a) and \ref{fig10}(b)
reveals that the incident noiseless squeezed states generate
entangled states for an arbitrary value of the transmissivity $ T
$, but the incident noiseless twin beams are capable of the
generation of the output squeezed states only in a certain
interval of the transmissivity $ T $ depending on the intensity.

\begin{figure}  
 \includegraphics[width=0.45\textwidth]{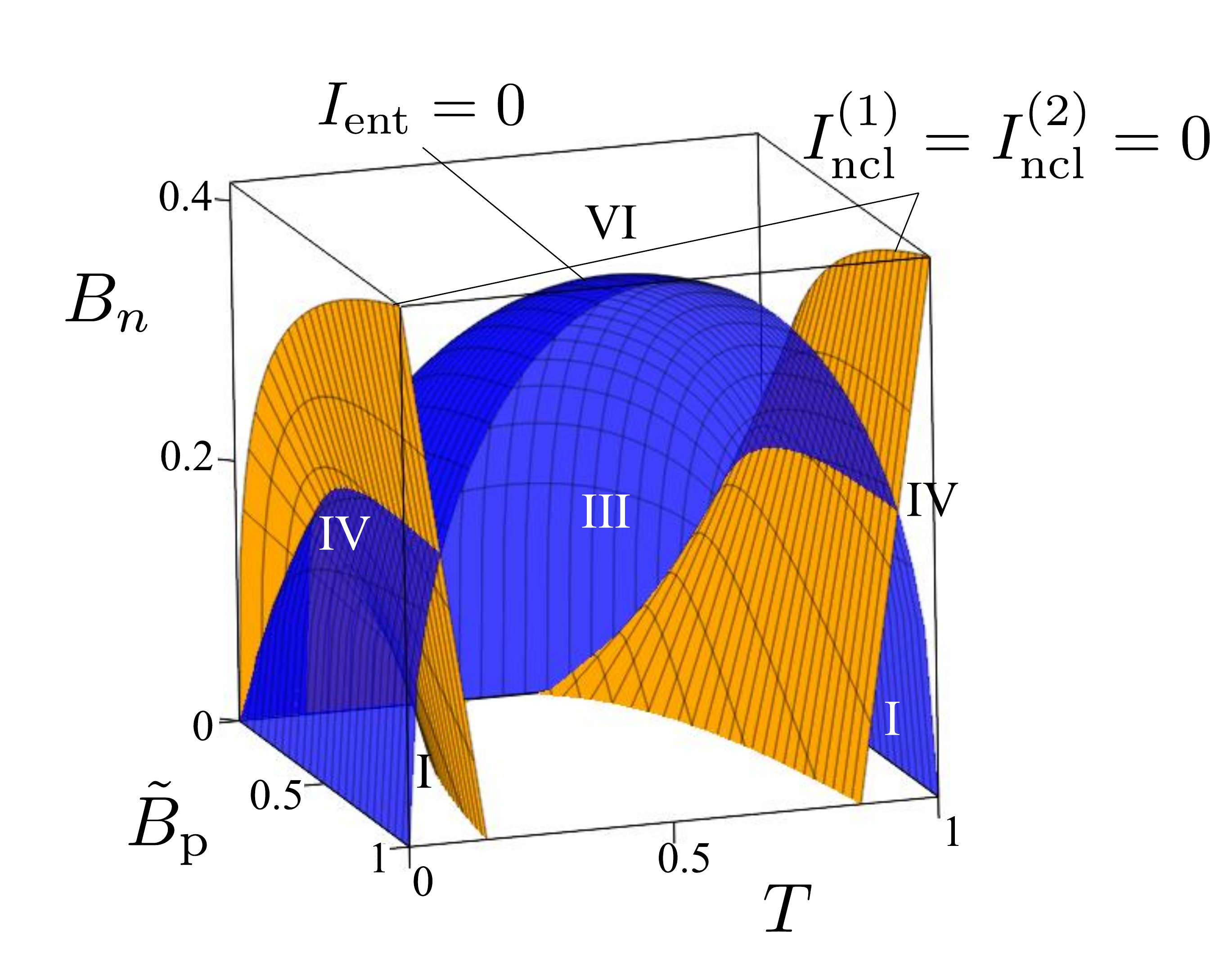}
\centerline{(a)}
 \includegraphics[width=0.45\textwidth]{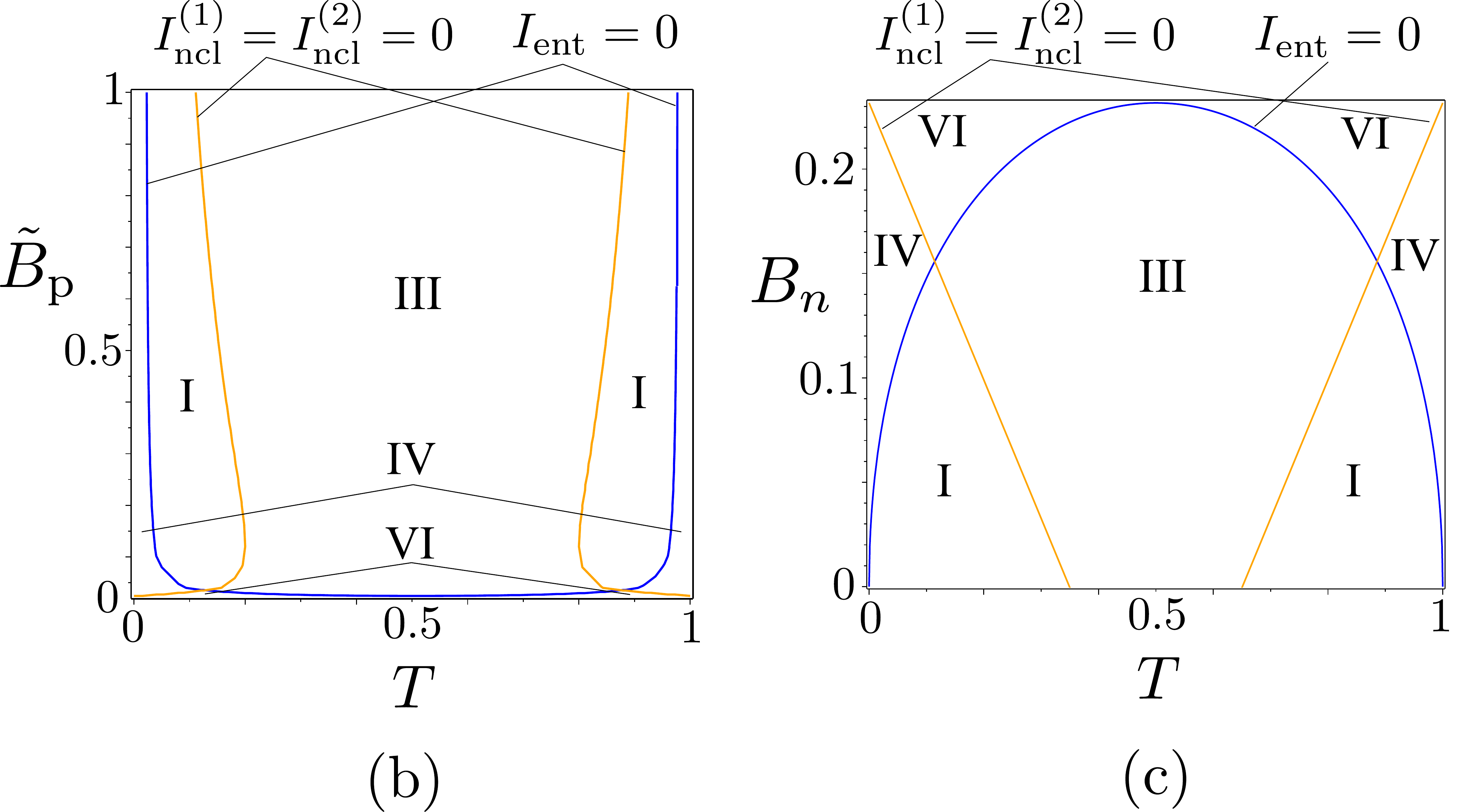}
 \caption{(Color online) Diagram (a) shows the
nonclassicality and entanglement invariants for the two squeezed
vacua occurring at the output ports of the beam splitter versus
the mean noise photon number $ B_{\rm n} $, mean number $ \tilde
B_{\rm p}$ of squeezed photons and transmissivity $ T $ assuming $
B_{\rm n}\equiv B_{\rm s}=B_{\rm i} $ and $ \tilde B_{\rm p}\equiv
\tilde B_{\rm p}^{\rm s} = \tilde B_{\rm p}^{\rm i} $ and
$\Delta\theta=\pi$. Surfaces at $ \LNj(B_{\rm n},\tilde B_{\rm
p},T) =  0 $ ($j=1,2$) (orange light gray) and $ \EI(B_{\rm
n},\tilde B_{\rm p},T) = 0 $ (blue dark gray) are shown
surrounding different regions specified in Tab.~\ref{table2}.
Diagrams~(b) and (c) show the perpendicular cross-sections of
diagram~(a) taken at given values of $B_{\rm n}=0.1$ and $\tilde
B_{\rm p}=0.1$, respectively. These cross-sections can be compared
with those in panels (b) and (c) in Figs.~\ref{fig6}
and~\ref{fig7}. } \label{fig11}
\end{figure}

Similarly as for the twin beams, the noise diminishes the global
NI $ \GNI $ [see the formula for $ \GNI $ in
Eq.~(\ref{eq_two_sq})]. Considering the incident states with $
\tilde B_{\rm p}^{\rm s} = \tilde B_{\rm p}^{\rm i} $ and $ B_{\rm
s}= B_{\rm i} $, the presence of noise leads to the occurrence of
the three different types of globally nonclassical states already
discussed in the connection with the noisy twin beams with
symmetric noise. Regions corresponding to different types of the
output states are shown in the diagram in Fig.~\ref{fig11}(a) that
can be compared with that of Fig.~\ref{fig6}(a).

\section{Twin beam mixed with squeezed states}

Finally, we analyze an interplay of noiseless twin beams and
equally populated noiseless squeezed states ($ \Delta\theta = 0 $)
in forming the output state at the beam splitter with phase $ \phi
$. Such state is generated by the Hamiltonian (\ref{hamil})
assuming $g_{11}=g_{22}=g$ and described by the following elements
of the evolution matrices $ {\bf U} $ and {\bf V}:
\begin{eqnarray} 
 U_{11} &= U_{22} &= \cosh(g_{12}t)\cosh(2gt), \nonumber \\
 V_{11} &= V_{22} &= i\cosh(g_{12}t)\sinh(2gt), \nonumber \\
 U_{12} &= U_{21} &= \sinh(g_{12}t)\sinh(2gt), \nonumber \\
 V_{12} &= V_{21} &= i\sinh(g_{12}t)\cosh(2gt).
\end{eqnarray}
Introducing the mean photon-pair number $ B_{\rm p} $ as $B_{\rm
p} = \sinh^2(g_{12}t)$ and mean number $\tilde B_{\rm p} $ of
squeezed photons per mode, $\tilde B_{\rm p} = \sinh^2(2gt) $, the
coefficients of the covariance matrix $ {\bf A}_{\cal N} $ are
found in the form:
\begin{eqnarray}  
 B_1&&=B_2 = B_{\rm p}+\tilde B_{\rm p}+2B_{\rm p}\tilde B_{\rm p}, \nonumber \\
 C_1&&=C_2 = i\sqrt{\tilde B_{\rm p}(\tilde B_{\rm p}+1)}(2B_{\rm p}+1), \nonumber \\
 D_{12} &&= i\sqrt{ B_{\rm p}( B_{\rm p}+1)}(2\tilde B_{\rm p}+1), \nonumber \\
 \bar D_{12} &&= -2\sqrt{ B_{\rm p}( B_{\rm p}+1)\tilde B_{\rm p}(\tilde B_{\rm p}+1)}.
\end{eqnarray}

\begin{figure} 
\includegraphics[width=0.45\textwidth]{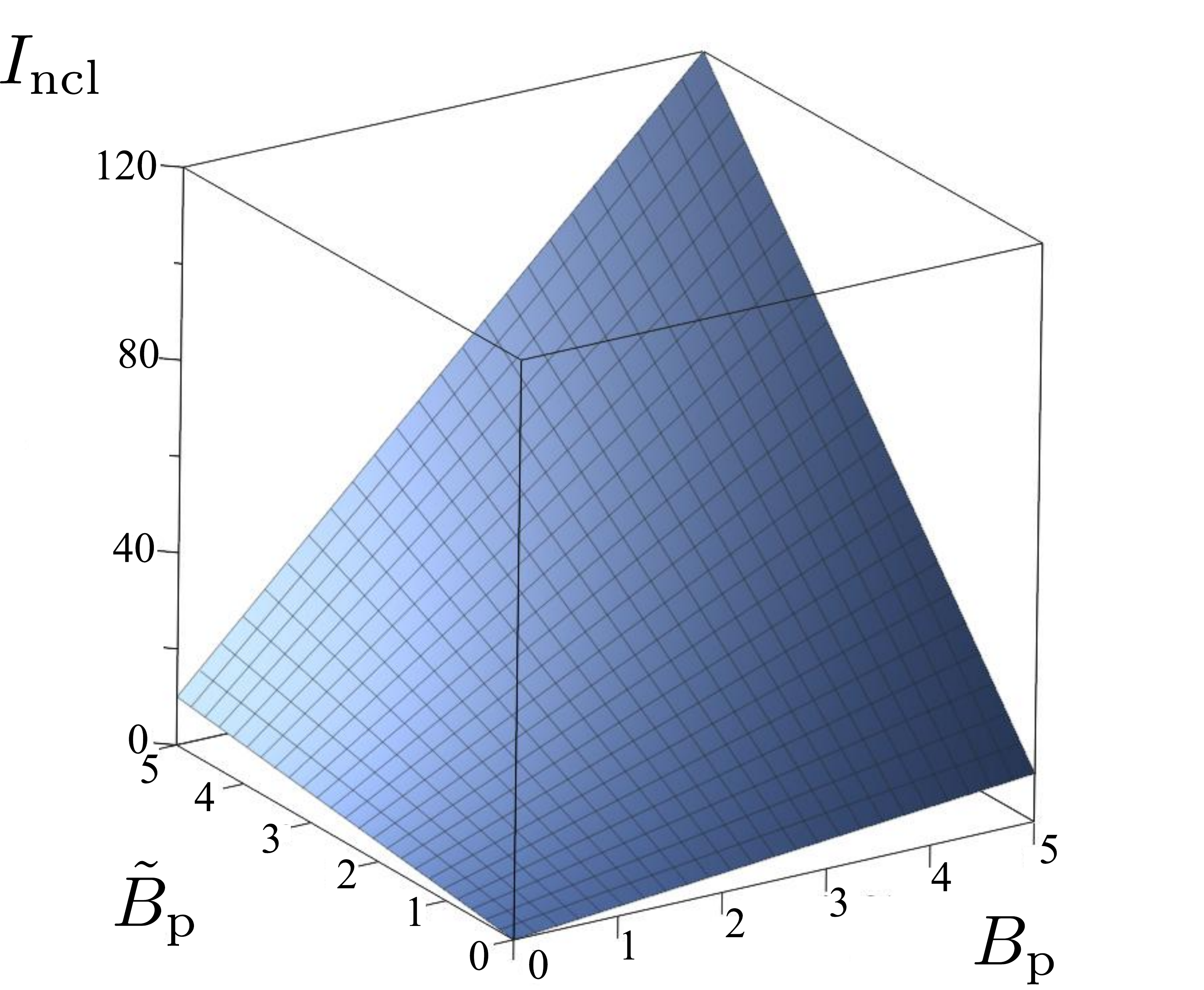}
 \caption{(Color online) Global nonclassicality invariant $ \GNI$ as a function of the mean photon-pair number $
  B_{\rm p}$ and mean number $ \tilde B_{\rm p}$ of squeezed
  photons considering the noiseless twin beams and squeezed states.}
\label{fig12}
\end{figure}

The local NIs $ I^{(j)}_{\rm ncl} $, EI $ \EI $, and global NI $
\GNI $ are then derived as follows:
\begin{eqnarray}\label{tw_sqz}          
 I^{(1,2)}_{\rm ncl} &=& [1-4TR\sin^2(\phi)]\tilde B_{\rm p}(\tilde B_{\rm p}+1)+4TR B_{\rm p}(B_{\rm p}+1) \nonumber \\
&&-(\tilde B_{\rm p}-B_{\rm p})^2 \pm K ,\nonumber \\
  K &=& 4\sqrt{TR}\cos(\phi) \sqrt{ B_{\rm p}( B_{\rm p}+1)\tilde B_{\rm p}(\tilde B_{\rm p}+1)},\nonumber \\
  \EI &=& (T-R)^2 B_{\rm p}( B_{\rm p}+1)+4TR\sin^2(\phi) \tilde B_{\rm p}(\tilde B_{\rm p}+1), \nonumber \\
\GNI &=& 2(B_{\rm p}+\tilde B_{\rm p}+2B_{\rm p}\tilde B_{\rm p}).
\end{eqnarray}
The formula for the global NI $ \GNI $, given in Eq.~(\ref{tw_sqz}), shows
that both parametric down-conversion and second subharmonic
generation contribute to the global NI making $ \GNI $ always
positive. Moreover, both processes enhance each other in producing
larger values of the global NI. The greater is the mean
photon-pair number $ B_{\rm p} $ and also the greater is the mean
number $ \tilde B_{\rm p} $ of squeezed photons, the greater is
the global NI $ \GNI $ (see Fig.~\ref{fig12}). Additionally, both
 LNI $ I^{(j)}_{\rm ncl} $ and EI $\EI$ become dependent on the
phase $ \phi $ of the beam splitter.

\begin{figure} 
\includegraphics[width=0.45\textwidth]{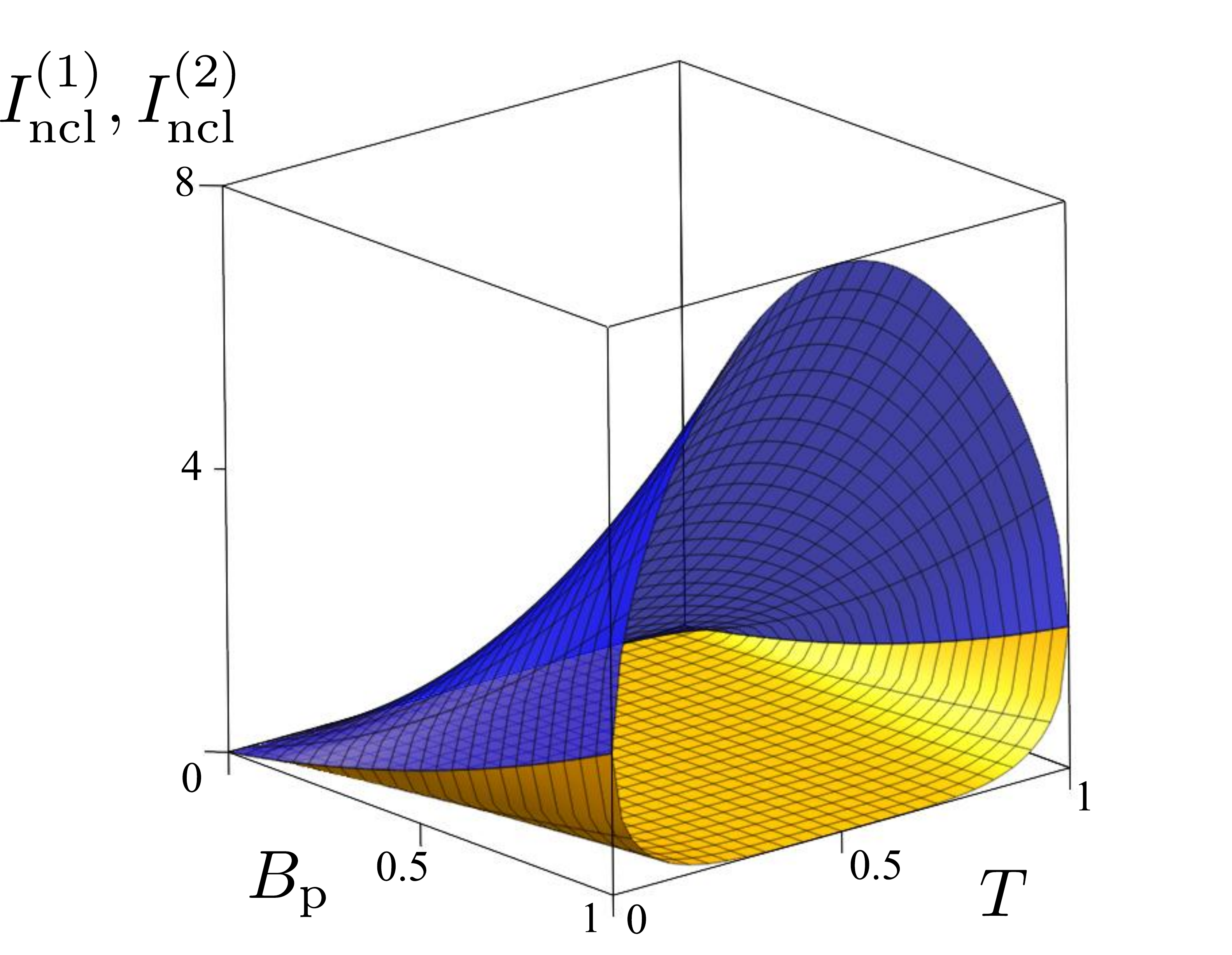}
 \caption{(Color online) Local nonclassicality invariants
  $\LNI$ (blue dark upper surface) and $\LNII$
  (orange light coloured lower surface) versus the beam-splitter transmissivity $T$
  and mean photon-pair number $ B_{\rm p}$ assuming $ B_{\rm
  p}=\tilde B_{\rm p}$ appropriate for the noiseless twin beams and
  squeezed states according to Eq.~(\ref{tw_sqz}) assuming $\phi=0$.}
\label{fig13}
\end{figure}

Provided that the phases of the incident squeezed states equal ($
\phi = n\pi $, $ n\in Z $), photons in pairs stick together (bunch) completely
when propagating through the beam splitter and so they cannot
contribute to the entanglement in the output state. In this case,
the entanglement originates only in photon pairs of the incident
twin beam. When $ T=1/2 $ all photons in pairs from the twin beam are
glued and so the output state is separable. Contrary to this, the
local NIs $ I^{(j)}_{\rm ncl} $ depend on both mean
photon-pair number $ B_{\rm p} $ and mean number $ \tilde B_{\rm
p} $ of squeezed photons. The fields characterizing photon pairs
in individual output ports and originating in the incident
squeezed states and the incident twin beam  interfere
causing the asymmetry between the output ports. Depending on the
parity of $n$ one obtains the maximal local NI $\LNI$ ($\LNII$) if
$n=2k$ $(n=2k+1)$, $k\in\mathbb Z$. This asymmetry is the largest
for $ T=1/2 $. In this case, the bunched photon pairs are completely
missing in one output port due to completely destructive
interference. On the other hand, constructive interference
provides the greatest number of the bunched photon pairs in the
other output port guaranteeing the largest attainable value of its
local NI $ I^{(j)}_{\rm ncl} $. This behavior is quantified in the
graph in Fig.~\ref{fig13}.

If $\phi=\frac{\pi}{2} + n\pi $, the local NIs are equal
($\LNI=\LNII$) and the state at the beam-splitter output ports
acquires a symmetry. Under these phase relations, also the incident
squeezed photon pairs contribute, together with the twin-beam
photon pairs, to the entanglement. It is worth noting that for
$B_{\rm p}=\tilde B_{\rm p}$ all the state quantifiers are the
same: $\LNI=\LNII=\EI=B_{\rm p}(B_{\rm p}+1)$.

\section{Conclusions}

Local and global invariants of the general two-mode Gaussian
states have been used to construct a specific local
nonclassicality quantifier and entanglement quantifier. These
quantifiers applied, respectively, to the single-mode marginal
states and the whole two-mode state add together to give a
quantity that is invariant under global linear unitary
transformations. This invariant then quantifies the
nonclassicality resources of Gaussian states. Remarkably, this
invariant is linearly proportional to the number of photon pairs
in the noiseless Gaussian states. The general results have been
used to study the beam-splitter transformations of fields composed
of photon pairs and additional noisy photons. Twin beams, squeezed
states as well as their combinations have been considered as
important examples. The behavior of photon pairs at the beam
splitter causing their breaking or gluing (i.e., antibinching or
bunching) has been used to explain the flow of nonclassical
resources between  local nonclassicalities (implying squeezing)
and entanglement. A complete transfer of the entanglement of
incident twin beams into the squeezing of the output modes has
been observed. Also the complete transfer of the incident
squeezing into the entanglement of the output fields can be
reached. The role of noise in the transfer of the nonclassicality
invariant via the beam splitter has been elucidated on several
examples.

\acknowledgments The authors thank Ji\v r\'i Svozil\'ik and
Anirban Pathak for discussions. This work was supported by the
projects No.~P205/12/0382 of the GA CR and No.~LO1305 of the MSMT
CR. I.A. acknowledges the project IGA\_PrF\_2016\_002 of IGA UP
Olomouc. A.M. acknowledges the support of a grant from the John
Templeton Foundation.

\appendix

\section{Quasiprobability distributions and characteristic functions}

For the completeness and clarity of our presentation, here we give
a few well-known formulas relating the quantities given in
Tab.~\ref{table2}, as derived by Cahill and
Glauber~\cite{Cahill69}. This approach is a generalization of the
standard Wigner and Glauber formalisms.

The Cahill-Glauber $s$-parametrized (or $s$-ordered)
quasiprobability distribution (QPD, quasidistribution), ${\cal
W}^{(s)}(\bm{\alpha})$ for an $N$-mode bosonic state $\hat\rho$
can be defined for a real parameter $s\in[-1,1]$ as
\begin{equation}
{\cal W}^{(s)}(\bm{\alpha}) = {\rm Tr} \left[
\hat{\rho}\hat{T}^{(s)}(\bm{\alpha}) \right], \label{185}
\end{equation}
which is the mean value of the operator $\hat{T}(\bm{\alpha})$
defined as the Fourier transform,
\begin{equation}
\hat{T}^{(s)}(\bm{\alpha}) =\int\hat{D}^{(s)}(\bm{\beta})\:
\exp\!\left( \sum_{n}\alpha_{n}\beta_{n}^*-{\rm c.c.}\right) {\rm
d}^2 \bm{\beta}', \label{183}
\end{equation}
of the $s$-parametrized multimode displacement operator given by
\begin{eqnarray}
\hat{D}^{(s)} (\bm{\beta}) &=& \prod_{n} \hat{D}^{(s)}(\beta_{n}) \nonumber \\
 &=& \prod_{n} \exp \left( \beta_{n} \hat{a}_{n}^\dagger -
\beta_{n}^* \hat{a}_{n} +\frac{s}{2} \left| \beta_{n} \right|^2
\right). \label{019}
\end{eqnarray}
Here, $\hat{a}_{n}$ ($\hat{{a}}_n^\dagger$) is the bosonic
annihilation (creation) operator for the $n$-th mode
($n=1,2,...,N$). The complex multivariable
$\bm{\alpha}\equiv\{\alpha_{n}\}=(\alpha_1,\alpha_2,...,\alpha_{N})$
is applied here as in Eq.~(\ref{Pfunction}), and, analogously
$\bm{\beta}\equiv\{\beta_{n}\}=(\beta_1,\beta_2,...,\beta_{N})$.
The symbol c.c. denotes the complex conjugate term, and the
integration is performed over $d^2\bm{\beta}'  \equiv
d^2\{\beta_{n}/\pi\}= \pi^{-N} \prod_{n} d^2 \beta_n = \pi^{-N}
\prod_{n}d({\rm Re} \beta_n) d({\rm Im} \beta_n)$.

In the special cases for $s=1,0,-1$, the QPD ${\cal
W}^{(s)}(\bm{\alpha})$ reduces to the popular Glauber-Sudarshan
$P$, Wigner $W$, and Husimi $Q$ functions corresponding to the
normal, symmetric, and antinormal orderings, respectively. Our
analysis in the paper is focused on the normally and
symmetrically-ordered functions. We recall that the standard
definition of nonclassicality is based on the non-positivity of
the $P$ function.

The statistical operator $\hat{\rho}$ corresponding to a given QPD
can be calculated as follows
\begin{eqnarray}
\hat{\rho}=\int{\cal W}^{(s)}(\bm{\alpha})\:
\hat{T}^{(-s)}(\bm{\alpha})\: {\rm d}^2\bm{\alpha}', \label{184}
\end{eqnarray}
which, in the special case for $s=1$ reduces to
Eq.~(\ref{Pfunction}) describing the $P$ representation of a given
state $\hat{\rho}$.

The $N$-mode $s$-parameterized characteristic function ${\cal
C}^{(s)}(\bm{\beta})$ for a given state $\hat{\rho}$ can be
defined as
\begin{eqnarray}\label{A5}
{\cal C}^{(s)} (\bm{\beta}) = {\rm Tr} \left[ \hat{\rho}
\hat{D}^{(s)} (\bm{\beta}) \right], \label{020}
\end{eqnarray}
which is the mean value of the multimode displacement operator
$\hat{D}^{(s)} (\bm{\beta})$. We recall that our description of
nonclassicality and entanglement is based on two special cases of
these characteristic functions. Specifically, the normal
characteristic function $ C_{\mathcal N} $, defined in
Eq.~(\ref{CF_normal}), is the special case of Eq.~(\ref{020}) for
$s=1$ assuming two-mode ($N=2$) field. While the symmetric characteristic function
$C_{\cal S}$ is given by Eq.~(\ref{A5}) for $s=0$ and $N=2$.

For Gaussian states, which are solely analyzed in this paper, the
characteristic function ${\cal C}^{(s)} (\bm{\beta})$ can uniquely
be defined via the covariance matrices, which are given by
Eq.~(\ref{CM}) for normal ordering ($s=1$) and by Eq.~(\ref{CMps})
for symmetric ordering ($s=0$) for a two-mode case.

By comparing the definitions in Eqs.~(\ref{185}) and~(\ref{020}),
it is easy to conclude that the $s$-parameterized QPD and
characteristic function for any $s\in[-1,1]$ are related via the
Fourier transform, i.e.,
\begin{eqnarray}
{\cal W}^{(s)} (\bm{\alpha}) = \int {\cal C}^{(s)} (\bm{\beta})
\prod_{n}\exp   \left( \alpha_{n} \beta_{n}^* - \alpha_{n}^*
\beta_{n} \right)  {\rm d}^2 \bm{\beta}',\quad \label{021}
\\
{\cal C}^{(s)} (\bm{\beta}) = \int {\cal W}^{(s)} (\bm{\alpha})
\prod_n \exp  \left( \alpha_{n}^* \beta_{n} - \alpha_{n}
\beta_{n}^* \right)  {\rm d}^2 \bm{\alpha}', \quad \label{022}
\end{eqnarray}
where the integration over ${\rm d}^2 \bm{\alpha}'$ is defined
analogously to ${\rm d}^2 \bm{\beta}'$, as in Eq.~(\ref{183}).
The normalization conditions are as follows
\begin{eqnarray}
\int {\cal W}^{(s)} (\bm{\alpha}) {\rm d}^2 \bm{\alpha}' = {\cal
C}^{(s)}(\bm{\beta}=\bm{0}) = 1. \label{023}
\end{eqnarray}

The relation between the QPDs ${\cal W}^{(s_1)} (\bm{\alpha})$ and
${\cal W}^{(s_2)} (\bm{\alpha})$, assuming $s_2<s_1$, is simply
given by
\begin{eqnarray}
{\cal W}^{(s_2)} (\bm{\alpha}) \:=\: \left( \frac{2}{s_1-s_2}
\right)^M \int  {\cal W}^{(s_1)} (\bm{\beta})
\hspace{1.5cm}\nonumber\\
\times\exp\left( - \frac{2}{s_1-s_2} \sum_{n} \left| \alpha_n -
\beta_n \right|^2 \right) {\rm d}^2 \bf{\beta}'. \label{026}
\end{eqnarray}
This means that the QPD ${\cal W}^{(s_2)} (\bm{\alpha})$ with any
parameter $s_2\in[-1,1]$ can easily be obtained by mixing the
$P$~function (corresponding to $s_1=1$) with the proper amount of
 Gaussian noise. The relation between the characteristic
functions corresponding to different parameters $s_1$ and $s_2$
reads as
\begin{eqnarray}
{\cal C}^{(s_2)} (\bm{\beta}) =\: {\cal C}^{(s_1)} (\bm{\beta})
\exp \left( \frac{s_2-s_1}{2} \sum_{n} \left|\beta_n
\right|^2\right). \label{027}
\end{eqnarray}
It is valid for any $s_1,s_2\in[-1,1]$, contrary to the analogous
relation in Eq.~(\ref{026}) for the QPDs. We applied
Eq.~(\ref{027}) to calculate the symmetrically-ordered
characteristic function $C_{\cal S}\equiv {\cal C}^{(0)}$ from the
normally-ordered characteristic function $C_{\cal N}\equiv {\cal
C}^{(1)}$, given by Eq.~(\ref{CF_normal}), for two-mode states.


\begin{thebibliography}{52}%
\makeatletter
\providecommand \@ifxundefined [1]{%
 \@ifx{#1\undefined}
}%
\providecommand \@ifnum [1]{%
 \ifnum #1\expandafter \@firstoftwo
 \else \expandafter \@secondoftwo
 \fi
}%
\providecommand \@ifx [1]{%
 \ifx #1\expandafter \@firstoftwo
 \else \expandafter \@secondoftwo
 \fi
}%
\providecommand \natexlab [1]{#1}%
\providecommand \enquote  [1]{``#1''}%
\providecommand \bibnamefont  [1]{#1}%
\providecommand \bibfnamefont [1]{#1}%
\providecommand \citenamefont [1]{#1}%
\providecommand \href@noop [0]{\@secondoftwo}%
\providecommand \href [0]{\begingroup \@sanitize@url \@href}%
\providecommand \@href[1]{\@@startlink{#1}\@@href}%
\providecommand \@@href[1]{\endgroup#1\@@endlink}%
\providecommand \@sanitize@url [0]{\catcode `\\12\catcode
`\$12\catcode
  `\&12\catcode `\#12\catcode `\^12\catcode `\_12\catcode `\%12\relax}%
\providecommand \@@startlink[1]{}%
\providecommand \@@endlink[0]{}%
\providecommand \url  [0]{\begingroup\@sanitize@url \@url }%
\providecommand \@url [1]{\endgroup\@href {#1}{\urlprefix }}%
\providecommand \urlprefix  [0]{URL }%
\providecommand \Eprint [0]{\href }%
\providecommand \doibase [0]{http://dx.doi.org/}%
\providecommand \selectlanguage [0]{\@gobble}%
\providecommand \bibinfo  [0]{\@secondoftwo}%
\providecommand \bibfield  [0]{\@secondoftwo}%
\providecommand \translation [1]{[#1]}%
\providecommand \BibitemOpen [0]{}%
\providecommand \bibitemStop [0]{}%
\providecommand \bibitemNoStop [0]{.\EOS\space}%
\providecommand \EOS [0]{\spacefactor3000\relax}%
\providecommand \BibitemShut  [1]{\csname bibitem#1\endcsname}%
\let\auto@bib@innerbib\@empty
\bibitem [{\citenamefont {Einstein}(1905)}]{Einstein05}%
  \BibitemOpen
  \bibfield  {author} {\bibinfo {author} {\bibfnamefont {A.}~\bibnamefont
  {Einstein}},\ }\bibfield  {title} {\enquote {\bibinfo {title} {{Ü}ber einen
  die {E}rzeugung und {V}erwandlung des {L}ichtes betreffenden heuristischen
  {G}esichtspunkt},}\ }\href@noop {} {\bibfield  {journal} {\bibinfo  {journal}
  {Ann. Phys.}\ }\textbf {\bibinfo {volume} {17}},\ \bibinfo {pages} {132}
  (\bibinfo {year} {1905})}\BibitemShut {NoStop}%
\bibitem [{\citenamefont {Einstein}\ \emph {et~al.}(1935)\citenamefont
  {Einstein}, \citenamefont {Podolsky},\ and\ \citenamefont
  {Rosen}}]{Einstein35}%
  \BibitemOpen
  \bibfield  {author} {\bibinfo {author} {\bibfnamefont {A.}~\bibnamefont
  {Einstein}}, \bibinfo {author} {\bibfnamefont {B.}~\bibnamefont {Podolsky}},
  \ and\ \bibinfo {author} {\bibfnamefont {N.}~\bibnamefont {Rosen}},\
  }\bibfield  {title} {\enquote {\bibinfo {title} {Can quantum-mechanical
  description of physical reality be considered complete?}}\ }\href@noop {}
  {\bibfield  {journal} {\bibinfo  {journal} {Phys. Rev.}\ }\textbf {\bibinfo
  {volume} {47}},\ \bibinfo {pages} {777} (\bibinfo {year} {1935})}\BibitemShut
  {NoStop}%
\bibitem [{\citenamefont {Schr{\"o}dinger}(1935)}]{Schrodinger35}%
  \BibitemOpen
  \bibfield  {author} {\bibinfo {author} {\bibfnamefont {E.}~\bibnamefont
  {Schr{\"o}dinger}},\ }\bibfield  {title} {\enquote {\bibinfo {title} {Die
  gegenwartige {S}ituation in der {Q}uantenmechanik},}\ }\href@noop {}
  {\bibfield  {journal} {\bibinfo  {journal} {Naturwissenschaften}\ }\textbf
  {\bibinfo {volume} {23}},\ \bibinfo {pages} {807} (\bibinfo {year}
  {1935})}\BibitemShut {NoStop}%
\bibitem [{\citenamefont {Pe\v{r}ina}\ \emph {et~al.}(1994)\citenamefont
  {Pe\v{r}ina}, \citenamefont {Hradil},\ and\ \citenamefont
  {Jur\v{c}o}}]{Perina1994Book}%
  \BibitemOpen
  \bibfield  {author} {\bibinfo {author} {\bibfnamefont {J.}~\bibnamefont
  {Pe\v{r}ina}}, \bibinfo {author} {\bibfnamefont {Z.}~\bibnamefont {Hradil}},
  \ and\ \bibinfo {author} {\bibfnamefont {B.}~\bibnamefont {Jur\v{c}o}},\
  }\href@noop {} {\emph {\bibinfo {title} {Quantum Optics and Fundamentals of
  Physics}}}\ (\bibinfo  {publisher} {Kluwer, Dordrecht},\ \bibinfo {year}
  {1994})\BibitemShut {NoStop}%
\bibitem [{\citenamefont {Mandel}\ and\ \citenamefont
  {Wolf}(1995)}]{MandelBook}%
  \BibitemOpen
  \bibfield  {author} {\bibinfo {author} {\bibfnamefont {L.}~\bibnamefont
  {Mandel}}\ and\ \bibinfo {author} {\bibfnamefont {E.}~\bibnamefont {Wolf}},\
  }\href@noop {} {\emph {\bibinfo {title} {Optical Coherence and Quantum
  Optics}}}\ (\bibinfo  {publisher} {Cambridge University Press},\ \bibinfo
  {address} {Cambridge},\ \bibinfo {year} {1995})\BibitemShut {NoStop}%
\bibitem [{\citenamefont {Glauber}(2007)}]{GlauberBook}%
  \BibitemOpen
  \bibfield  {author} {\bibinfo {author} {\bibfnamefont {R.~J.}\ \bibnamefont
  {Glauber}},\ }\href@noop {} {\emph {\bibinfo {title} {Quantum Theory of
  Optical Coherence: Selected Papers and Lectures}}}\ (\bibinfo  {publisher}
  {Wiley-VCH},\ \bibinfo {address} {Weinheim},\ \bibinfo {year}
  {2007})\BibitemShut {NoStop}%
\bibitem [{\citenamefont {Dodonov}\ and\ \citenamefont
  {Man'ko}(2003)}]{DodonovBook}%
  \BibitemOpen
  \bibfield  {author} {\bibinfo {author} {\bibfnamefont {V.}~\bibnamefont
  {Dodonov}}\ and\ \bibinfo {author} {\bibfnamefont {V.}~\bibnamefont
  {Man'ko}},\ }\href@noop {} {\emph {\bibinfo {title} {Theory of Nonclassical
  States of Light}}}\ (\bibinfo  {publisher} {Taylor \& Francis, New York},\
  \bibinfo {year} {2003})\BibitemShut {NoStop}%
\bibitem [{\citenamefont {Vogel}\ and\ \citenamefont
  {Welsch}(2006)}]{VogelBook}%
  \BibitemOpen
  \bibfield  {author} {\bibinfo {author} {\bibfnamefont {W.}~\bibnamefont
  {Vogel}}\ and\ \bibinfo {author} {\bibfnamefont {D.}~\bibnamefont {Welsch}},\
  }\href@noop {} {\emph {\bibinfo {title} {Quantum Optics}}}\ (\bibinfo
  {publisher} {Wiley-VCH, Weinheim},\ \bibinfo {year} {2006})\BibitemShut
  {NoStop}%
\bibitem [{\citenamefont {Pe\v{r}ina}(1991)}]{Perina1991Book}%
  \BibitemOpen
  \bibfield  {author} {\bibinfo {author} {\bibfnamefont {J.}~\bibnamefont
  {Pe\v{r}ina}},\ }\href@noop {} {\emph {\bibinfo {title} {Quantum Statistics
  of Linear and Nonlinear Optical Phenomena}}}\ (\bibinfo  {publisher} {Kluwer,
  Dordrecht},\ \bibinfo {year} {1991})\BibitemShut {NoStop}%
\bibitem [{\citenamefont {Glauber}(1963)}]{Glauber63}%
  \BibitemOpen
  \bibfield  {author} {\bibinfo {author} {\bibfnamefont {R.~J.}\ \bibnamefont
  {Glauber}},\ }\bibfield  {title} {\enquote {\bibinfo {title} {Coherent and
  incoherent states of the radiation field},}\ }\href@noop {} {\bibfield
  {journal} {\bibinfo  {journal} {Phys. Rev.}\ }\textbf {\bibinfo {volume}
  {131}},\ \bibinfo {pages} {2766} (\bibinfo {year} {1963})}\BibitemShut
  {NoStop}%
\bibitem [{\citenamefont {Sudarshan}(1963)}]{Sudarshan63}%
  \BibitemOpen
  \bibfield  {author} {\bibinfo {author} {\bibfnamefont {E.~C.~G.}\
  \bibnamefont {Sudarshan}},\ }\bibfield  {title} {\enquote {\bibinfo {title}
  {Equivalence of semiclassical and quantum mechanical descriptions of
  statistical light beams},}\ }\href@noop {} {\bibfield  {journal} {\bibinfo
  {journal} {Phys. Rev. Lett.}\ }\textbf {\bibinfo {volume} {10}},\ \bibinfo
  {pages} {277} (\bibinfo {year} {1963})}\BibitemShut {NoStop}%
\bibitem [{\citenamefont {Richter}\ and\ \citenamefont
  {Vogel}(2002)}]{Richter02}%
  \BibitemOpen
  \bibfield  {author} {\bibinfo {author} {\bibfnamefont {T.}~\bibnamefont
  {Richter}}\ and\ \bibinfo {author} {\bibfnamefont {W.}~\bibnamefont
  {Vogel}},\ }\bibfield  {title} {\enquote {\bibinfo {title} {Nonclassicality
  of quantum states: A hierarchy of observable conditions},}\ }\href@noop {}
  {\bibfield  {journal} {\bibinfo  {journal} {Phys. Rev. Lett.}\ }\textbf
  {\bibinfo {volume} {89}},\ \bibinfo {eid} {283601} (\bibinfo {year}
  {2002})}\BibitemShut {NoStop}%
\bibitem [{\citenamefont {Miranowicz}\ \emph {et~al.}(2010)\citenamefont
  {Miranowicz}, \citenamefont {Bartkowiak}, \citenamefont {Wang}, \citenamefont
  {Liu},\ and\ \citenamefont {Nori}}]{Miranowicz10}%
  \BibitemOpen
  \bibfield  {author} {\bibinfo {author} {\bibfnamefont {A.}~\bibnamefont
  {Miranowicz}}, \bibinfo {author} {\bibfnamefont {M.}~\bibnamefont
  {Bartkowiak}}, \bibinfo {author} {\bibfnamefont {X.}~\bibnamefont {Wang}},
  \bibinfo {author} {\bibfnamefont {Y.-X.}\ \bibnamefont {Liu}}, \ and\
  \bibinfo {author} {\bibfnamefont {F.}~\bibnamefont {Nori}},\ }\bibfield
  {title} {\enquote {\bibinfo {title} {Testing nonclassicality in multimode
  fields: A unified derivation of classical inequalities},}\ }\href@noop {}
  {\bibfield  {journal} {\bibinfo  {journal} {Phys. Rev. A}\ }\textbf {\bibinfo
  {volume} {82}},\ \bibinfo {pages} {013824} (\bibinfo {year}
  {2010})}\BibitemShut {NoStop}%
\bibitem [{\citenamefont {Bartkowiak}\ \emph {et~al.}(2011)\citenamefont
  {Bartkowiak}, \citenamefont {Miranowicz}, \citenamefont {Wang}, \citenamefont
  {Liu}, \citenamefont {Leo\ifmmode~\acute{n}\else \'{n}\fi{}ski},\ and\
  \citenamefont {Nori}}]{Bartkowiak11}%
  \BibitemOpen
  \bibfield  {author} {\bibinfo {author} {\bibfnamefont {M.}~\bibnamefont
  {Bartkowiak}}, \bibinfo {author} {\bibfnamefont {A.}~\bibnamefont
  {Miranowicz}}, \bibinfo {author} {\bibfnamefont {X.}~\bibnamefont {Wang}},
  \bibinfo {author} {\bibfnamefont {Y.~X.}\ \bibnamefont {Liu}}, \bibinfo
  {author} {\bibfnamefont {W.}~\bibnamefont {Leo\ifmmode~\acute{n}\else
  \'{n}\fi{}ski}}, \ and\ \bibinfo {author} {\bibfnamefont {F.}~\bibnamefont
  {Nori}},\ }\bibfield  {title} {\enquote {\bibinfo {title} {Sudden vanishing
  and reappearance of nonclassical effects: General occurrence of finite-time
  decays and periodic vanishings of nonclassicality and entanglement
  witnesses},}\ }\href {\doibase 10.1103/PhysRevA.83.053814} {\bibfield
  {journal} {\bibinfo  {journal} {Phys. Rev. A}\ }\textbf {\bibinfo {volume}
  {83}},\ \bibinfo {pages} {053814} (\bibinfo {year} {2011})}\BibitemShut
  {NoStop}%
\bibitem [{\citenamefont {Allevi}\ \emph {et~al.}(2013)\citenamefont {Allevi},
  \citenamefont {Lamperti}, \citenamefont {Bondani}, \citenamefont
  {{Pe\v{r}ina~Jr.}}, \citenamefont {Mich\'alek}, \citenamefont {Haderka},\
  and\ \citenamefont {Machulka}}]{Allevi2013}%
  \BibitemOpen
  \bibfield  {author} {\bibinfo {author} {\bibfnamefont {A.}~\bibnamefont
  {Allevi}}, \bibinfo {author} {\bibfnamefont {M.}~\bibnamefont {Lamperti}},
  \bibinfo {author} {\bibfnamefont {M.}~\bibnamefont {Bondani}}, \bibinfo
  {author} {\bibfnamefont {J.}~\bibnamefont {{Pe\v{r}ina~Jr.}}}, \bibinfo
  {author} {\bibfnamefont {V.}~\bibnamefont {Mich\'alek}}, \bibinfo {author}
  {\bibfnamefont {O.}~\bibnamefont {Haderka}}, \ and\ \bibinfo {author}
  {\bibfnamefont {R.}~\bibnamefont {Machulka}},\ }\bibfield  {title} {\enquote
  {\bibinfo {title} {Characterizing the nonclassicality of mesoscopic optical
  twin-beam states},}\ }\href@noop {} {\bibfield  {journal} {\bibinfo
  {journal} {Phys. Rev. A}\ }\textbf {\bibinfo {volume} {88}},\ \bibinfo
  {pages} {063807} (\bibinfo {year} {2013})}\BibitemShut {NoStop}%
\bibitem [{\citenamefont {Vogel}(2008)}]{Vogel08}%
  \BibitemOpen
  \bibfield  {author} {\bibinfo {author} {\bibfnamefont {W.}~\bibnamefont
  {Vogel}},\ }\bibfield  {title} {\enquote {\bibinfo {title} {Nonclassical
  correlation properties of radiation fields},}\ }\href@noop {} {\bibfield
  {journal} {\bibinfo  {journal} {Phys. Rev. Lett.}\ }\textbf {\bibinfo
  {volume} {100}},\ \bibinfo {eid} {013605} (\bibinfo {year}
  {2008})}\BibitemShut {NoStop}%
\bibitem [{\citenamefont {Ryl}\ \emph {et~al.}(2015)\citenamefont {Ryl},
  \citenamefont {Sperling}, \citenamefont {Agudelo}, \citenamefont {Mraz},
  \citenamefont {K{\"o}hnke}, \citenamefont {Hage},\ and\ \citenamefont
  {Vogel}}]{Ryl2015}%
  \BibitemOpen
  \bibfield  {author} {\bibinfo {author} {\bibfnamefont {S.}~\bibnamefont
  {Ryl}}, \bibinfo {author} {\bibfnamefont {J.}~\bibnamefont {Sperling}},
  \bibinfo {author} {\bibfnamefont {E.}~\bibnamefont {Agudelo}}, \bibinfo
  {author} {\bibfnamefont {M.}~\bibnamefont {Mraz}}, \bibinfo {author}
  {\bibfnamefont {S.}~\bibnamefont {K{\"o}hnke}}, \bibinfo {author}
  {\bibfnamefont {B.}~\bibnamefont {Hage}}, \ and\ \bibinfo {author}
  {\bibfnamefont {W.}~\bibnamefont {Vogel}},\ }\bibfield  {title} {\enquote
  {\bibinfo {title} {Unified nonclassicality criteria},}\ }\href@noop {}
  {\bibfield  {journal} {\bibinfo  {journal} {Phys. Rev. A}\ }\textbf {\bibinfo
  {volume} {92}},\ \bibinfo {eid} {011801(R)} (\bibinfo {year}
  {2015})}\BibitemShut {NoStop}%
\bibitem [{\citenamefont {Haderka}\ \emph {et~al.}(2005)\citenamefont
  {Haderka}, \citenamefont {{Pe\v{r}ina~Jr.}}, \citenamefont {Hamar},\ and\
  \citenamefont {Pe\v{r}ina}}]{Haderka2005a}%
  \BibitemOpen
  \bibfield  {author} {\bibinfo {author} {\bibfnamefont {O.}~\bibnamefont
  {Haderka}}, \bibinfo {author} {\bibfnamefont {J.}~\bibnamefont
  {{Pe\v{r}ina~Jr.}}}, \bibinfo {author} {\bibfnamefont {M.}~\bibnamefont
  {Hamar}}, \ and\ \bibinfo {author} {\bibfnamefont {J.}~\bibnamefont
  {Pe\v{r}ina}},\ }\bibfield  {title} {\enquote {\bibinfo {title} {Direct
  measurement and reconstruction of nonclassical features of twin beams
  generated in spontaneous parametric down-conversion},}\ }\href@noop {}
  {\bibfield  {journal} {\bibinfo  {journal} {Phys. Rev. A}\ }\textbf {\bibinfo
  {volume} {71}},\ \bibinfo {pages} {033815} (\bibinfo {year}
  {2005})}\BibitemShut {NoStop}%
\bibitem [{\citenamefont {{Pe\v{r}ina~Jr.}}\ \emph {et~al.}(2012)\citenamefont
  {{Pe\v{r}ina~Jr.}}, \citenamefont {Hamar}, \citenamefont {Mich\'{a}lek},\
  and\ \citenamefont {Haderka}}]{PerinaJr2012}%
  \BibitemOpen
  \bibfield  {author} {\bibinfo {author} {\bibfnamefont {J.}~\bibnamefont
  {{Pe\v{r}ina~Jr.}}}, \bibinfo {author} {\bibfnamefont {M.}~\bibnamefont
  {Hamar}}, \bibinfo {author} {\bibfnamefont {V.}~\bibnamefont {Mich\'{a}lek}},
  \ and\ \bibinfo {author} {\bibfnamefont {O.}~\bibnamefont {Haderka}},\
  }\bibfield  {title} {\enquote {\bibinfo {title} {Photon-number distributions
  of twin beams generated in spontaneous parametric down-conversion and
  measured by an intensified {CCD} camera},}\ }\href@noop {} {\bibfield
  {journal} {\bibinfo  {journal} {Phys. Rev. A}\ }\textbf {\bibinfo {volume}
  {85}},\ \bibinfo {pages} {023816} (\bibinfo {year} {2012})}\BibitemShut
  {NoStop}%
\bibitem [{\citenamefont {{Pe\v{r}ina~Jr.}}\ \emph {et~al.}(2013)\citenamefont
  {{Pe\v{r}ina~Jr.}}, \citenamefont {Haderka}, \citenamefont {Mich\'{a}lek},\
  and\ \citenamefont {Hamar}}]{PerinaJr2013a}%
  \BibitemOpen
  \bibfield  {author} {\bibinfo {author} {\bibfnamefont {J.}~\bibnamefont
  {{Pe\v{r}ina~Jr.}}}, \bibinfo {author} {\bibfnamefont {O.}~\bibnamefont
  {Haderka}}, \bibinfo {author} {\bibfnamefont {V.}~\bibnamefont
  {Mich\'{a}lek}}, \ and\ \bibinfo {author} {\bibfnamefont {M.}~\bibnamefont
  {Hamar}},\ }\bibfield  {title} {\enquote {\bibinfo {title} {State
  reconstruction of a multimode twin beam using photodetection},}\ }\href@noop
  {} {\bibfield  {journal} {\bibinfo  {journal} {Phys. Rev. A}\ }\textbf
  {\bibinfo {volume} {87}},\ \bibinfo {pages} {022108} (\bibinfo {year}
  {2013})}\BibitemShut {NoStop}%
\bibitem [{\citenamefont {Lee}(1990)}]{Lee1990a}%
  \BibitemOpen
  \bibfield  {author} {\bibinfo {author} {\bibfnamefont {C.~T.}\ \bibnamefont
  {Lee}},\ }\bibfield  {title} {\enquote {\bibinfo {title} {Higher-order
  criteria for nonclassical effects in photon statistics},}\ }\href@noop {}
  {\bibfield  {journal} {\bibinfo  {journal} {Phys. Rev. A}\ }\textbf {\bibinfo
  {volume} {41}},\ \bibinfo {pages} {1721---1723} (\bibinfo {year}
  {1990})}\BibitemShut {NoStop}%
\bibitem [{\citenamefont {Verma}\ and\ \citenamefont {Pathak}(2010)}]{Verma10}%
  \BibitemOpen
  \bibfield  {author} {\bibinfo {author} {\bibfnamefont {A.}~\bibnamefont
  {Verma}}\ and\ \bibinfo {author} {\bibfnamefont {A.}~\bibnamefont {Pathak}},\
  }\bibfield  {title} {\enquote {\bibinfo {title} {Generalized structure of
  higher order nonclassicality},}\ }\href@noop {} {\bibfield  {journal}
  {\bibinfo  {journal} {Phys. Lett. A}\ }\textbf {\bibinfo {volume} {374}},\
  \bibinfo {pages} {1009} (\bibinfo {year} {2010})}\BibitemShut {NoStop}%
\bibitem [{\citenamefont {Hong}\ \emph {et~al.}(1987)\citenamefont {Hong},
  \citenamefont {Ou},\ and\ \citenamefont {Mandel}}]{Hong1987}%
  \BibitemOpen
  \bibfield  {author} {\bibinfo {author} {\bibfnamefont {C.~K.}\ \bibnamefont
  {Hong}}, \bibinfo {author} {\bibfnamefont {Z.~Y.}\ \bibnamefont {Ou}}, \ and\
  \bibinfo {author} {\bibfnamefont {L.}~\bibnamefont {Mandel}},\ }\bibfield
  {title} {\enquote {\bibinfo {title} {Measurement of subpicosecond time
  intervals between two photons by interference},}\ }\href@noop {} {\bibfield
  {journal} {\bibinfo  {journal} {Phys. Rev. Lett}\ }\textbf {\bibinfo {volume}
  {59}},\ \bibinfo {eid} {2044} (\bibinfo {year} {1987})}\BibitemShut {NoStop}%
\bibitem [{\citenamefont {Braunstein}(2005)}]{Braunstein05}%
  \BibitemOpen
  \bibfield  {author} {\bibinfo {author} {\bibfnamefont {S.~L.}\ \bibnamefont
  {Braunstein}},\ }\bibfield  {title} {\enquote {\bibinfo {title} {Squeezing as
  an irreducible resource},}\ }\href@noop {} {\bibfield  {journal} {\bibinfo
  {journal} {Phys. Rev. A}\ }\textbf {\bibinfo {volume} {71}},\ \bibinfo {eid}
  {055801} (\bibinfo {year} {2005})}\BibitemShut {NoStop}%
\bibitem [{\citenamefont {Adesso}\ \emph {et~al.}(2006)\citenamefont {Adesso},
  \citenamefont {Serafini},\ and\ \citenamefont {Illuminati}}]{Adesso06}%
  \BibitemOpen
  \bibfield  {author} {\bibinfo {author} {\bibfnamefont {G.}~\bibnamefont
  {Adesso}}, \bibinfo {author} {\bibfnamefont {A.}~\bibnamefont {Serafini}}, \
  and\ \bibinfo {author} {\bibfnamefont {F.}~\bibnamefont {Illuminati}},\
  }\bibfield  {title} {\enquote {\bibinfo {title} {Multipartite entanglement in
  three-mode {G}aussian states of continuous-variable systems: Quantification,
  sharing structure, and decoherence},}\ }\href@noop {} {\bibfield  {journal}
  {\bibinfo  {journal} {Phys. Rev. A}\ }\textbf {\bibinfo {volume} {73}},\
  \bibinfo {pages} {032345} (\bibinfo {year} {2006})}\BibitemShut {NoStop}%
\bibitem [{\citenamefont {Vidal}\ and\ \citenamefont {Werner}(2002)}]{Vidal02}%
  \BibitemOpen
  \bibfield  {author} {\bibinfo {author} {\bibfnamefont {G.}~\bibnamefont
  {Vidal}}\ and\ \bibinfo {author} {\bibfnamefont {R.~F.}\ \bibnamefont
  {Werner}},\ }\bibfield  {title} {\enquote {\bibinfo {title} {Computable
  measure of entanglement},}\ }\href@noop {} {\bibfield  {journal} {\bibinfo
  {journal} {Phys. Rev. A}\ }\textbf {\bibinfo {volume} {65}},\ \bibinfo
  {pages} {032314} (\bibinfo {year} {2002})}\BibitemShut {NoStop}%
\bibitem [{\citenamefont {Plenio}(2005)}]{Plenio05}%
  \BibitemOpen
  \bibfield  {author} {\bibinfo {author} {\bibfnamefont {M.~B.}\ \bibnamefont
  {Plenio}},\ }\bibfield  {title} {\enquote {\bibinfo {title} {Logarithmic
  negativity: A full entanglement monotone that is not convex},}\ }\href
  {\doibase 10.1103/PhysRevLett.95.090503} {\bibfield  {journal} {\bibinfo
  {journal} {Phys. Rev. Lett.}\ }\textbf {\bibinfo {volume} {95}},\ \bibinfo
  {pages} {090503} (\bibinfo {year} {2005})}\BibitemShut {NoStop}%
\bibitem [{\citenamefont {Horodecki}\ \emph {et~al.}(2009)\citenamefont
  {Horodecki}, \citenamefont {Horodecki}, \citenamefont {Horodecki},\ and\
  \citenamefont {Horodecki}}]{Horodecki09review}%
  \BibitemOpen
  \bibfield  {author} {\bibinfo {author} {\bibfnamefont {R.}~\bibnamefont
  {Horodecki}}, \bibinfo {author} {\bibfnamefont {P.}~\bibnamefont
  {Horodecki}}, \bibinfo {author} {\bibfnamefont {M.}~\bibnamefont
  {Horodecki}}, \ and\ \bibinfo {author} {\bibfnamefont {K.}~\bibnamefont
  {Horodecki}},\ }\bibfield  {title} {\enquote {\bibinfo {title} {Quantum
  entanglement},}\ }\href@noop {} {\bibfield  {journal} {\bibinfo  {journal}
  {Rev. Mod. Phys.}\ }\textbf {\bibinfo {volume} {81}},\ \bibinfo {eid} {865}
  (\bibinfo {year} {2009})}\BibitemShut {NoStop}%
\bibitem [{\citenamefont {Peres}(1996)}]{Peres96}%
  \BibitemOpen
  \bibfield  {author} {\bibinfo {author} {\bibfnamefont {Asher}\ \bibnamefont
  {Peres}},\ }\bibfield  {title} {\enquote {\bibinfo {title} {Separability
  criterion for density matrices},}\ }\href@noop {} {\bibfield  {journal}
  {\bibinfo  {journal} {Phys. Rev. Lett}\ }\textbf {\bibinfo {volume} {77}},\
  \bibinfo {eid} {1413} (\bibinfo {year} {1996})}\BibitemShut {NoStop}%
\bibitem [{\citenamefont {Horodecki}(1997)}]{Horodecki97}%
  \BibitemOpen
  \bibfield  {author} {\bibinfo {author} {\bibfnamefont {P.}~\bibnamefont
  {Horodecki}},\ }\bibfield  {title} {\enquote {\bibinfo {title} {Separability
  criterion and inseparable mixed states with positive partial
  transposition},}\ }\href@noop {} {\bibfield  {journal} {\bibinfo  {journal}
  {Phys. Lett. A}\ }\textbf {\bibinfo {volume} {232}},\ \bibinfo {eid} {333}
  (\bibinfo {year} {1997})}\BibitemShut {NoStop}%
\bibitem [{\citenamefont {Marian}\ and\ \citenamefont
  {Marian}(2008)}]{Marian08}%
  \BibitemOpen
  \bibfield  {author} {\bibinfo {author} {\bibfnamefont {P.}~\bibnamefont
  {Marian}}\ and\ \bibinfo {author} {\bibfnamefont {T.~A.}\ \bibnamefont
  {Marian}},\ }\bibfield  {title} {\enquote {\bibinfo {title} {Bures distance
  as a measure of entanglement for symmetric two-mode {G}aussian states},}\
  }\href@noop {} {\bibfield  {journal} {\bibinfo  {journal} {Phys. Rev. A}\
  }\textbf {\bibinfo {volume} {77}},\ \bibinfo {eid} {062319} (\bibinfo {year}
  {2008})}\BibitemShut {NoStop}%
\bibitem [{\citenamefont {Asb\'oth}\ \emph {et~al.}(2005)\citenamefont
  {Asb\'oth}, \citenamefont {Calsamiglia},\ and\ \citenamefont
  {Ritsch}}]{Asboth05}%
  \BibitemOpen
  \bibfield  {author} {\bibinfo {author} {\bibfnamefont {J.~K.}\ \bibnamefont
  {Asb\'oth}}, \bibinfo {author} {\bibfnamefont {J.}~\bibnamefont
  {Calsamiglia}}, \ and\ \bibinfo {author} {\bibfnamefont {H.}~\bibnamefont
  {Ritsch}},\ }\bibfield  {title} {\enquote {\bibinfo {title} {Squeezing as an
  irreducible resource},}\ }\href@noop {} {\bibfield  {journal} {\bibinfo
  {journal} {Phys. Rev. Lett.}\ }\textbf {\bibinfo {volume} {94}},\ \bibinfo
  {eid} {173602} (\bibinfo {year} {2005})}\BibitemShut {NoStop}%
\bibitem [{\citenamefont {Brunelli}\ \emph {et~al.}(2015)\citenamefont
  {Brunelli}, \citenamefont {Benedetti}, \citenamefont {Olivares},
  \citenamefont {Ferraro},\ and\ \citenamefont {Paris}}]{Brunelli15}%
  \BibitemOpen
  \bibfield  {author} {\bibinfo {author} {\bibfnamefont {M.}~\bibnamefont
  {Brunelli}}, \bibinfo {author} {\bibfnamefont {C.}~\bibnamefont {Benedetti}},
  \bibinfo {author} {\bibfnamefont {S.}~\bibnamefont {Olivares}}, \bibinfo
  {author} {\bibfnamefont {A.}~\bibnamefont {Ferraro}}, \ and\ \bibinfo
  {author} {\bibfnamefont {M.~G.~A.}\ \bibnamefont {Paris}},\ }\bibfield
  {title} {\enquote {\bibinfo {title} {Single- and two-mode quantumness at a
  beam splitter},}\ }\href@noop {} {\bibfield  {journal} {\bibinfo  {journal}
  {Phys. Rev. A}\ }\textbf {\bibinfo {volume} {91}},\ \bibinfo {pages} {062315}
  (\bibinfo {year} {2015})}\BibitemShut {NoStop}%
\bibitem [{\citenamefont {Miranowicz}\ \emph
  {et~al.}(2015{\natexlab{a}})\citenamefont {Miranowicz}, \citenamefont
  {Bartkiewicz}, \citenamefont {Pathak}, \citenamefont {Pe\v{r}ina},
  \citenamefont {Chen},\ and\ \citenamefont {Nori}}]{Miran15a}%
  \BibitemOpen
  \bibfield  {author} {\bibinfo {author} {\bibfnamefont {A.}~\bibnamefont
  {Miranowicz}}, \bibinfo {author} {\bibfnamefont {K.}~\bibnamefont
  {Bartkiewicz}}, \bibinfo {author} {\bibfnamefont {A.}~\bibnamefont {Pathak}},
  \bibinfo {author} {\bibfnamefont {J.}~\bibnamefont {Pe\v{r}ina}}, \bibinfo
  {author} {\bibfnamefont {Y.N.}\ \bibnamefont {Chen}}, \ and\ \bibinfo
  {author} {\bibfnamefont {F.}~\bibnamefont {Nori}},\ }\bibfield  {title}
  {\enquote {\bibinfo {title} {Statistical mixtures of states can be more
  quantum than their superpositions: {C}omparison of nonclassicality measures
  for single-qubit states},}\ }\href {\doibase 10.1103/PhysRevA.91.042309}
  {\bibfield  {journal} {\bibinfo  {journal} {Phys. Rev. A}\ }\textbf {\bibinfo
  {volume} {91}},\ \bibinfo {pages} {042309} (\bibinfo {year}
  {2015}{\natexlab{a}})}\BibitemShut {NoStop}%
\bibitem [{\citenamefont {Miranowicz}\ \emph
  {et~al.}(2015{\natexlab{b}})\citenamefont {Miranowicz}, \citenamefont
  {Bartkiewicz}, \citenamefont {Lambert}, \citenamefont {Chen},\ and\
  \citenamefont {Nori}}]{Miran15b}%
  \BibitemOpen
  \bibfield  {author} {\bibinfo {author} {\bibfnamefont {A.}~\bibnamefont
  {Miranowicz}}, \bibinfo {author} {\bibfnamefont {K.}~\bibnamefont
  {Bartkiewicz}}, \bibinfo {author} {\bibfnamefont {N.}~\bibnamefont
  {Lambert}}, \bibinfo {author} {\bibfnamefont {Y.N.}\ \bibnamefont {Chen}}, \
  and\ \bibinfo {author} {\bibfnamefont {F.}~\bibnamefont {Nori}},\ }\bibfield
  {title} {\enquote {\bibinfo {title} {Increasing relative nonclassicality
  quantified by standard entanglement potentials by dissipation and unbalanced
  beam splitting},}\ }\href {\doibase 10.1103/PhysRevA.92.062314} {\bibfield
  {journal} {\bibinfo  {journal} {Phys. Rev. A}\ }\textbf {\bibinfo {volume}
  {92}},\ \bibinfo {pages} {062314} (\bibinfo {year}
  {2015}{\natexlab{b}})}\BibitemShut {NoStop}%
\bibitem [{\citenamefont {Arkhipov}\ \emph {et~al.}(2015)\citenamefont
  {Arkhipov}, \citenamefont {Jr.}, \citenamefont {Pe\v{r}ina},\ and\
  \citenamefont {Miranowicz}}]{arkhipov15}%
  \BibitemOpen
  \bibfield  {author} {\bibinfo {author} {\bibfnamefont {I.~I.}\ \bibnamefont
  {Arkhipov}}, \bibinfo {author} {\bibfnamefont {J.~Pe\v{r}ina}\ \bibnamefont
  {Jr.}}, \bibinfo {author} {\bibfnamefont {J.}~\bibnamefont {Pe\v{r}ina}}, \
  and\ \bibinfo {author} {\bibfnamefont {A.}~\bibnamefont {Miranowicz}},\
  }\bibfield  {title} {\enquote {\bibinfo {title} {Comparative study of
  nonclassicality, entanglement, and dimensionality of multimode noisy twin
  beams},}\ }\href@noop {} {\bibfield  {journal} {\bibinfo  {journal} {Phys.
  Rev. A}\ }\textbf {\bibinfo {volume} {91}},\ \bibinfo {eid} {033837}
  (\bibinfo {year} {2015})}\BibitemShut {NoStop}%
\bibitem [{\citenamefont {Vogel}\ and\ \citenamefont
  {Sperling}(2014)}]{Vogel14}%
  \BibitemOpen
  \bibfield  {author} {\bibinfo {author} {\bibfnamefont {W.}~\bibnamefont
  {Vogel}}\ and\ \bibinfo {author} {\bibfnamefont {J.}~\bibnamefont
  {Sperling}},\ }\bibfield  {title} {\enquote {\bibinfo {title} {Quantum optics
  in the phase space},}\ }\href@noop {} {\bibfield  {journal} {\bibinfo
  {journal} {Phys. Rev. A}\ }\textbf {\bibinfo {volume} {89}},\ \bibinfo {eid}
  {052302} (\bibinfo {year} {2014})}\BibitemShut {NoStop}%
\bibitem [{\citenamefont {Ge}\ \emph {et~al.}(2015)\citenamefont {Ge},
  \citenamefont {Tasgin},\ and\ \citenamefont {Zubairy}}]{Zubairy15}%
  \BibitemOpen
  \bibfield  {author} {\bibinfo {author} {\bibfnamefont {W.}~\bibnamefont
  {Ge}}, \bibinfo {author} {\bibfnamefont {M.~E.}\ \bibnamefont {Tasgin}}, \
  and\ \bibinfo {author} {\bibfnamefont {M.~S.}\ \bibnamefont {Zubairy}},\
  }\bibfield  {title} {\enquote {\bibinfo {title} {Conservation relation of
  nonclassicality and entanglement for {G}aussian states in a beam splitter},}\
  }\href {\doibase 10.1103/PhysRevA.92.052328} {\bibfield  {journal} {\bibinfo
  {journal} {Phys. Rev. A}\ }\textbf {\bibinfo {volume} {92}},\ \bibinfo
  {pages} {052328} (\bibinfo {year} {2015})}\BibitemShut {NoStop}%
\bibitem [{\citenamefont {Arkhipov}\ \emph {et~al.}(2016)\citenamefont
  {Arkhipov}, \citenamefont {{Pe\v{r}ina Jr.}}, \citenamefont {Svozil\'ik},\
  and\ \citenamefont {Miranowicz}}]{arkhipov_letter16}%
  \BibitemOpen
  \bibfield  {author} {\bibinfo {author} {\bibfnamefont {I.}~\bibnamefont
  {Arkhipov}}, \bibinfo {author} {\bibfnamefont {J.}~\bibnamefont {{Pe\v{r}ina
  Jr.}}}, \bibinfo {author} {\bibfnamefont {J.}~\bibnamefont {Svozil\'ik}}, \
  and\ \bibinfo {author} {\bibfnamefont {A.}~\bibnamefont {Miranowicz}},\
  }\bibfield  {title} {\enquote {\bibinfo {title} {Nonclassicality invariant of
  general two-mode {G}aussian states},}\ }\href
  {http://arxiv.org/abs/1601.04868} {\bibfield  {journal} {\bibinfo  {journal}
  {Sci. Rep.}\ }\textbf {\bibinfo {volume} {6}},\ \bibinfo {pages} {26523}
  (\bibinfo {year} {2016})}\BibitemShut {NoStop}%
\bibitem [{\citenamefont {Nielsen}\ and\ \citenamefont
  {Chuang}(2000)}]{NielsenBook}%
  \BibitemOpen
  \bibfield  {author} {\bibinfo {author} {\bibfnamefont {M.~A.}\ \bibnamefont
  {Nielsen}}\ and\ \bibinfo {author} {\bibfnamefont {I.~L.}\ \bibnamefont
  {Chuang}},\ }\href@noop {} {\emph {\bibinfo {title} {Quantum Computation and
  Quantum Information}}}\ (\bibinfo  {publisher} {Cambridge University Press},\
  \bibinfo {address} {Cambridge},\ \bibinfo {year} {2000})\BibitemShut
  {NoStop}%
\bibitem [{\citenamefont {Polzik}\ \emph {et~al.}(1992)\citenamefont {Polzik},
  \citenamefont {Carri},\ and\ \citenamefont {Kimble}}]{Polzik92}%
  \BibitemOpen
  \bibfield  {author} {\bibinfo {author} {\bibfnamefont {E.~S.}\ \bibnamefont
  {Polzik}}, \bibinfo {author} {\bibfnamefont {J.}~\bibnamefont {Carri}}, \
  and\ \bibinfo {author} {\bibfnamefont {H.~J.}\ \bibnamefont {Kimble}},\
  }\bibfield  {title} {\enquote {\bibinfo {title} {Spectroscopy with squeezed
  light},}\ }\href@noop {} {\bibfield  {journal} {\bibinfo  {journal} {Phys.
  Rev. Lett.}\ }\textbf {\bibinfo {volume} {68}},\ \bibinfo {eid} {3020}
  (\bibinfo {year} {1992})}\BibitemShut {NoStop}%
\bibitem [{\citenamefont {Ralph}\ and\ \citenamefont {Lam}(1998)}]{Ralph98}%
  \BibitemOpen
  \bibfield  {author} {\bibinfo {author} {\bibfnamefont {T.~C.}\ \bibnamefont
  {Ralph}}\ and\ \bibinfo {author} {\bibfnamefont {P.~K.}\ \bibnamefont
  {Lam}},\ }\bibfield  {title} {\enquote {\bibinfo {title} {Teleportation with
  bright squeezed light},}\ }\href@noop {} {\bibfield  {journal} {\bibinfo
  {journal} {Phys. Rev. Lett.}\ }\textbf {\bibinfo {volume} {81}},\ \bibinfo
  {eid} {5668} (\bibinfo {year} {1998})}\BibitemShut {NoStop}%
\bibitem [{\citenamefont {Wolfgramm}\ \emph {et~al.}(2010)\citenamefont
  {Wolfgramm}, \citenamefont {Cer\'e}, \citenamefont {Beduini}, \citenamefont
  {Predojevi\'c}, \citenamefont {Koschorreck},\ and\ \citenamefont
  {Mitchell}}]{Wolfgramm10}%
  \BibitemOpen
  \bibfield  {author} {\bibinfo {author} {\bibfnamefont {F.}~\bibnamefont
  {Wolfgramm}}, \bibinfo {author} {\bibfnamefont {A.}~\bibnamefont {Cer\'e}},
  \bibinfo {author} {\bibfnamefont {F.~A.}\ \bibnamefont {Beduini}}, \bibinfo
  {author} {\bibfnamefont {A.}~\bibnamefont {Predojevi\'c}}, \bibinfo {author}
  {\bibfnamefont {M.}~\bibnamefont {Koschorreck}}, \ and\ \bibinfo {author}
  {\bibfnamefont {M.~W.}\ \bibnamefont {Mitchell}},\ }\bibfield  {title}
  {\enquote {\bibinfo {title} {Squeezed-light optical magnetometry},}\
  }\href@noop {} {\bibfield  {journal} {\bibinfo  {journal} {Phys. Rev. Lett.}\
  }\textbf {\bibinfo {volume} {105}},\ \bibinfo {eid} {053601} (\bibinfo {year}
  {2010})}\BibitemShut {NoStop}%
\bibitem [{\citenamefont {{J. Pe\v rina, Jr.}}\ and\ \citenamefont {{J. Pe\v
  rina}}(2000)}]{PerinaJr2000}%
  \BibitemOpen
  \bibfield  {author} {\bibinfo {author} {\bibnamefont {{J. Pe\v rina, Jr.}}}\
  and\ \bibinfo {author} {\bibnamefont {{J. Pe\v rina}}},\ }\bibfield  {title}
  {\enquote {\bibinfo {title} {Quantum statistics of nonlinear optical
  couplers},}\ }\href@noop {} {\bibfield  {journal} {\bibinfo  {journal} {Prog.
  Opt.}\ }\textbf {\bibinfo {volume} {41}},\ \bibinfo {eid} {361} (\bibinfo
  {year} {2000})}\BibitemShut {NoStop}%
\bibitem [{\citenamefont {{Pe\v{r}ina}}\ and\ \citenamefont
  {{K\v{r}epelka}}(2011)}]{PerinaKrepelka11}%
  \BibitemOpen
  \bibfield  {author} {\bibinfo {author} {\bibfnamefont {J.}~\bibnamefont
  {{Pe\v{r}ina}}}\ and\ \bibinfo {author} {\bibfnamefont {J.}~\bibnamefont
  {{K\v{r}epelka}}},\ }\bibfield  {title} {\enquote {\bibinfo {title} {Joint
  probability distribution and entanglement in optical parametric processes},}\
  }\href@noop {} {\bibfield  {journal} {\bibinfo  {journal} {Opt. Commun.}\
  }\textbf {\bibinfo {volume} {284}},\ \bibinfo {pages} {4941} (\bibinfo {year}
  {2011})}\BibitemShut {NoStop}%
\bibitem [{\citenamefont {Simon}(2000)}]{Simon00}%
  \BibitemOpen
  \bibfield  {author} {\bibinfo {author} {\bibfnamefont {R.}~\bibnamefont
  {Simon}},\ }\bibfield  {title} {\enquote {\bibinfo {title} {Peres-{H}orodecki
  separability criterion for continuous variable systems},}\ }\href@noop {}
  {\bibfield  {journal} {\bibinfo  {journal} {Phys. Rev. Lett}\ }\textbf
  {\bibinfo {volume} {84}},\ \bibinfo {eid} {2726} (\bibinfo {year}
  {2000})}\BibitemShut {NoStop}%
\bibitem [{\citenamefont {Marian}\ \emph {et~al.}(2001)\citenamefont {Marian},
  \citenamefont {Marian},\ and\ \citenamefont {Scutaru}}]{Marian01}%
  \BibitemOpen
  \bibfield  {author} {\bibinfo {author} {\bibfnamefont {P.}~\bibnamefont
  {Marian}}, \bibinfo {author} {\bibfnamefont {T.~A}\ \bibnamefont {Marian}}, \
  and\ \bibinfo {author} {\bibfnamefont {H.}~\bibnamefont {Scutaru}},\
  }\bibfield  {title} {\enquote {\bibinfo {title} {Inseparability of mixed
  two-mode {G}aussian states generated with a {SU} (1,1) interferometer},}\
  }\href {http://stacks.iop.org/0305-4470/34/i=35/a=322} {\bibfield  {journal}
  {\bibinfo  {journal} {Journal of Physics A: Mathematical and General}\
  }\textbf {\bibinfo {volume} {34}},\ \bibinfo {pages} {6969} (\bibinfo {year}
  {2001})}\BibitemShut {NoStop}%
\bibitem [{\citenamefont {Olivares}(2012)}]{Olivares12}%
  \BibitemOpen
  \bibfield  {author} {\bibinfo {author} {\bibfnamefont {S.}~\bibnamefont
  {Olivares}},\ }\bibfield  {title} {\enquote {\bibinfo {title} {Quantum optics
  in the phase space},}\ }\href@noop {} {\bibfield  {journal} {\bibinfo
  {journal} {Eur. Phys. J. Special Topics}\ }\textbf {\bibinfo {volume}
  {203}},\ \bibinfo {eid} {3} (\bibinfo {year} {2012})}\BibitemShut {NoStop}%
\bibitem [{\citenamefont {Klyshko}(1994)}]{Klyshko1994}%
  \BibitemOpen
  \bibfield  {author} {\bibinfo {author} {\bibfnamefont {D.~N.}\ \bibnamefont
  {Klyshko}},\ }\bibfield  {title} {\enquote {\bibinfo {title} {Quantum optics:
  {Q}uantum, classical, and metaphysical aspects},}\ }\href@noop {} {\bibfield
  {journal} {\bibinfo  {journal} {Physics-Uspekhi}\ }\textbf {\bibinfo {volume}
  {37}},\ \bibinfo {eid} {1097} (\bibinfo {year} {1994})}\BibitemShut {NoStop}%
\bibitem [{\citenamefont {Iskhakov}\ \emph {et~al.}(2013)\citenamefont
  {Iskhakov}, \citenamefont {Spasibko},\ and\ \citenamefont
  {Chekhova}}]{Iskhakov13}%
  \BibitemOpen
  \bibfield  {author} {\bibinfo {author} {\bibfnamefont {T.~Sh.}\ \bibnamefont
  {Iskhakov}}, \bibinfo {author} {\bibfnamefont {K.~Yu.}\ \bibnamefont
  {Spasibko}}, \ and\ \bibinfo {author} {\bibfnamefont {M.~V.}\ \bibnamefont
  {Chekhova}},\ }\bibfield  {title} {\enquote {\bibinfo {title} {Macroscopic
  {H}ong-{O}u-{M}andel interference},}\ }\href@noop {} {\bibfield  {journal}
  {\bibinfo  {journal} {New J. Phys.}\ }\textbf {\bibinfo {volume} {15}},\
  \bibinfo {eid} {093036} (\bibinfo {year} {2013})}\BibitemShut {NoStop}%
\bibitem [{\citenamefont {Paris}(1997)}]{Paris97}%
  \BibitemOpen
  \bibfield  {author} {\bibinfo {author} {\bibfnamefont {M.~G.~A.}\
  \bibnamefont {Paris}},\ }\bibfield  {title} {\enquote {\bibinfo {title}
  {Joint generation of identical squeezed states},}\ }\href@noop {} {\bibfield
  {journal} {\bibinfo  {journal} {Phys. Lett. A}\ }\textbf {\bibinfo {volume}
  {225}},\ \bibinfo {eid} {28} (\bibinfo {year} {1997})}\BibitemShut {NoStop}%
\bibitem [{\citenamefont {Luk\v{s}}\ \emph {et~al.}(1988)\citenamefont
  {Luk\v{s}}, \citenamefont {Pe\v{r}inov\'{a}},\ and\ \citenamefont
  {Pe\v{r}ina}}]{Luks1988}%
  \BibitemOpen
  \bibfield  {author} {\bibinfo {author} {\bibfnamefont {A.}~\bibnamefont
  {Luk\v{s}}}, \bibinfo {author} {\bibfnamefont {V.}~\bibnamefont
  {Pe\v{r}inov\'{a}}}, \ and\ \bibinfo {author} {\bibfnamefont
  {J.}~\bibnamefont {Pe\v{r}ina}},\ }\bibfield  {title} {\enquote {\bibinfo
  {title} {Principal squeezing of vacuum fluctuations},}\ }\href@noop {}
  {\bibfield  {journal} {\bibinfo  {journal} {Opt. Commun.}\ }\textbf {\bibinfo
  {volume} {67}},\ \bibinfo {pages} {149} (\bibinfo {year} {1988})}\BibitemShut
  {NoStop}%
\bibitem [{\citenamefont {Cahill}\ and\ \citenamefont
  {Glauber}(1969)}]{Cahill69}%
  \BibitemOpen
  \bibfield  {author} {\bibinfo {author} {\bibfnamefont {K.~E.}\ \bibnamefont
  {Cahill}}\ and\ \bibinfo {author} {\bibfnamefont {R.~J.}\ \bibnamefont
  {Glauber}},\ }\bibfield  {title} {\enquote {\bibinfo {title} {Ordered
  expansions in boson amplitude operators},}\ }\href {\doibase
  10.1103/PhysRev.177.1857} {\bibfield  {journal} {\bibinfo  {journal} {Phys.
  Rev.}\ }\textbf {\bibinfo {volume} {177}},\ \bibinfo {pages} {1857} (\bibinfo
  {year} {1969})}\BibitemShut {NoStop}%
\end{thebibliography}

%

\end{document}